\documentclass{article}

\usepackage[english]{babel} 
\usepackage[T1]{fontenc}
\usepackage[utf8]{inputenc}
\usepackage{lmodern}

\usepackage{csquotes}
\DeclareUnicodeCharacter{2212}{-}

\usepackage[backend=biber,style=phys, eprint=true, doi=false]{biblatex}
\usepackage{amsmath, amsthm, amsfonts, amssymb}
\usepackage{mathtools}
\usepackage{mathrsfs}
\usepackage{dsfont}
\usepackage{physics}
\usepackage{braket}
\usepackage{mathdots}
\usepackage{yhmath}
\usepackage{cancel}
\usepackage{tensor}
\usepackage{esvect}

\usepackage[a4paper, total={6in, 8in}]{geometry}
\usepackage{float}
\usepackage{indentfirst}
\usepackage{enumerate}
\usepackage{verbatim}
\usepackage{microtype} 
\usepackage{relsize}
\usepackage{tocloft}
\usepackage{pdflscape}
\usepackage{adjustbox}
\usepackage[normalem]{ulem}
\usepackage{caption}
\usepackage{url}
\usepackage[misc,clock,geometry]{ifsym} 

\usepackage{array}
\usepackage{multirow}
\usepackage{tabularx}
\usepackage{booktabs}
\usepackage{extarrows}
\usepackage{xcolor}
\usepackage[most]{tcolorbox}
\definecolor{MyRed}{HTML}{D0021B}

\usepackage{graphicx}
\usepackage{epsfig,epstopdf}
\usepackage[final]{pdfpages}
\usepackage{tikz}
\usetikzlibrary{fadings, patterns, shadows.blur, shapes, arrows}

\usepackage{hyperref}
\hypersetup{
    colorlinks=true, 
    linktoc=all,     
    linkcolor=magenta,  
    citecolor=magenta,
}

\tikzset{mathterm/.style={draw=white,fill=white,rectangle,anchor=base}}
\tikzstyle{every picture}+=[remember picture]
\everymath{\displaystyle}

\makeatletter

\newcommand\indicate[2][black]{%
   \tikz [baseline] \node [inner sep=0pt,anchor=base] (i#2) {\vphantom|};
   \@ifnextchar[{\@indicateopts{#1}{#2}}{\@indicatenoopts{#1}{#2}}}
\def\@indicatenoopts#1#2{%
   {\color{#1} \tikz[overlay] \path[line width=1pt,draw=#1,-stealth] (i#2) edge (#2);}}
\def\@indicateopts#1#2[#3]{%
   {\color{#1} \tikz[overlay] \path[line width=1pt,draw=#1,-stealth] (i#2) [#3] edge (#2);}}
\makeatother

\makeatletter
\AtBeginDocument{\let\LS@rot\@undefined}
\makeatother

\restylefloat*{figure}

\theoremstyle{plain}

\newcommand{\iu}{{\mathrm{i}\mkern1mu}}

\usepackage{orcidlink}
\usepackage{authblk}

\author[1,2]{Bruna Sahdo\thanks{bruna.sahdo@oeaw.ac.at}}
\author[1]{Esteban Castro-Ruiz\thanks{esteban.castroruiz@oeaw.ac.at}}
\affil[1]{Institute for Quantum Optics and Quantum Information (IQOQI), Austrian Academy of Sciences, Boltzmanngasse 3, 1090 Vienna, Austria}
\affil[2]{University of Vienna, Faculty of Physics, Boltzmanngasse 5, A-1090 Vienna, Austria}
\date{}

\begin{document}

\title{\vspace{-2.8cm} Compositionality in quantum reference frame perspectives}

\begin{titlepage}
\thispagestyle{empty}
\maketitle
\noindent Understanding a composite system through its constituents is a fundamental practice in physics. In the context of quantum reference frames (QRFs), however, combining the usual quantum-theoretic notion of composition with QRF perspectives gives rise to subtle issues, such as the `paradox of the third particle'. Here we study in depth how to compose subsystems in QRF perspectives, building on a recent formalism for QRFs~[\href{https://doi.org/10.1038/s42005-025-02036-x}{E.Castro-Ruiz and O.Oreshkov, 2025}]. We first show how the frames of external observers can be internalised and treated within the framework. This establishes a consistent hierarchy of QRF perspectives, which we use to define the adding and removing of subsystems. We then explain how, owing to the so-called `extra-particle' degrees of freedom, the formalism gives a consistent treatment of subsystems, avoiding any paradoxes by construction. Consistency then implies that composing subsystems in a QRF perspective differs from, and generalises, the standard quantum-theoretic case. In particular, not every state can be appended in tensor-product form relative to a QRF, and we characterise the set of compatible states. Finally, we introduce `classicalisation', a procedure that recovers a classical-like perspective from a general QRF. This procedure circumvents the aforementioned state restrictions but carries a different operational meaning, which we study through a concrete example.

\tableofcontents

\end{titlepage}

\makeatletter
\let\toc@pre\relax
\let\toc@post\relax
\makeatother

\section{Introduction}
\noindent Our ability to understand the world hinges on our ability to decompose it. We understand the whole through its parts, and we assemble the parts to reconstruct the whole. Physical theories exploit this at every step: we describe a subsystem without accounting for the rest of the universe, and we combine subsystems into larger ones without having to start the description afresh.

In quantum theory, the compositional structure is encoded in the tensor product of subsystems~\cite{NielsenChuang} and its underlying principles are to this day a research topic in quantum foundations~\cite{Hardy2001,Chiribella2011,Masanes2011,Carcassi2021,Erba2026}. To describe a subsystem in the first place, however, one usually relies, often implicitly, on a reference frame. In practice, reference frames are physical systems like any other, and are therefore subject to quantum theory. Taking this simple observation seriously requires us to formulate the description of physical systems relative to quantum reference frames (QRFs)~\cite{Eddington1949,Aharonov1967,Aharonov1984,Bartlett2007,Angelo2011}. Recent years have seen significant progress in this direction, with particular emphasis on the transformations between different QRFs~\cite{Merriam2005,Giacomini2019,Vanrietvelde2020,Hamette2020,CastroRuiz2020,Hoehn2021,Ballesteros2021,Krumm2021,CastroRuiz2025a,Hamette2021,Glowacki2023,Carette2025,Kabel2025,Garmier2025,Mekonnen2026}.

Nevertheless, composition presents a challenge in the QRF programme: can transformations between QRF perspectives act consistently on arbitrary subsystems, which may be composed of smaller constituents or embedded into larger portions of the universe? In the following we illustrate the subtle issues that may arise when composing and decomposing subsystems in QRFs. Our formulation follows that of Ref.~\cite{CastroRuiz2025a}, and includes an instance of the `paradox of the third particle', first presented in~\cite{Angelo2011} and further studied in~\cite{Krumm2021,CastroRuiz2025a,Renato2026,Brukner2026,Palumbo2026}.

\subsection{The problem of composition} \label{SEC:TheProblem}
\begin{figure}[ht]
    \centering
\includegraphics[scale=0.7]{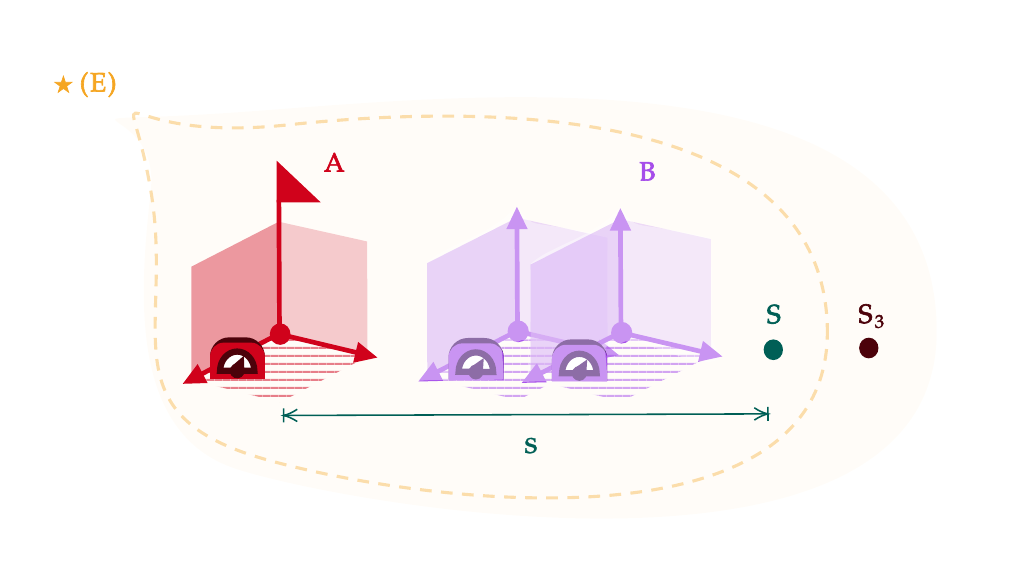}   
\caption{A reduced setup, formed by reference frames $\mathbf{A,B}$ and system $\mathbf{S}$ (inside the dashed contour) is seen as part of an extended setup, containing the additional system $\mathbf{S_3}$. If the internal perspectives of $\mathbf{A}$ and $\mathbf{B}$ include only the relative degrees of freedom, reversible transformations between perspectives do not behave adequately under composition. This generates issues such as the `paradox of the third particle'. The formalism used here restores compositionality due to the `extra particle' degrees of freedom present in each perspective.
}
    \label{FIG:3rdParticleParadox}
\end{figure}

\noindent Consider a collection of quantum systems $\mathbf{A}, \mathbf{B}$ and $\mathbf{S}$, which can be thought of as particles on a line. 
Let Alice, an observer, use $\mathbf{A}$ as a reference frame with respect to which she determines the positions of $\mathbf{B}$ and $\mathbf S$. Likewise, suppose Bob, another observer, uses $\mathbf B$ as his reference frame. Relative to an external observer $(\mathbf{E})$, frame $\mathbf{A}$ is sharply-localised in space (dashed contour of Fig.~\ref{FIG:3rdParticleParadox}), and the state of $\mathbf{B}$ and $\mathbf{S}$ relative to $\mathbf A$ is given by  
\begin{equation}\label{quantumcatBstate}
    \ket{\rho_\text{red}} = \frac{1}{\sqrt{2}}\left(\ket{b_1}+e^{\iu \theta}\ket{b_2}\right)_{\mathbf{B|A}}\otimes\ket{s}_{\mathbf{S|A}} , 
\end{equation}
for positions $s,b_1, b_2 \in\mathbb{R}$, with $b_1 \neq b_2$, and $0\leq\theta< 2\pi$. 

Given the state above, we may ask what is the state of $\mathbf A$ and $\mathbf S$ relative to Bob's frame. According to Refs.~\cite{Giacomini2019,Vanrietvelde2020,Hamette2020}, we may apply a reversible QRF transformation between Alice's description of $\mathbf B$ and $\mathbf S$ and Bob's description of $\mathbf A$ and $\mathbf S$. The resulting state is given by  
\begin{equation}\label{eq:reducedstatewithout}
  \frac{1}{\sqrt{2}}\left(\ket{-b_1}\ket{s-b_1}+ e^{\iu \theta}\ket{-b_2}\ket{s-b_2}\right)_{\mathbf{AS|B}},
\end{equation} 
and the phase $\theta$ is in principle accessible through suitable measurements. 

Suppose that, beyond the reduced setup containing $\mathbf A$, $\mathbf B$ and $\mathbf S$, there exists an independent system $\mathbf{S_3}$ defining an extended setup (outer contour of 
Fig.~\ref{FIG:3rdParticleParadox}). What is the state of $\mathbf{A}$ and $\mathbf S$ relative to $\mathbf{B}$ in this case? Concretely, suppose a state of $\mathbf{S_3}$, sharply localised at position $s_3$ relative to $\mathbf{A}$, is added in tensor-product form (henceforth `tensorised'). The resulting extended state is given by
\begin{equation}\label{eq:extendedstate}
\ket{\rho_\text{ext}} =  \frac{1}{\sqrt{2}}\left(\ket{b_1}+e^{\iu \theta}\ket{b_2}\right)_{\mathbf{B|A}} \otimes \ket{s}_{\mathbf{S|A}}\otimes \ket{s_3}_{\mathbf {S_3|A}}.
\end{equation}
Applying the QRF transformation and computing the reduced state of $\mathbf{AS}$ relative to Bob we obtain the state
\begin{equation}\label{eq:reducedstatewith}
     \frac{1}{2}\left(\ket{-b_1}\bra{-b_1}\otimes\ket{s-b_1} \bra{s-b_1}+\ket{-b_2}\bra{-b_2}\otimes\ket{s-b_2}\bra{s-b_2}\right)_{\mathbf{AS|B}},
\end{equation} which no longer contains information about $\theta$. The ambiguity between~\eqref{eq:reducedstatewithout} and~\eqref{eq:reducedstatewith} seems to indicate that the choice to initially include $\mathbf S_3$ in the description affects the results that Bob would obtain when measuring $\mathbf{AS|B}$. This is the paradox of the third particle. One may argue that Eq.~\eqref{quantumcatBstate} is not the correct description of the setup, since it fails to acknowledge the presence of $\mathbf{S_3}$. Instead, state~\eqref{eq:extendedstate} should give the correct account of the situation. However, the use of this extended state yields a problem with the reversibility of QRF transformations: the map sending the \emph{reduced state} of~\eqref{eq:extendedstate} to the state~\eqref{eq:reducedstatewith} cannot be reversible, as the former state is pure and the latter is mixed.

We argue that the paradox is in fact a symptom of a deeper tension between
reversible QRF transformations defined solely in terms of relative
subsystems~\cite{Giacomini2019,Hamette2020,Vanrietvelde2020,Krumm2021} and
\emph{compositionality}. By the latter we mean, in general terms, the requirement that a framework behave coherently under the composition and decomposition of subsystems. Focusing on a reduced part of the setup and applying a QRF transformation  should be actions that may be performed in either order with the same result. Moreover, how the framework describes a collection of subsystems should not depend on whether these subsystems are embedded in any larger setup.

Consequently, compositionality in the present example would require that applying a QRF transformation on the reduced setup or on the extended setup yield the same reduced state of $\mathbf{AS}$ relative to $\mathbf{B}$. More concretely, a necessary condition for the reduced and extended QRF transformations,
$\mathcal{S}^{\text{red}}_{\mathbf{A\to B}}$ and
$\mathcal{S}^{\text{ext}}_{\mathbf{A\to B}}$, to be \emph{compositional} is, schematically,
\begin{equation}\label{Compositionality}
\operatorname{Tr}_{\left(\text{all but } \mathbf{AS|B}\right)} \circ \,
\mathcal{S}^\text{ext}_{\mathbf{A\to B}} \left[\rho_{\text{ext}}\right]
= \operatorname{Tr}_{\left(\text{all but } \mathbf{AS|B}\right)} \circ \,
\mathcal{S}^\text{red}_{\mathbf{A\to B}} \left[\rho_{\text{red}}\right].
\end{equation}
We elaborate on this notion in
Subsection~\ref{SUBSEC:ParadoxesDoNotEmerge}, once the necessary tools are in place. 

Property~\eqref{Compositionality} does not hold for the mentioned QRF transformations. In the extended setup, the state of $\mathbf{ASS_3}$ relative to Bob is obtained from Alice's perspective and the left-hand side of~\eqref{Compositionality} then requires a partial trace over $\mathbf{S_3|B}$, leading to~\eqref{eq:reducedstatewith}. In the reduced setup, there is no system other than $\mathbf{AS|B}$, and the right-hand side of~\eqref{Compositionality} trivially leads to~\eqref{eq:reducedstatewithout}. This failure is physically expected:
$\mathbf{S_3|A}$ and $\mathbf{S_3|B}$ are physically distinct subsystems, and therefore tracing them out before and after the QRF transformation should not in general lead to the same final state. 

In view of the above, most treatments of the paradox, including the original work~\cite{Angelo2011}, argue that the ordinary partial trace is not the right way to obtain reduced states in QRF perspectives. For instance, Refs.~\cite{Krumm2021, Hoehn2022} introduce an alternative ``relational partial trace'' based on which a solution to the paradox is proposed, and Ref.~\cite{Brukner2026} points out the inequivalence of the subsystems, thereby arguing against the partial trace of $\mathbf{S_3|B}$. These proposals are consistent at the level of physical predictions and in this sense provide genuine solutions. 

However, the paradox of the third particle is still subject to discussion. See, for example, Ref.~\cite{Palumbo2026}, where solutions to the paradox employing modified traces are critically analysed, and Ref.~\cite{Hausmann2025}, which points out that the paradox is also present for classical reference frames and highlights the tension between reversibility and the use of ordinary partial traces. 

In our view, the debate remains open in part because the aforementioned tension between QRF transformations and compositionality is structural. In particular, it remains unclear whether, in the solutions mentioned above, a subsystem within a perspective (such as `$\mathbf{AS|B}$') can be identified and treated independently of potential external systems one could in principle include in the description.

\subsection{Our contribution}
\noindent Building upon the original work, here we analyse the compositional structure of the framework presented in~\cite{CastroRuiz2025a}. We show in detail that it provides a consistent and general resolution to the problem of composing subsystems in quantum reference frames, with the so-called `extra particle' subsystems playing a crucial role. The framework features reversible QRF transformations that respect a well-defined notion of relative subsystems, which is independent of the number of systems
considered and remains stable under their addition or removal\footnote{This should not be confused with the relativity of subsystems across perspectives~\cite{AhmadAli2022}, which remains present in this framework.}. The
resulting compositional structure enables the analysis of richer scenarios, which we explore in this work: comparisons between the descriptions given by external observers with implicit frames and internal observers, and comparisons between QRF perspectives before and after subsystems are added or removed. We restrict the analysis to finite Abelian groups, but many of our results can be generalised directly. 

After reviewing the framework in Section~\ref{SEC:IntroToFormalism}, we show in Section~\ref{SEC:InternalisingExternalFrames} that external observers can be treated on the same footing as internal ones, leading to a consistent hierarchy of QRF perspectives. In particular, different observers can describe each other's use of the formalism and arrive at the same conclusions. Independence of the external frame allows us to define, in Section~\ref{SEC:BipToTripEmbedding}, how subsystems compose in QRF perspectives. Specifically, we define what it means to extend a QRF setup to include a new subsystem. The same structure defines subsystem removal and, based on that, we explain how the paradox of the third particle does not arise by construction. In Section~\ref{SEC:Adding and removing relative subsystems}, we focus on state updates, within a QRF perspective, when a setup is extended. We show that, in contrast to standard quantum theory, not every state can be appended in tensor-product form relative to a QRF. We characterise which relative states admit such an extension given the initial state in the QRF perspective. Finally, in Section~\ref{SEC:AddingWClassicalisation}, we introduce \emph{classicalisation}, a procedure that provides a `classical-like' description of a QRF perspective. This makes it possible to add relative subsystems while circumventing, in a specific sense, the restrictions above. We discuss the operational meaning of such an adding procedure through a concrete example and show that classicalisation provides a link between two inequivalent approaches to QRFs~\cite{CastroRuiz2025a,Giacomini2019}. 

\section{Introducing the formalism}\label{SEC:IntroToFormalism}

\noindent In this section, we introduce the basics of the formalism developed in Ref.~\cite{CastroRuiz2025a}. We use a simple setting where the reference frames are associated with the group of discrete translations on a circle. 
The reader who is already familiar with the formalism can skip this section and jump directly to Section~\ref{SEC:InternalisingExternalFrames}.

We use the term \emph{setup} to refer to a composite physical system, together with a partition into subsystems and an action of the symmetry group on each of them, as specified by a potential external observer. An external observer is one who possesses an implicit external frame and who uses standard quantum theory to describe the subsystems and their symmetries. Given a setup, the goal is to promote each suitable subsystem to a QRF and define its associated perspective. The perspective is the description of the setup available to an \emph{internal} observer, meaning one who has no access to the external frame and instead uses the given QRF to construct physical quantities relationally. We will refer interchangeably to the perspective of a QRF (e.g. $\mathbf{A}$) and that of the observer who uses it (e.g. Alice).

Starting from the external reference frame, the task of defining the perspective of an internal QRF can be roughly divided into two steps:
\begin{itemize}
\item[1.]{Identifying the relative quantities that specify the system relative to the QRF. This induces a description which is natural to the chosen QRF};
\item[2.]{Discarding quantities that are not in principle accessible to the internal observer who uses the QRF.}
\end{itemize}

We now illustrate how the steps above are implemented formally in a specific setup. 
Consider two $N$-dimensional systems, $\mathbf A$ and $\mathbf S$. We first describe them from the point of view of an external observer, whom we call Eve and denote by $\mathbf{(E)}$. Eve assigns to $\mathbf A$ and $\mathbf S$ the joint Hilbert space \({(\mathcal{H}_\mathbf{A} \otimes \mathcal{H}_\mathbf{S})^\mathbf{(E)}}\), $\mathcal{H}^{(\mathbf E)}_\mathbf{A} \simeq \mathcal{H}^{(\mathbf E)}_\mathbf{S}\simeq \mathbb C^N$. Although the construction begins with Eve, it does not depend on her. The internal perspectives and transformations have a meaning of their own and can be adopted from the start once the framework is developed. Moreover, the internal description is compatible with any potential external observer~\cite{CastroRuiz2025a}. We expand on this point in the next section. For simplicity of notation, we omit for now the superscript $\mathbf{(E)}.$

\begin{figure}[ht]   
\begin{center}
\tikzset{every picture/.style={line width=0.75pt}} 

\begin{tikzpicture}[x=0.6pt,y=0.6pt,yscale=-1,xscale=1]

\draw [color={rgb, 255:red, 245; green, 166; blue, 35 }  ,draw opacity=1 ][line width=1.5]    (134.67,251.57) .. controls (-0.12,249.12) and (3.2,53.75) .. (136.88,55.64) .. controls (270.57,57.53) and (266.69,252.89) .. (134.67,251.57)(81.38,242.26) -- (85.24,235.26)(43.94,204.66) -- (50.95,200.81)(31.1,151.92) -- (39.1,152.03)(44.63,100.94) -- (51.6,104.88)(83.72,63.23) -- (87.38,70.34)(134.89,51.62) -- (134.89,59.62)(188.26,63.94) -- (184.52,71.01)(226.05,100.86) -- (219.08,104.79)(239.43,153.35) -- (231.43,153.33)(226.69,204.54) -- (219.65,200.74)(188.71,242.97) -- (184.87,235.95)(136.66,255.57) -- (136.62,247.57) ;
\draw  [draw opacity=0][fill={rgb, 255:red, 0; green, 95; blue, 86 }  ,fill opacity=1 ] (182.44,232.2) .. controls (186.06,230.44) and (190.46,232.1) .. (192.25,235.9) .. controls (194.04,239.71) and (192.56,244.22) .. (188.94,245.98) .. controls (185.32,247.75) and (180.93,246.09) .. (179.13,242.28) .. controls (177.34,238.48) and (178.82,233.97) .. (182.44,232.2) -- cycle ;
\draw  [draw opacity=0][fill={rgb, 255:red, 208; green, 2; blue, 27 }  ,fill opacity=1 ] (181.41,69.25) .. controls (181.37,65.33) and (184.7,62.12) .. (188.85,62.08) .. controls (192.99,62.04) and (196.38,65.19) .. (196.41,69.12) .. controls (196.45,73.04) and (193.12,76.25) .. (188.97,76.29) .. controls (184.83,76.33) and (181.44,73.18) .. (181.41,69.25) -- cycle ;
\draw [color={rgb, 255:red, 208; green, 2; blue, 27 }  ,draw opacity=1 ][fill={rgb, 255:red, 208; green, 2; blue, 27 }  ,fill opacity=1 ][line width=1.5]    (192,63.95) -- (210.1,33.2) ;
\draw  [color={rgb, 255:red, 208; green, 2; blue, 27 }  ,draw opacity=1 ][fill={rgb, 255:red, 208; green, 2; blue, 27 }  ,fill opacity=1 ][line width=1.5]  (216.44,21.99) -- (221.66,39.73) -- (210.1,33.2) -- cycle ;

\draw  [draw opacity=0][fill={rgb, 255:red, 255; green, 255; blue, 255 }  ,fill opacity=0.5 ] (181.41,69.25) .. controls (181.37,65.04) and (184.75,61.59) .. (188.97,61.55) .. controls (193.18,61.51) and (196.63,64.9) .. (196.67,69.11) .. controls (196.7,73.33) and (193.32,76.78) .. (189.11,76.81) .. controls (184.89,76.85) and (181.45,73.47) .. (181.41,69.25) -- cycle ;
\draw  [draw opacity=0][fill={rgb, 255:red, 255; green, 255; blue, 255 }  ,fill opacity=0.5 ] (190.68,61.63) -- (215.17,17.47) -- (232.59,27.14) -- (208.1,71.3) -- cycle ;

\draw  [draw opacity=0][fill={rgb, 255:red, 208; green, 2; blue, 27 }  ,fill opacity=1 ] (216.76,101.42) .. controls (218.27,97.8) and (222.6,96.16) .. (226.42,97.75) .. controls (230.24,99.34) and (232.12,103.57) .. (230.61,107.19) .. controls (229.11,110.81) and (224.78,112.46) .. (220.96,110.87) .. controls (217.13,109.27) and (215.25,105.05) .. (216.76,101.42) -- cycle ;
\draw [color={rgb, 255:red, 208; green, 2; blue, 27 }  ,draw opacity=1 ][fill={rgb, 255:red, 208; green, 2; blue, 27 }  ,fill opacity=1 ][line width=1.5]    (228.59,100.71) -- (257.31,79.54) ;
\draw  [color={rgb, 255:red, 208; green, 2; blue, 27 }  ,draw opacity=1 ][fill={rgb, 255:red, 208; green, 2; blue, 27 }  ,fill opacity=1 ][line width=1.5]  (267.54,71.72) -- (265.37,90.09) -- (257.31,79.54) -- cycle ;

\draw  [draw opacity=0][fill={rgb, 255:red, 255; green, 255; blue, 255 }  ,fill opacity=0.5 ] (216.76,101.42) .. controls (218.38,97.53) and (222.85,95.69) .. (226.74,97.31) .. controls (230.63,98.93) and (232.47,103.4) .. (230.85,107.29) .. controls (229.23,111.18) and (224.76,113.02) .. (220.87,111.4) .. controls (216.98,109.78) and (215.14,105.31) .. (216.76,101.42) -- cycle ;
\draw  [draw opacity=0][fill={rgb, 255:red, 255; green, 255; blue, 255 }  ,fill opacity=0.5 ] (228.28,98.06) -- (268.15,67.07) -- (280.37,82.8) -- (240.5,113.79) -- cycle ;

\draw (287.33,98.73) node [anchor=north west][inner sep=0.75pt]  [color={rgb, 255:red, 208; green, 2; blue, 27 }  ,opacity=1 ]  {$\frac{\ket{1} +e^{\iu\theta }\ket{2}}{\sqrt{2}}$};
\draw (260.83,99.94) node [anchor=north west][inner sep=0.75pt]  [font=\normalsize,color={rgb, 255:red, 208; green, 2; blue, 27 }  ,opacity=1 ,rotate=-1.89] [align=left] {$\displaystyle \mathbf{A} :$};
\draw (208.35,239.73) node [anchor=north west][inner sep=0.75pt]  [font=\normalsize,color={rgb, 255:red, 65; green, 117; blue, 5 }  ,opacity=1 ] [align=left] {$\displaystyle \textcolor[rgb]{0,0.37,0.34}{\mathbf{S} :}$};
\draw (229.67,239.07) node [anchor=north west][inner sep=0.75pt]  [color={rgb, 255:red, 0; green, 95; blue, 86 }  ,opacity=1 ]  {$\ket{5}$};
\draw (173.33,78.4) node [anchor=north west][inner sep=0.75pt]  [color={rgb, 255:red, 245; green, 166; blue, 35 }  ,opacity=1 ]  {$1$};
\draw (200.33,106.4) node [anchor=north west][inner sep=0.75pt]  [color={rgb, 255:red, 245; green, 166; blue, 35 }  ,opacity=1 ]  {$2$};
\draw (72.33,81.4) node [anchor=north west][inner sep=0.75pt]  [color={rgb, 255:red, 245; green, 166; blue, 35 }  ,opacity=1 ]  {$N-1$};
\draw (211.33,144.4) node [anchor=north west][inner sep=0.75pt]  [color={rgb, 255:red, 245; green, 166; blue, 35 }  ,opacity=1 ]  {$3$};
\draw (11,26.67) node [anchor=north west][inner sep=0.75pt]  [font=\normalsize,color={rgb, 255:red, 234; green, 77; blue, 77 }  ,opacity=1 ] [align=left] {\textcolor[rgb]{0.96,0.65,0.14}{$\displaystyle {\displaystyle \mathbf{\textcolor[rgb]{0.96,0.65,0.14}{\star}\textcolor[rgb]{0.96,0.65,0.14}{(}\textcolor[rgb]{0.96,0.65,0.14}{E}\textcolor[rgb]{0.96,0.65,0.14}{)}\textcolor[rgb]{0.96,0.65,0.14}{\ }}}$}};
\draw (172.33,212.4) node [anchor=north west][inner sep=0.75pt]  [color={rgb, 255:red, 245; green, 166; blue, 35 }  ,opacity=1 ]  {$5$};
\draw (202.33,185.4) node [anchor=north west][inner sep=0.75pt]  [color={rgb, 255:red, 245; green, 166; blue, 35 }  ,opacity=1 ]  {$4$};
\draw (130.33,65.4) node [anchor=north west][inner sep=0.75pt]  [color={rgb, 255:red, 245; green, 166; blue, 35 }  ,opacity=1 ]  {$0$};
\draw (112.33,223.4) node [anchor=north west][inner sep=0.75pt]  [color={rgb, 255:red, 245; green, 166; blue, 35 }  ,opacity=1 ]  {$...\ 6$};

\end{tikzpicture}
\end{center}
    \caption{Illustration of a setup containing two systems described by an external observer through the Hilbert space $(\mathcal{H}_\mathbf{A}\otimes \mathcal{H}_\mathbf{S})^{\mathbf{(E)}}$. Each of the systems carries the regular representation of the group $G=\mathbb{Z}_N$ and the global external state in the picture is $\ket{\psi}^{\mathbf{(E)}}_{\mathbf{AS}}=1/\sqrt{2}\left(\ket{a_1}+e^{\iu \theta}\ket{a_2}\right)_{\mathbf{A}}\otimes\ket{s}_{\mathbf{S}},$ with $a_1=1, a_2=2$ and $s=5$. In Appendix~\ref{appendixA} we calculate the state obtained in this case when $\mathbf{A}$ is used as reference frame for the position of $\mathbf{S}$, as well as other elementary examples.}\label{FIG:CircleDiscreteTranslations}
\end{figure}
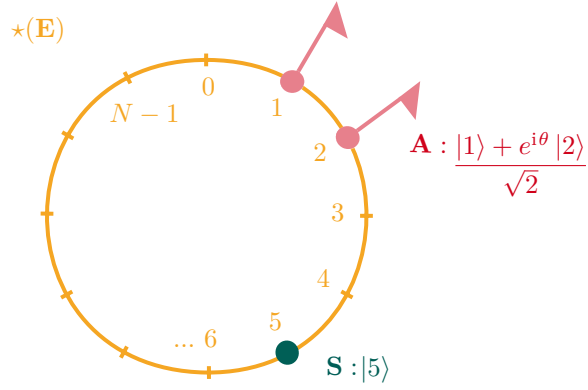

Each Hilbert space $\mathcal{H}_\mathbf{A}$ and $\mathcal{H}_\mathbf{S}$ has an orthonormal basis $\{\ket{x}\}_{x \in \mathbb Z_N}$, corresponding to positions on a circle (see Fig.~\ref{FIG:CircleDiscreteTranslations}). Each space carries the regular representation of the symmetry group $G=\mathbb Z_N = (\{0,1,\dots,N-1\},+),$ where $+$ denotes addition modulo $N$. 
For system $\mathbf{A}$, which we later promote to a QRF, we denote the left and right-regular representations by $\{L_x \in L(\mathcal{H}_{\mathbf A})\}_{x \in  \mathbb{Z}_{N}}$ and $\{R_x \in L(\mathcal{H}_{\mathbf A})\}_{x \in  \mathbb{Z}_{N}}$, where $L(\mathcal{H})$ is the set of linear operators on the Hilbert space $\mathcal{H}$. Their action on the position basis is given by
\begin{equation}
	L_x \ket{x'}_{\mathbf A}= \ket{x + x'}_{\mathbf A}, \qquad \text{and} \qquad R_x \ket{x'}_{\mathbf A} = \ket{x'-x}_{\mathbf A}.
\end{equation}  By convention, $G$ acts on $\mathcal H_\mathbf A$ via the left-regular representation. This action can be interpreted as a finite set of displacements on the circle. The right-regular representation is not so relevant for Abelian groups like $\mathbb Z_N$, as it coincides with the inverse of the left-regular representation. However, for non-Abelian groups it is crucial to distinguish between left- and right-regular actions. 

Being equipped with the regular representation, $\mathbf{A}$ is an ideal QRF~\footnote{
	In this work, we consider only ideal QRFs. However, one can also define QRF perspectives and the corresponding reversible transformations relative to non-ideal QRFs, which obey weaker assumptions. For instance, Ref.~\cite{Garmier2025} extends the formalism of Ref.~\cite{CastroRuiz2025a} to the non-ideal case.}. 
    In principle, only systems that are promoted to QRFs need to have this structure, but for simplicity we assume it for all systems throughout. We denote the regular representation acting on $\mathbf S$ by $\{U_x \in L(\mathcal{H}_\mathbf{S})\}_{x \in  \mathbb{Z}_{N}}$ to distinguish $\mathbf S$ as a subsystem which is not necessarily promoted to a QRF.

Despite its simplicity, this toy model captures much of the conceptual subtleties that come with considering QRF perspectives. The same toy model is used in Refs.~\cite{Krumm2021,Hoehn2022} in an approach to QRF transformations that is closely related to the `perspective neutral' approach~\cite{Vanrietvelde2020, Hamette2021}.

We define the operator $\hat{x}_\mathbf{A}$ through 
$\hat{x}_\mathbf{A}\ket{x}_\mathbf{A} = x\ket{x}_\mathbf{A}$, in analogy with the continuous position operator for a particle. We also define a discrete momentum basis, conjugate to $\{\ket{x}\}_{x\in \mathbb{Z}_N}$, via the discrete Fourier transform:
\begin{equation}\label{momentumbasisdef}\ket{p}_{\mathbf{A}} := \frac{1}{\sqrt{N}} \sum^{N-1}_{x=0} e^{\iu \frac{2\pi}{N} xp} \ket{x}_{\mathbf{A}}.
\end{equation} 
Since each $L_x$ acts as a cyclic shift on the basis $\{\ket{x}\}_{x\in \mathbb{Z}_N}$, with $L_N =L_0 = \mathds{1}$, the elements $L_x$ are diagonal in $\{\ket{p}\}_{p\in \mathbb{Z}_N}$ and may be written as
\begin{equation}\label{momentumoperatordef}
    L_x = \sum_{p=0}^{N-1} e^{-\iu\frac{2\pi}{N}xp} \ket{p}\bra{p}_{\mathbf{A}} = e^{-\iu\frac{2\pi}{N}x\hat{p}_{\mathbf{A}}}, 
\end{equation}
where the operator $\hat{p}_{\mathbf{A}}$ is defined by $\hat{p}_{\mathbf{A}}\ket{p}_{\mathbf{A}} = p\ket{p}_{\mathbf{A}}$. The operators $\hat{x}_{\mathbf{A}}$ and $\hat{p}_{\mathbf{A}}$ thus define a structure analogous to that of a particle with conjugate position and momentum observables.

We are now in a position to implement step 1: describe the setup relative to system $\mathbf{A}$, which can be used as a QRF by the internal observer Alice. The main idea is to introduce the notion of `the system $\mathbf S$ relative to QRF $\mathbf{A}$', denoted by $\mathbf{S|A}$, as an algebra of relative operators on $(\mathcal H_\mathbf A \otimes \mathcal H_\mathbf S)^{(\mathbf E)}$. Specifically, 
\begin{equation}\label{S|A}
    \mathbf{S|A}  :=  \left\{ \sum_x \ket{x}\bra{x}_{\mathbf{A}}\otimes \left(U_x M_\mathbf S U^{\dagger}_x \right)_\mathbf{S}\Big|  M_\mathbf S\in\mathcal{L}(\mathcal{H}_\mathbf{S})\right\},
\end{equation}
where the sum on $x$ runs from $0$ to $N-1$ throughout the paper. The definition above can be motivated by how a general operator $M$, applied on $\mathbf{S}$ from the perspective of Alice, is expected to `look like' for Eve. For instance, plugging $M_\mathbf S=\hat{x}_\mathbf{S}$ in~(\ref{S|A}) gives $\hat x_{\mathbf{S|A}} = \hat{x}_\mathbf{S}-\hat{x}_\mathbf{A}$.
This is a direct generalisation of how relative position coordinates are defined classically from `absolute' position coordinates. Eq.~\eqref{S|A} defines implicitly a map from $\mathcal{L}(\mathcal{H}_\mathbf{S})$ to  $\mathcal{L}(\mathcal{H}_\mathbf{AS})$, known as the `relativisation' map~\cite{Kitaev2004, Bartlett2007,LoveridgeThesis2018, Loveridge2017}. Because $\mathbf A$ is ideal, the algebra $\mathbf{S|A}$ is isomorphic to $\mathbf{S}:=L(\mathcal{H}_\mathbf{S})$.

Once the relative operators are identified, it is always possible to define a refactorisation of the setup's Hilbert space such that operators in $\mathbf{S|A}$ are naturally represented~\cite{CastroRuiz2025a,Garmier2025}, that is, such that they act non-trivially in only one factor of the factorisation. For our case, this is done through the unitary map~\cite{Vanrietvelde2020,Hamette2020,CastroRuiz2025a}: 
\begin{equation}\label{jumpingmap}
   \mathsf{U}^{\mathbf{AS}}_{\to \mathbf{A}}:=   
 \sum_x \ket{x}\bra{x}_\mathbf{A}\otimes \left(U^{\dagger}_x\right)_\mathbf{S},
\end{equation} 
which induces a refactorisation 
\begin{equation}\label{Refactorisation}
 \mathcal{H}_\mathbf{S|A'}\otimes\mathcal{H}_\mathbf{S|A} \; := \;  \iota \circ \mathsf U^{\mathbf{AS}}_{\to \mathbf A} \Big( \mathcal{H}_\mathbf{A}\otimes\mathcal{H}_\mathbf{S}  \Big),
\end{equation} 
where $\iota \ket{x}_\mathbf{A} \otimes \ket {x'}_{\mathbf {S}} = \ket{x}_\mathbf{S|A'} \otimes \ket{x'}_{\mathbf{S|A}}$. Throughout, the map $\iota$ and similar label-changing maps will be used implicitly to simplify notation, and we will write  
$
\mathcal{H}_\mathbf{S|A'}\otimes\mathcal{H}_\mathbf{S|A} \; \simeq \mathcal{H}_\mathbf{A}\otimes\mathcal{H}_\mathbf{S}.  $

The subsystem labels in $\mathcal{H}_\mathbf{S|A'}\otimes\mathcal{H}_\mathbf{S|A}$ have operational meaning. For example, as $\mathbf{S|A} \simeq L(\mathcal{H}_\mathbf{S|A})$, the operators in~(\ref{S|A}) correspond to all operators of the form $\mathds{1}_\mathbf{S|A'}\otimes M_\mathbf{S|A}$ when expressed in the factorisation natural to $\mathbf{A}$. From Alice's perspective, $\mathbf{S|A}$ corresponds to a tensor factor of the Hilbert space, in the same way that $\mathbf S$ corresponds to a factor with respect to Eve. Note, however, that we have denoted the Hilbert space from the perspective of Eve by the superscript $(\mathbf E)$ rather than the subscript $\mathbf{|E}$. This is because Eve plays the role of an external observer with an implicit reference frame. In the next section we will see how our framework allows us to internalise Eve's QRF, treating it on the same footing as $\mathbf{A}$. 

The algebra $L(\mathcal{H}_\mathbf{S|A'})$ is, by construction, the commutant of $\mathbf{S|A}$. The refactorisation~\eqref{Refactorisation} is what we mean by a description of the setup natural to the reference frame (step 1). One can rewrite all relevant states and operators of the setup through the map 
\begin{equation}\label{RelationalisationMap}
    \mathcal{U}^\mathbf{AS}_{\to \mathbf{A}}:=(\mathsf{U}^\mathbf{AS}_{\to \mathbf{A}})\cdot(\mathsf{U}^\mathbf{AS}_{\to \mathbf{A}})^{\dagger}, 
    \end{equation}
    obtaining a relational description with respect to $\mathbf{A}$. 

Now we proceed to step 2: discarding operators that are not physically meaningful to internal observers. These are the operators which are not invariant under the action of the group. From Eve's perspective, $\mathbb Z_N$ acts globally via the representation $\{L_x\otimes U_x\}_{x\in \mathbb{Z}_N}$, and an operator $M \in L(\mathcal{H}_\mathbf{A}\otimes \mathcal{H}_\mathbf{S})$ is invariant if it commutes with every element of that representation, namely,
\begin{equation}\label{eq:gaugewrtalice}
\left(L_x\otimes U_x\right)_\mathbf{A,S}  M \left(L^{\dagger}_x\otimes U^{\dagger}_x\right)_\mathbf{A,S} = M \qquad \forall x\in \mathbb{Z}_N.
\end{equation} 
As the subsystem on which the translations~\eqref{eq:gaugewrtalice} act non-trivially is physically meaningless to Alice, we refer to it as the gauge subsystem.

Invariant operators also form an algebra, which we denote by $\mathcal{B}^\mathbf{AS}_{\text{inv}}$. If an operator is not invariant, we can extract its invariant part by projecting onto $\mathcal{B}^\mathbf{AS}_{\text{inv}}$. The projection superoperator is given by
\begin{equation}\label{Gtwirl}
   \mathcal{G}_\mathbf{A,S}=\frac{1}{N}\sum_x\left(L_x\otimes U_x\right)_\mathbf{A,S}  \left(\cdot \right)\left(L^{\dagger}_x\otimes U^{\dagger}_x\right)_\mathbf{A,S},
\end{equation}
a map known as incoherent G-twirl (see, for instance,~\cite{Poulin2006,Bartlett2007}). It is easy to see that the action of the group after the refactorisation~\eqref{Refactorisation} is simply given by $\left(L_x\otimes \mathds{1}\right)_\mathbf{S|A',S|A}$, which implies that the G-twirl in~(\ref{Gtwirl}) is reexpressed in the relative factorisation as
\begin{equation}\label{GTwirlInRelativeFactorisation}
   \mathcal{G}_\mathbf{S|A'}=\frac{1}{N}\sum_x \left(L_x\otimes \mathds{1}\right)_\mathbf{S|A',S|A}  \left(\cdot \right)\left(L^{\dagger}_x\otimes \mathds{1}\right)_\mathbf{S|A',S|A},
\end{equation} where $L_x$ are the elements of the left-regular representation of $L(\mathcal{H}_\mathbf{S|A'})$.

One can immediately check that all operators in $\mathbf{S|A}$ belong to $\mathcal{B}^\mathbf{AS}_{\text{inv}}$. The inclusion is, however, a strict one. In particular, the algebra 
\begin{equation}\label{ExtraParticleAlgebra}
     \mathbf{\overline{S|A}}:= \quad \left\{ \sum_{xx'} \bra{x}M_R\ket{x'}\ket{x}\bra{x'}_ \mathbf{A}\otimes \left(U_x  U^{\dagger}_{x'}\right)_ \mathbf{S}  \Big| M_R \in \mathcal{L}(\mathcal{H}_\mathbf{A}) \text{  s.t.  } [M_R,L_y]=0 \text{  }\forall y\right\}
\end{equation}
is also invariant under the action of $\mathbb Z_N$. In fact, it is the commutant of $\mathbf{S|A}$ in the invariant algebra $\mathcal{B}^\mathbf{AS}_{\text{inv}}$. Together, $\mathbf{S|A}$ and $\mathbf{\overline{S|A}}$ generate the full invariant subalgebra: $\mathbf{S|A} \vee \mathbf{\overline{S|A}} = \mathcal{B}^\mathbf{AS}_{\text{inv}}$. This is the case whenever we have ideal QRFs, but is not true a priori for non-ideal ones~\cite{Garmier2025}. The subsystem $\mathbf{\overline{S|A}}$ is the so-called \emph{extra particle}~\cite{CastroRuiz2025a}, and we will explain how it plays a crucial role in defining compositional and reversible QRF transformations. Being invariant, operators in $\mathbf{\overline{S|A}}$ are in principle measurable from the perspective of Alice. An operational scheme for doing so in terms of a collaborative strategy between internal observers is presented in \cite{delaHamette2026, Doat2026}.

 In the factorisation relative to $\mathbf A$, the extra particle algebra~(\ref{ExtraParticleAlgebra}) is represented by operators of the form $M_{\mathbf{\overline{S|A}}} \otimes \mathds{1}_{\mathbf{S|A}}$, where we use the notation $\mathbf{\overline{S|A}}$ as a subindex to denote operators on $\mathcal{H}_{\mathbf{S|A'}}$ that are invariant under the left-regular representation acting on this Hilbert space. Note that, unlike the other algebras defined in this section, the extra particle operators cannot in general be written as $L(\mathcal{H})$ for some Hilbert space $\mathcal{H}\subseteq \mathcal{H}_\mathbf{S|A'}.$ Rather, they define a generalised subsystem at the level of algebras~\cite{Knill2000, Barnum2004, Zanardi2001, Zanardi2004,Rio2015,Chiribella2018,Vanrietvelde2025,Garmier2025}. In the simple case we are considering, the extra particle is given by functions of the momentum operator, $\hat{p}_{\mathbf{\overline{S|A}}}$~\footnote{We use the convention $\hat{p}_\mathbf{\overline{S|A}}=-\hat{p}_{\mathbf{S|A'}}$, where the latter operator is the momentum in $L(\mathcal{H}_\mathbf{S|A'})$ defined analogously to~(\ref{momentumoperatordef}). This is done in view of the generalisation to non-Abelian groups, since the label $\mathbf{\overline{S|A}}$ characterises the extra particle, which carries the right-regular representation, $R_x$. For the Abelian case, we have the identification $R_x=L_x^\dagger$  $\forall x \in \mathbb{Z}_N$, which causes the momentum operators to differ by a sign only. In general, however, these operators belong to different spaces, with the extra particle living in the left-invariant part.}. 

In summary, we now have a factorisation that is natural to a given QRF, implemented through~(\ref{jumpingmap}) (step 1), and a means to eliminate all non-invariant operators, through~(\ref{GTwirlInRelativeFactorisation}) (step 2). The invariant algebra and the specification of relative factorisation defines what we call the perspective of QRF $\mathbf{A}$. Importantly, this perspective includes not only the relative degrees of freedom $\mathbf{S\vert A}$ but also the extra particle $\mathbf{\overline{S\vert A}}$. To `jump' into the perspective of $\mathbf{A}$, we may simply apply $\mathcal{G}_\mathbf{S|A'}\circ \mathcal{U}^\mathbf{AS}_{\to \mathbf{A}}$ to all operators in $L(\mathcal{H}_\mathbf{A}\otimes \mathcal{H}_\mathbf{S})$. 
Alternatively, we can first apply the G-twirl in the form of~\eqref{Gtwirl} and then change factorisations, resulting in the map $\mathcal{U}^\mathbf{AS}_\mathbf{\to A}\circ\mathcal{G}_\mathbf{A,S}$. 

The general form of an operator from $\mathbf A$'s perspective is simply
\begin{equation}\label{MomentumSuperselectedOp}
    \sum_{p=0}^{N-1} \ket{p}\bra{p}_{\mathbf{\overline{S|A}}}\otimes \left(M_p\right)_\mathbf{S|A},
\end{equation} 
where $\ket{p}_\mathbf{\overline{S|A}}$ are the eigenstates of $\hat{p}_\mathbf{\overline{S|A}}$ and $M_p$ are general operators on $L(\mathcal{H}_\mathbf{S|A}).$ The extra particle subsystem is effectively classical, assuming only diagonal momentum states (possibly correlated with the relative subsystem). Furthermore, because $\mathsf{U}^{\mathbf{AS}}_{\to \mathbf{A}}$ in Eq.~\eqref{jumpingmap} maps the global group action to the left-regular action on $\mathcal{H}_{\mathbf{S|A'}}$, these momentum degrees of freedom correspond to the total momentum degrees of freedom of the $\mathbf{AS}$ system. The expression above for a general operator in the perspective then reflects the superselection rule for total momentum induced by global translation invariance~\cite{Bartlett2007}. 

In a setup containing another reference frame $\mathbf{B}$, we can adapt the general definitions to apply to $\mathbf{A, B}$ and $\mathbf{S}$, and define the factorisation relative to $\mathbf{B}$. If we wish to jump from the perspective of $\mathbf{A}$ to that of $\mathbf{B}$, it suffices to apply 
\begin{equation}
\mathcal S_{\mathbf A \to \mathbf B } \; = \; \mathcal{U}^\mathbf{ABS}_{\to \mathbf{B}}\circ \left(\mathcal{U}^\mathbf{ABS}_{\to \mathbf{A}}\right)^\dagger.
\end{equation}
This map is the transformation between two QRF perspectives. Operationally, given a state from Alice's perspective, it allows her to compute probabilities for measurements relative to QRF $\mathbf{B}.$  Mathematically, it merely reexpresses the same invariant operators in a subsystem decomposition natural to $\mathbf{B}$. In equations,
\begin{equation}
\mathcal B^{\mathbf{ABS}}_{\text{inv}} \simeq \mathbf{\overline{BS|A}}\vee \mathbf{BS|A} \simeq \mathbf{\overline{AS|B}}\vee\mathbf{AS|B}.
\end{equation}
Roughly speaking, the equation above means that the information contained in $\mathbf{\overline{BS|A}}\vee\mathbf{BS|A}$ is necessary and sufficient to recover the information about $\mathbf{\overline{AS|B}}\vee\mathbf{AS|B}.$ Moreover, because $[\mathbf{\overline{BS|A}}, \mathbf{A|B}] \neq 0$, the information of $\mathbf{\overline{BS|A}}$ is necessary to recover $\mathbf{AS|B}$. This fact is at the heart of the paradox of the third particle and will be important in Subsection~\ref{SUBSEC:ParadoxesDoNotEmerge}.

 We end this section with an important remark: the extra-particle degrees of freedom carry information about the `quantumness' of the reference frame.
 Specifically, suppose that the state of $\mathbf{AS}$ relative to Eve is of the form 
 \begin{equation}\label{classical-quantum-statewrteve}
    \psi=\sum_x \mu_x \ket{x}\bra{x}_{\mathbf{A}}\otimes \left(\sigma_x\right)_\mathbf{S},
     \end{equation} 
for some quantum states $\sigma_x$ of $\mathbf{S}$ and a classical probability distribution $\mu_x$. This is the most general form of a classical-quantum state, i.e. composite states for which one subsystem is necessarily diagonal in the $x$-basis and has at most classical correlations with the other system. Then, it can be shown that, in the perspective of QRF $\mathbf{A}$, the state $\rho$ on $\mathcal H_{\mathbf{S|A'}} \otimes \mathcal H_\mathbf{S|A}$ reads
\begin{equation}\label{classical-quantum-statewrtalice}
\rho = \mathcal{G}_\mathbf{A,S}\circ \mathcal{U}^\mathbf{AS}_{\to \mathbf{A}}(\psi)=\left[\frac{\mathds{1}}{N}\otimes \sum_x \mu_x U^{\dagger}_x\sigma_x U_x \right]_\mathbf{\overline{S|A},S|A}. 
\end{equation} 
 Thus, if the extra particle in the perspective of $\mathbf{A}$ is not the tensorised maximally mixed state, the external state of system $\mathbf{A}$ cannot be of the classical-quantum form~\eqref{classical-quantum-statewrteve}. Interestingly, this information is in principle available from $\mathbf A$'s internal perspective. In Appendix~\ref{appendixA} we illustrate this point with concrete examples. As states of the form~\eqref{classical-quantum-statewrteve} are always mapped to states of the type~\eqref{classical-quantum-statewrtalice}, we call the latter `classical-like'. These states will be conceptually important in the following sections.

\section{A consistent hierarchy of QRF perspectives} \label{SEC:InternalisingExternalFrames}
\noindent The formalism introduced in Section~\ref{SEC:IntroToFormalism} assumes the compatibility of a QRF perspective with a potential external description, unlike perspectival approaches which are internal from the start~\cite{Giacomini2019,Hamette2020}. As noted in Ref.~\cite{CastroRuiz2025a},  the definition of the extra particle and the relative subsystem algebras is independent of the implicit external frame. One may think that, despite this independence, external observers have a special status. In this section, we expand on the results of Ref.~\cite{CastroRuiz2025a} and show that this is not the case: the perspective of an external observer can always be treated internally within the framework, leading to a consistent hierarchy of QRF perspectives. In this sense, the formalism can ``describe the use of itself''\footnote{We borrow this phrase from the Frauchiger-Renner thought experiment~\cite{frauchiger2018quantum}. Although the hierarchy of QRF perspectives presented here has a certain `Wigner's friend flavor', and it may be interesting to study it in these contexts, we observe that the conceptual challenges of Wigner's friend scenarios~\cite{brukner2018no, frauchiger2018quantum, bong2020strong} are \emph{a priori} independent of the notion of QRFs. For related discussions in this regard, see~\cite{Adlam2025,vanrietvelde2026specifying,di2026adlam}. }. This section is more abstract and largely self-contained; the reader may skip ahead to the next section without affecting their understanding of the rest of the paper. Detailed proofs are given in Appendix~\ref{appendixInternalisation}.

\subsection{Internalising Eve's reference} 
\noindent
To define Alice's internal perspective, we may start from an initial global structure assigned by the external observer Eve. Eve's description, in turn, can always be understood as relative to a physical reference frame described by yet another observer, with respect to whom her reference frame becomes explicit. Concretely, consider extending the setup of Section~\ref{SEC:IntroToFormalism}, with systems $\mathbf{A}$ and $\mathbf{S}$, to also contain Eve’s reference frame $\mathbf{E}$. The three systems are now described by Francis, an external observer who possesses an implicit reference frame $\mathbf{(F)}$. Global discrete translations relative to $\mathbf{(F)}$ act on $\mathbf{E,A}$ and $\mathbf{S}$, and the G-twirl takes the form
\begin{equation}
    \mathcal{G}_\mathbf{E,A,S}:= \frac{1}{N}\sum_x \left(L_x\otimes L_x\otimes U_x\right)_\mathbf{E,A,S}  \left(\cdot\right)\left(L^\dagger_x\otimes L^\dagger_x\otimes U^\dagger_x\right)_\mathbf{E,A,S},
\end{equation} where $(L_x)_\mathbf{E}$ are the elements of the left-regular representation of $\mathbf{E}$. Similarly, the maps $\mathcal{U}^\mathbf{EAS}_\mathbf{\to A}$ and $\mathcal{U}^\mathbf{EAS}_\mathbf{\to E}$ that Francis uses to describe the perspectives of $\mathbf{A}$ and $\mathbf{E}$ can be defined analogously to~\eqref{RelationalisationMap}. The description of $\mathbf{AS}$ with respect to the external $\mathbf{(E)}$ is naturally identified with the corresponding subsystem relative to $\mathbf E$. Namely,
\begin{equation}\label{eq:identification}
\mathbf{AS}^\mathbf{(E)} \cong \mathbf{(AS|E)^{(F)}},
\end{equation} through the identification of $\ket{x,x'}^{(\mathbf E)}_\mathbf{A,S}$ with $ \ket{x,x'}^\mathbf{(F)}_\mathbf{A|E,S|E}.$  Relatedly, the operators formerly regarded as `global translations' acting on a gauge subsystem are now identified with relative operators, $(L_x\otimes U_x)_\mathbf{A|E,S|E}$. (Since we are dealing with many different perspectives, in this section we explicitly write subindices in all operators to indicate the spaces where they act.)

If Eve assigns the state $\psi^\mathbf{(E)}_\mathbf{AS}\cong \psi_\mathbf{AS|E}$,
we say that a state assignment by Francis,  $\phi^\mathbf{(F)}_\mathbf{EAS}$, \emph{internalises} Eve's description if the following compatibility condition holds:
\begin{equation}\label{InternalisationCondition}
\operatorname{Tr}_\mathbf{AS|E'} \circ \,   \mathcal{U}^\mathbf{EAS}_\mathbf{\to E} \circ \mathcal{G}_\mathbf{E,A,S} \left(\phi^\mathbf{(F)}_\mathbf{EAS}\right) = \psi_\mathbf{AS|E}. 
\end{equation}  
 This condition naturally generalises to an arbitrary setup, containing any number of QRFs and systems. While a potential state assigned by Francis always exists, it is in general not unique: several states could satisfy the condition above for a given state assigned by Eve.
  
 Given the identification~\eqref{eq:identification}, Francis can describe Eve's use of the formalism, in the sense of steps 1 and 2 of the previous section. Specifically, Francis can write down the $G$-twirl 
 \begin{equation}
     \mathcal{G}_\mathbf{A|E,S|E}:= \frac1N\sum_x (L_x\otimes U_x)_\mathbf{A|E,S|E} \, (\cdot) \,  (L^\dagger_x\otimes U^\dagger_x)_\mathbf{A|E,S|E}
 \end{equation} which corresponds to~\eqref{Gtwirl},  and the map used to define Alice's perspective from Eve,
 \begin{equation}
     \mathcal{U}^\mathbf{AS|E}_\mathbf{\to A|E}:=  \left(\mathsf{U}^\mathbf{AS|E}_{\to \mathbf{A|E}}\right)\cdot\left(\mathsf{U}^\mathbf{AS|E}_{\to \mathbf{A|E}}\right)^{\dagger}, \quad \text{where}   \quad \mathsf{U}^{\mathbf{AS|E}}_{\to \mathbf{A|E}}:=   
 \sum_x \ket{x}\bra{x}_\mathbf{A|E}\otimes \left(U^{\dagger}_x\right)_\mathbf{S|E},
 \end{equation} which corresponds to~\eqref{RelationalisationMap}.
 
 Internalisation then allows us to treat external observers on the same footing as internal ones. In the next subsection, we show that this procedure is consistent: any two potential external observers $\mathbf{E_1}$ and $\mathbf{E_2}$ may use the formalism to describe a fixed invariant setup and always agree on the representation of the state associated with Alice's perspective inside such setup. This holds true even if, e.g., $\mathbf{E_2}$ is in a quantum superposition of position eigenstates relative to $\mathbf{E_1}$. The results presented in this section build upon the result of~\cite{CastroRuiz2025a}, which showed that $\mathbf{E_1}$ and $\mathbf{E_2}$ both agree on the definition of $\mathcal{B}_{\text{inv}}^{\mathbf{AS}}$, making this subsystem independent of the external frame.

\subsection{All observers agree on the perspective and state representations}
\begin{figure}[h] 
\centering
    \includegraphics[width=0.5\linewidth]{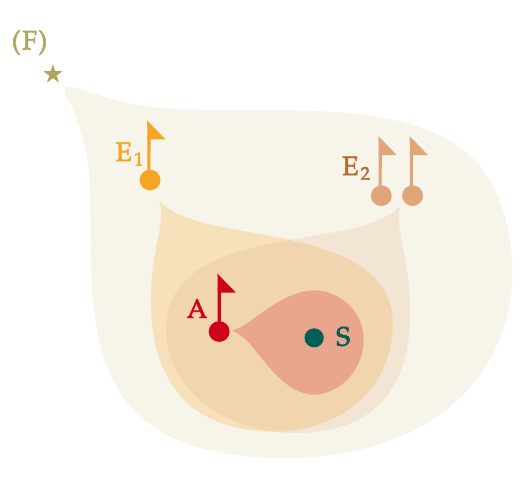}
    \caption{Illustration of how the present QRF formalism consistently accounts for observers' descriptions. External observers can use the formalism to describe QRF perspectives following the prescription of Section~\ref{SEC:IntroToFormalism}. At the same time, the frame of such an observer can itself be internalised, so that the observer's use of the formalism is described by a new external observer, or by another observer at the same level with an explicit frame. Across this hierarchy of perspectives, with implicit and explicit frames, everyone agrees on the invariant information contained in a subsystem and the form of the perspective internal to it. In the picture, the descriptions of two observers Eve 1 and Eve 2 are internalised through the reference frames  $\mathbf{E_1}$ and $\mathbf{E_2}$ in the description of the external observer Francis, $\mathbf{(F)}.$ All of them agree on the representation of Alice's perspective inside the subsystem $\mathbf{AS}$. The agreement holds even though what each of them calls `Alice's frame' and `the system S' correspond to different relative subsystems. As a result, the states and operators associated with a perspective are independent of the choice of external frame and of any additional frames or observers that may be included in the hierarchy.}
\label{FIG:OgnionOfExternalObservers}
\end{figure}

\noindent
Suppose $\rho^\mathbf{(E_1)}_\mathbf{\overline{S|A},S|A}$ is the state of Alice's perspective in a setup containing $\mathbf{AS}$ according to an external observer Eve 1, with implicit frame $\mathbf{(E_1)}$. Suppose that another observer, Francis, internalises the description of Eve 1 in a setup that contains, beyond $\mathbf{AS}$ and the internalised frame $\mathbf{E_1}$, another reference frame $\mathbf{E_2}$. The perspective of $\mathbf{E_2}$ can be thought of as the internalised description of another observer, Eve 2. In general, the internalisation could be made such that the state of $\mathbf{E_2}$ relative to $\mathbf{E_1}$ is quantum, as depicted in Fig.~\ref{FIG:OgnionOfExternalObservers}.

Let $\tilde{\phi}^\mathbf{(F)}_\mathbf{E_1E_2AS}$ be \emph{any} total state compatible  with $\rho^\mathbf{(E_1)}_\mathbf{\overline{S|A},S|A}$ being the state of Alice's perspective according to Eve 1. 
We denote the state of $\mathbf{E_1}$'s perspective by $\tilde{\psi}_\mathbf{\overline{E_2AS|E_1},E_2AS|E_1}$. We may then apply the standard jumping map to obtain the state in $\mathbf{E_2}$'s perspective:
\begin{equation}\label{etaTildeDef}
\tilde{\eta}_\mathbf{\overline{E_1AS|E_2},E_1AS|E_2}:=\mathcal{S}_\mathbf{E_1\to E_2}\left(\tilde{\psi}_\mathbf{\overline{E_2AS|E_1},E_2AS|E_1}\right).
\end{equation} 

If Eve 2 uses the formalism from her perspective according to steps 1 and 2 of Section~\ref{SEC:IntroToFormalism}, to derive the state in Alice's perspective inside the subsystem $\mathbf{AS}^\mathbf{(E_2)}$, we can describe the resulting state (through an identification analogous to~\eqref{eq:identification}) as
\begin{equation}\label{etaDef}
    \rho^\mathbf{(E_2)}_\mathbf{\overline{S|A},S|A}\cong \mathcal{U}^\mathbf{AS|E_2}_\mathbf{\to A|E_2} \circ \,\mathcal{G}_\mathbf{A|E_2, S|E_2}\left(\eta_\mathbf{AS|E_2}\right),
\end{equation}
where $\eta_\mathbf{AS|E_2}$ is the reduced state of $\mathbf{AS|E_2}$ obtained from $\tilde{\eta}_\mathbf{\overline{E_1AS|E_2},E_1AS|E_2}.$

On the other hand, Francis can directly apply the formalism to obtain the perspective of Alice in the subsystem $\mathbf{AS}^\mathbf{(F)}$:
\begin{equation}\label{xiDef}
    \rho^\mathbf{(F)}_\mathbf{\overline{S|A},S|A}:= \mathcal{U}^\mathbf{AS}_\mathbf{\to A}\,\circ \mathcal{G}_\mathbf{A,S} \left(\phi^\mathbf{(F)}_\mathbf{AS}\right),
\end{equation}
where $\phi^\mathbf{(F)}_\mathbf{AS}$ is the reduced state of $\mathbf{AS}^\mathbf{(F)}$ obtained from $\tilde{\phi}^\mathbf{(F)}_\mathbf{E_1E_2AS}.$

In Appendix~\ref{appendixInternalisation} we show that the states above have exactly the same matrix representation and coincide with the state of Alice's perspective assigned by Eve 1 in the corresponding canonical bases $\{\ket{x,y}\}_\mathbf{S|A',S|A}$. Namely, we have
\begin{equation}\label{AgreementStatement}
\bra{x',y'}\rho^\mathbf{(E_1)}_\mathbf{\overline{S|A},S|A}\ket{x, y}
=\bra{x',y'}\rho^\mathbf{(E_2)}_\mathbf{\overline{S|A},S|A}\ket{x, y}
=\bra{x',y'}\rho^\mathbf{(F)}_\mathbf{\overline{S|A},S|A}\ket{x, y},
\end{equation}
and the analogous statement holds for the matrix representation of any operator acting on Alice's perspective.

In conclusion, we observe that, although the subsystem that Eve~1 calls `$\mathbf{AS}$' and the one that Eve~2 calls `$\mathbf{AS}$' are the distinct subsystems $\mathbf{AS|E_1}$ and $\mathbf{AS|E_2}$ according to Francis, what Francis, Eve~1 and Eve~2 conclude regarding Alice's perspective depends on none of: (i) which internalised frame performs the computation, (ii) whether that frame is classical or quantum relative to the others, (iii) which layer of the observer tower one works at, or (iv) which internalisation $\tilde{\phi}^\mathbf{(F)}$ among the many consistent with $\rho^\mathbf{(E_1)}_\mathbf{\overline{S|A},S|A}$ was used to embed the description of Eve 1 in the first place. Eve 1, Eve 2 and Francis are therefore justified, for the purpose of defining the perspective of $\mathbf A$, to apply the formalism at their level independently of one another. Hence, the upper indices $\mathbf{(E_1), (E_2), (F)}$ for invariant states of $\mathbf{AS}$ are redundant and disposable.

\section{Going from a bipartite to a tripartite setup}\label{SEC:BipToTripEmbedding}
\noindent In this section we study the compositional properties of our framework. Specifically, we ask how a QRF perspective extends to a setup containing additional subsystems. We address this first from the standpoint of a fixed but arbitrary external observer, Eve, before giving a fully internal account.  

Suppose that the setup of Section~\ref{SEC:IntroToFormalism}, consisting of $\mathbf A$ and $\mathbf S$,  is part of a larger setup, which includes an additional subsystem $\mathbf{S_{3}}$. As $\mathbf{S}$ and $\mathbf{S_3}$ are systems on equal footing, we denote their (regular) representations by $U_x$. Eve can use textbook quantum theory to describe the new system. That is, Eve can describe a larger system by extending the original Hilbert space with an additional tensor factor, such that operators of the original system act nontrivially only on the original factor. Then, to $\mathbf{S_3}$ corresponds a new Hilbert space $\mathcal{H}_\mathbf{S_3}$, and the canonical embedding for operators from Eve's point of view is $M_\mathbf{AS}\mapsto M_\mathbf{AS}\otimes \mathds{1}_\mathbf{S_3}.$ 

With respect to Alice's QRF $\mathbf A$, the relative subsystems and the extra particle of the extended setup read (in Eve's factorisation)
\begin{align}
        \mathbf{S|A}&=  \left\{ \sum_x \ket{x}\bra{x}_\mathbf{A}\otimes \left(U_x M U^{\dagger}_x\right)_\mathbf{S}\otimes\mathds{1}_{\mathbf{S_3}} \Big|  M\in\mathcal{L}(\mathcal{H}_\mathbf{S})\right\}\label{S|A3},
\\
 \mathbf{S_{3}|A}&:= \left\{ \sum_x \ket{x}\bra{x}_\mathbf{A}\otimes \mathds{1}_\mathbf{S} \otimes \left(U_x M U^{\dagger}_x \right)_{\mathbf{S_3}}\Big|  M\in\mathcal{L}(\mathcal{H}_\mathbf{S_3})\right\},
\\
\mathbf{\overline{SS_{3}|A}}&= \left\{\sum_{xx'} \bra{x}M_R\ket{x'}\ket{
x}\bra{x'}_\mathbf{A}\otimes \left(U_x  U^{\dagger}_{x'}\right)_\mathbf{S} \otimes \left(U_x  U^{\dagger}_{x'}\right)_\mathbf{S_{3}} \Big| M_R \in \mathcal{L}(\mathcal{H}_\mathbf{A}) \text{  s.t.  } [M_R,L_y]=0 \text{  }\forall y\right \},
    \end{align}
where the total relative algebra  factorises as $\mathbf{SS_{3}|A} = \mathbf{S|A} \otimes \mathbf{S_{3}|A}$. 

The extra particle $\mathbf{\overline{SS_{3}|A}}$ acts non-trivially on all subsystems in Eve's factorisation. Moreover, the extra particle of the extended setup $\mathbf{\overline{SS_{3}|A}}$ is not the same subsystem as the extra particle of the reduced setup $\mathbf{\overline{S|A}}$, which is embedded in the extended setup as
    \begin{equation}\label{oldEP}
        \mathbf{\overline{S|A}}:  \left\{ \sum_{xx'} \bra{x}M_R\ket{x'}\ket{x}\bra{x'}_\mathbf{A}\otimes \left(U_x  U^{\dagger}_{x'}\right)_\mathbf{S}\otimes \mathds{1}_\mathbf{S_3}
        \Big| M_R \in \mathcal{L}(\mathcal{H}_\mathbf{A}) \text{  s.t.  } [M_R,L_y]=0 \text{  }\forall y
        \right\}.
    \end{equation}
The extra particle of the reduced setup, $\mathbf{\overline{S|A}}$, belongs to the algebra of invariant operators in the extended setup. However, crucially, operators in the reduced extra particle do not commute in general with operators in the newly added relative subsystem: 
\begin{equation}\label{EQnoncommutativity}
\Big[\mathbf{\overline{S|A}}, \mathbf{S_3|A} \Big] \neq 0.
\end{equation}
In this sense, the extra particle $\mathbf{\overline{S|A}}$ can be said to contain information about any potential relative subsystem external to the setup defined by $\mathbf{A}$ and $\mathbf S$~\cite{CastroRuiz2025a}.

\begin{figure}[h]
\centering
\begin{center}
\tikzset{every picture/.style={line width=0.75pt}} 

\begin{tikzpicture}[x=0.58pt,y=0.58pt,yscale=-1,xscale=1]

\draw [color={rgb, 255:red, 156; green, 12; blue, 0 }  ,draw opacity=1 ]   (316.96,71.37) .. controls (333.05,75.34) and (351.82,54.18) .. (282.1,52.85) .. controls (70.29,50.21) and (117.21,173.21) .. (319.64,146.76) ;
\draw [shift={(319.64,146.76)}, rotate = 172.55] [fill={rgb, 255:red, 156; green, 12; blue, 0 }  ,fill opacity=1 ][line width=0.08]  [draw opacity=0] (8.93,-4.29) -- (0,0) -- (8.93,4.29) -- cycle    ;

\draw (83.8,32.4) node    {$L(\mathcal{H}_{\mathbf{A}} \otimes \mathcal{H}_{\mathbf{S}})\quad$};
\draw (83.8,173) node    {$L(\mathcal{H}_{\mathbf{A}} \otimes \mathcal{H}_{\mathbf{S}} \otimes \mathcal{H}_{\mathbf{S_{3}}})\quad$};
\draw (403.4,32.4) node    {$\quad L(\mathcal{H}_{\mathbf{S|A'}} \otimes \mathcal{H}_{\mathbf{S|A}})$};
\draw (403.4,173) node    {  $\quad L(\mathcal{H}_{\mathbf{SS_{3} |A'}} \otimes \mathcal{H}_{\mathbf{S|A}} \otimes \mathcal{H}_{\mathbf{S_{3} |A}})$};
\draw (215.9,5.4) node [anchor=north west][inner sep=0.75pt]  [font=\normalsize,color={rgb, 255:red, 0; green, 0; blue, 0 }  ,opacity=1 ] [align=left] {$\displaystyle \mathcal{U}_{\rightarrow \mathbf{A}}^{\mathbf{AS}}$};
\draw (26.8,92.5) node [anchor=north west][inner sep=0.75pt]  [font=\normalsize,color={rgb, 255:red, 0; green, 0; blue, 0 }  ,opacity=1 ] [align=left] {$\displaystyle \ \ \otimes \mathds{1}_{\mathbf{S_{3}}}$};
\draw (477,64) node [anchor=north west][inner sep=0.75pt]  [font=\footnotesize,color={rgb, 255:red, 0; green, 0; blue, 0 }  ,opacity=1 ] [align=left] {$\displaystyle \mathrm{Embedding}$};
\draw (213.4,181) node [anchor=north west][inner sep=0.75pt]  [font=\normalsize,color={rgb, 255:red, 0; green, 0; blue, 0 }  ,opacity=1 ] [align=left] {$\displaystyle \mathcal{U}_{\mathbf{\rightarrow A}}^{\mathbf{ASS_{3}}}$};
\draw (408.4,84.69) node [anchor=north west][inner sep=0.75pt]  [font=\normalsize,color={rgb, 255:red, 0; green, 0; blue, 0 }  ,opacity=1 ] [align=left] {$\displaystyle \mathcal{V}_{\mathbf{A}} =\mathcal{U}_{\mathbf{\rightarrow A}}^{\mathbf{ASS_{3}}} \circ  \otimes \mathds{1}_{\mathbf{S_3}} \circ \left(\mathcal{U}_{\mathbf{\rightarrow A}}^{\mathbf{AS}}\right)^{\dagger } \ $};
\draw    (136.3,32.4) -- (335.4,32.4) ;
\draw [shift={(338.4,32.4)}, rotate = 180] [fill={rgb, 255:red, 0; green, 0; blue, 0 }  ][line width=0.08]  [draw opacity=0] (8.93,-4.29) -- (0,0) -- (8.93,4.29) -- (5.93,0) -- cycle    ;
\draw [shift={(133.3,32.4)}, rotate = 0] [fill={rgb, 255:red, 0; green, 0; blue, 0 }  ][line width=0.08]  [draw opacity=0] (8.93,-4.29) -- (0,0) -- (8.93,4.29) -- (5.93,0) -- cycle    ;
\draw    (83.8,53.4) -- (83.8,156.5) ;
\draw [shift={(83.8,159.5)}, rotate = 270] [fill={rgb, 255:red, 0; green, 0; blue, 0 }  ][line width=0.08]  [draw opacity=0] (8.93,-4.29) -- (0,0) -- (8.93,4.29) -- (5.93,0) -- cycle    ;
\draw [shift={(83.8,53.4)}, rotate = 90] [color={rgb, 255:red, 0; green, 0; blue, 0 }  ][line width=0.75]      (0,-8.94) .. controls (-2.47,-8.94) and (-4.47,-6.94) .. (-4.47,-4.47) .. controls (-4.47,-2) and (-2.47,0) .. (0,0) ;
\draw    (161.3,173) -- (296.4,173) ;
\draw [shift={(299.4,173)}, rotate = 180] [fill={rgb, 255:red, 0; green, 0; blue, 0 }  ][line width=0.08]  [draw opacity=0] (8.93,-4.29) -- (0,0) -- (8.93,4.29) -- (5.93,0) -- cycle    ;
\draw [shift={(158.3,173)}, rotate = 0] [fill={rgb, 255:red, 0; green, 0; blue, 0 }  ][line width=0.08]  [draw opacity=0] (8.93,-4.29) -- (0,0) -- (8.93,4.29) -- (5.93,0) -- cycle    ;
\draw [color={rgb, 255:red, 0; green, 0; blue, 0 }  ,draw opacity=1 ]   (403.4,53.4) -- (403.4,156.5) ;
\draw [shift={(403.4,159.5)}, rotate = 270] [fill={rgb, 255:red, 0; green, 0; blue, 0 }  ,fill opacity=1 ][line width=0.08]  [draw opacity=0] (8.93,-4.29) -- (0,0) -- (8.93,4.29) -- (5.93,0) -- cycle    ;
\draw [shift={(403.4,53.4)}, rotate = 90] [color={rgb, 255:red, 0; green, 0; blue, 0 }  ,draw opacity=1 ][line width=0.75]      (0,-8.94) .. controls (-2.47,-8.94) and (-4.47,-6.94) .. (-4.47,-4.47) .. controls (-4.47,-2) and (-2.47,0) .. (0,0) ;

\end{tikzpicture}
\end{center}
\vspace{-0.6cm}
\caption{Mapping operators of a bipartite setup into a tripartite setup in the perspective of QRF $\mathbf{A}$ in a way that is consistent with a potential external quantum description.} \label{FIG:EmbeddingDiagramOperators}
\end{figure}
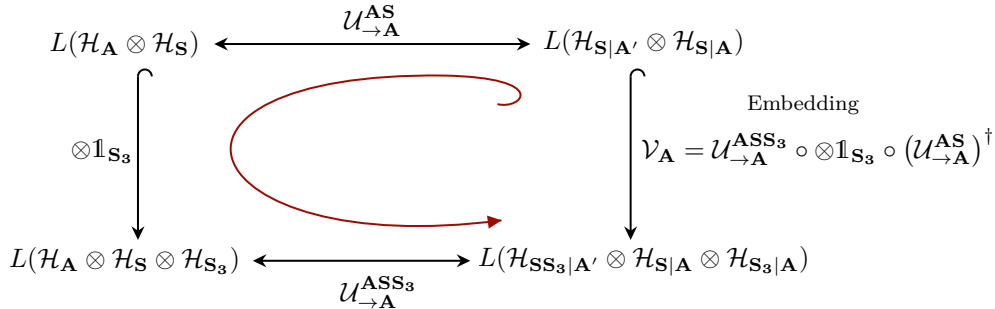

We now describe the algebras above in the factorisation natural to $\mathbf A$. 
First, we transform to this factorisation via the extended map
\begin{equation} \mathcal{U}^\mathbf{ASS_3}_{\to \mathbf{A}} := (\mathsf U^\mathbf{ASS_3}_{\to \mathbf{A}}) (\cdot) (\mathsf U_{\to \mathbf{A}}^\mathbf{ASS_3})^{\dagger}, \qquad \text{where} \qquad
   \mathsf U^\mathbf{ASS_3}_{\to \mathbf{A}}:= \sum_x \ket{x}\bra{x}_\mathbf{A} \otimes (U^{\dagger}_x)_\mathbf{S} \otimes (U^\dagger_x)_\mathbf{S_3},
\end{equation}
and we use the diagram of Fig.~\ref{FIG:EmbeddingDiagramOperators} to embed reduced setup into the extended one.

When the setup is extended, an operator $M\in L(\mathcal{H}_\mathbf{S|A', S|A})$ is therefore mapped to $\mathcal{V}_\mathbf{A}(M) =  V_\mathbf{A} (M\otimes \mathds{1}_\mathbf{S_3}) V_\mathbf{A}^{\dagger}$, where 
\begin{equation} \label{VUnitary}
V_\mathbf{A}:= \sum_x \ket{x}\bra{x}_\mathbf{S|A'} \otimes \mathds{1}_{\mathbf{S|A}}\otimes \left(U^{\dagger}_x\right)_\mathbf{S_{3}}
\end{equation} 
induces the refactorisation 
\begin{equation}
\mathcal H_{\mathbf{S|A'}} \otimes \mathcal H_{\mathbf{S|A}} \mathcal \otimes \mathcal H_{\mathbf{S_3}} \simeq \mathcal H_{\mathbf{S S_3|A'}} \otimes \mathcal H_{\mathbf{S|A}} \mathcal \otimes \mathcal H_{\mathbf{S_3 |A}}.
\end{equation}
The action of the group on $\mathcal H_{\mathbf{S|A'}} \otimes \mathcal H_{\mathbf{S|A}} \mathcal \otimes \mathcal H_{\mathbf{S_3}}$, given by $(L_x \otimes \mathds 1 \otimes U_x)_{\mathbf{S|A', S|A, S_3}}$, is transformed under $V_\mathbf A$ to $(L_x \otimes \mathds 1 \otimes \mathds 1)_{\mathbf{SS_3|A', S|A, S_3|A}}$. If we restrict  $\mathcal{V}_\mathbf{A}$ to act on the invariant bipartite algebra, we obtain an embedding of $\mathcal{B}^\mathbf{AS}_\text{inv}$ into $\mathcal{B}^\mathbf{ASS_3}_\text{inv}$. 

Observe that~\eqref{EQnoncommutativity} means there cannot exist any factorisation having both $\mathbf{\overline{S|A}}$ and $\mathbf{S_3|A}$ as factors. However, we may interpolate between factorisations including $\mathbf{\overline{S|A}}$ and $\mathbf{S_3|A}$ thanks to the map $V_\mathbf{A}$.
In Section~\ref{SEC:Adding and removing relative subsystems} we examine how the interplay between these two subsystems restricts the compositionality of relative systems from the perspective of  $\mathbf{A}$.

\subsection{State extensions}
\noindent
We can now understand how a state from the perspective of Alice is updated when subsystems are added or removed from the setup. Consider an invariant state $\rho \in L(\mathcal H_{\mathbf{S|A'}} \otimes \mathcal H_{\mathbf{S|A}})$. Imagine that the setup is now enlarged to contain an additional system $\mathbf{S_3}$, whose inclusion induces a state update 
\begin{equation}\label{UpdateRule}
\rho \longmapsto \tilde{\rho} \in L(\mathcal{H}_\mathbf{SS_3|A'}\otimes \mathcal{H}_\mathbf{S|A}\otimes \mathcal{H}_\mathbf{S_3|A}).
\end{equation}  We want to consider the cases in which~\eqref{UpdateRule} preserves the state $\rho$, which is embedded in the extended setup according to Diagram~\ref{FIG:EmbeddingDiagramOperators}. Demanding that $\rho$ is preserved captures the idea that no physical changes occur on the original setup: Alice is simply extending the mathematical description to incorporate a new system. With this in mind, we say that $\tilde{\rho}$ in~\eqref{UpdateRule} is an \textit{extension} of $\rho$ iff it is invariant under the action of $\mathbb Z _N$ on $L(\mathcal{H}_\mathbf{SS_3|A'}\otimes \mathcal{H}_\mathbf{S|A}\otimes \mathcal{H}_\mathbf{S_3|A})$ and respects  
\begin{equation} \label{ExtensionCondition}
\mathcal{V}^\dagger_\mathbf{A}(\tilde{\rho})=\operatorname{Tr}_\mathbf{S_3}\left[V_\mathbf{A}^\dagger \tilde{\rho} V_\mathbf{A}\right] = \rho,
\end{equation}  where $\mathcal{V}^\dagger_\mathbf{A}$ is the adjoint of the embedding $\mathcal{V}_\mathbf{A}$ in Fig.~\ref{FIG:EmbeddingDiagramOperators}. As shown in Section~\ref{SEC:InternalisingExternalFrames}, the construction leading to $\mathcal{V}_\mathbf{A}$ is independent of Eve and can therefore be fixed without loss of generality. 

\subsection{Compositionality and the paradox}
\label{SUBSEC:ParadoxesDoNotEmerge} 
\noindent
In this subsection, we go back to the scenario introduced in Subsection~\ref{SEC:TheProblem} (adapted to the $\mathbb{Z}_N$ example) to illustrate the compositional structure of the present framework. In particular, we explain why, by construction, the paradox of the third particle does not emerge.

\begin{figure}[h]
\centering
\tikzset{every picture/.style={line width=0.75pt}} 

\begin{tikzpicture}[x=0.6pt,y=0.6pt,yscale=-1,xscale=1]

\draw [color={rgb, 255:red, 156; green, 12; blue, 0 }  ,draw opacity=1 ][line width=0.75]    (215.34,162.47) .. controls (332.41,141.92) and (432.99,204.18) .. (429.72,86.43) ;
\draw [shift={(429.67,84.65)}, rotate = 87.93] [fill={rgb, 255:red, 156; green, 12; blue, 0 }  ,fill opacity=1 ][line width=0.08]  [draw opacity=0] (9.82,-4.72) -- (0,0) -- (9.82,4.72) -- cycle    ;
\draw [color={rgb, 255:red, 156; green, 12; blue, 0 }  ,draw opacity=1 ][line width=0.75]    (215.34,162.47) .. controls (222.96,54.36) and (272.5,61.74) .. (419.77,67.72) ;
\draw [shift={(422,67.81)}, rotate = 182.31] [fill={rgb, 255:red, 156; green, 12; blue, 0 }  ,fill opacity=1 ][line width=0.08]  [draw opacity=0] (9.82,-4.72) -- (0,0) -- (9.82,4.72) -- cycle    ;

\draw (201.22,48.39) node  [font=\large,color={rgb, 255:red, 0; green, 0; blue, 0 }  ,opacity=1 ]  {$\rho $};
\draw (201.22,174.08) node  [font=\large,color={rgb, 255:red, 0; green, 0; blue, 0 }  ,opacity=1 ]  {$\tilde{\rho }$};
\draw (450.34,48.39) node  [font=\large]  {$\sigma $};
\draw (450.34,174.08) node  [font=\large]  {$\tilde{\sigma }$};
\draw (298.67,17.51) node [anchor=north west][inner sep=0.75pt]  [font=\normalsize,color={rgb, 255:red, 0; green, 0; blue, 0 }  ,opacity=1 ] [align=left] {$\displaystyle \mathcal{S}_{\mathbf{A\rightarrow B}}^{\mathbf{ABS}}$};
\draw (297.17,182.08) node [anchor=north west][inner sep=0.75pt]  [font=\normalsize,color={rgb, 255:red, 0; green, 0; blue, 0 }  ,opacity=1 ] [align=left] {$\displaystyle \mathcal{S}_{\mathbf{A\rightarrow B}}^{\mathbf{ABSS}_{3}}$};
\draw (455.34,101.69) node [anchor=north west][inner sep=0.75pt]  [font=\normalsize,color={rgb, 255:red, 0; green, 0; blue, 0 }  ,opacity=1 ] [align=left] {$\displaystyle \mathcal{V}_{\mathbf{B}}^{\dagger } \ $};
\draw (157.18,197.84) node [anchor=north west][inner sep=0.75pt]  [font=\footnotesize,color={rgb, 255:red, 255; green, 158; blue, 0 }  ,opacity=1 ] [align=left] {$\displaystyle \ \ \ (\mathrm{extension}$)};
\draw (168.22,101.69) node [anchor=north west][inner sep=0.75pt]  [font=\normalsize,color={rgb, 255:red, 0; green, 0; blue, 0 }  ,opacity=1 ] [align=left] {$\displaystyle \mathcal{V}_{\mathbf{A}}^{\dagger } \ $};
\draw (43.34,112.19) node  [font=\large]  {$\rho _{\mathbf{BS|A}}$};
\draw (93.73,63.13) node [anchor=north west][inner sep=0.75pt]  [font=\normalsize,color={rgb, 255:red, 0; green, 0; blue, 0 }  ,opacity=1 ,rotate=-337.58] [align=left] {$\displaystyle \operatorname{Tr}_{\mathbf{BS|A'}}$};
\draw (84.47,132.61) node [anchor=north west][inner sep=0.75pt]  [font=\normalsize,color={rgb, 255:red, 0; green, 0; blue, 0 }  ,opacity=1 ,rotate=-21.68] [align=left] {$\displaystyle \operatorname{Tr}_{\mathbf{BSS_{3} |A',S_{3} |A}}$};
\draw (606.8,112.19) node  [font=\large]  {${\displaystyle \sigma }_{\mathbf{AS|B}}$};
\draw (498.79,40.55) node [anchor=north west][inner sep=0.75pt]  [font=\normalsize,color={rgb, 255:red, 0; green, 0; blue, 0 }  ,opacity=1 ,rotate=-22.84] [align=left] {$\displaystyle \operatorname{Tr}_{\mathbf{AS|B'}}$};
\draw (471.9,168.73) node [anchor=north west][inner sep=0.75pt]  [font=\normalsize,color={rgb, 255:red, 0; green, 0; blue, 0 }  ,opacity=1 ,rotate=-338.5] [align=left] {$\displaystyle \operatorname{Tr}_{\mathbf{ASS_{3} |B',S_{3} |B}}$};
\draw    (214.22,48.39) -- (437.84,48.39) ;
\draw [shift={(440.84,48.39)}, rotate = 180] [fill={rgb, 255:red, 0; green, 0; blue, 0 }  ][line width=0.08]  [draw opacity=0] (10.72,-5.15) -- (0,0) -- (10.72,5.15) -- (7.12,0) -- cycle    ;
\draw [shift={(211.22,48.39)}, rotate = 0] [fill={rgb, 255:red, 0; green, 0; blue, 0 }  ][line width=0.08]  [draw opacity=0] (10.72,-5.15) -- (0,0) -- (10.72,5.15) -- (7.12,0) -- cycle    ;
\draw [color={rgb, 255:red, 0; green, 0; blue, 0 }  ,draw opacity=1 ]   (201.22,65.39) -- (201.22,158.08) ;
\draw [shift={(201.22,62.39)}, rotate = 90] [fill={rgb, 255:red, 0; green, 0; blue, 0 }  ,fill opacity=1 ][line width=0.08]  [draw opacity=0] (10.72,-5.15) -- (0,0) -- (10.72,5.15) -- (7.12,0) -- cycle    ;
\draw    (214.22,174.08) -- (437.84,174.08) ;
\draw [shift={(440.84,174.08)}, rotate = 180] [fill={rgb, 255:red, 0; green, 0; blue, 0 }  ][line width=0.08]  [draw opacity=0] (10.72,-5.15) -- (0,0) -- (10.72,5.15) -- (7.12,0) -- cycle    ;
\draw [shift={(211.22,174.08)}, rotate = 0] [fill={rgb, 255:red, 0; green, 0; blue, 0 }  ][line width=0.08]  [draw opacity=0] (10.72,-5.15) -- (0,0) -- (10.72,5.15) -- (7.12,0) -- cycle    ;
\draw [color={rgb, 255:red, 0; green, 0; blue, 0 }  ,draw opacity=1 ]   (450.34,65.39) -- (450.34,158.08) ;
\draw [shift={(450.34,62.39)}, rotate = 90] [fill={rgb, 255:red, 0; green, 0; blue, 0 }  ,fill opacity=1 ][line width=0.08]  [draw opacity=0] (10.72,-5.15) -- (0,0) -- (10.72,5.15) -- (7.12,0) -- cycle    ;
\draw    (191.22,52.43) -- (74.62,99.55) ;
\draw [shift={(71.84,100.68)}, rotate = 337.99] [fill={rgb, 255:red, 0; green, 0; blue, 0 }  ][line width=0.08]  [draw opacity=0] (10.72,-5.15) -- (0,0) -- (10.72,5.15) -- (7.12,0) -- cycle    ;
\draw    (191.22,170.16) -- (74.63,124.46) ;
\draw [shift={(71.84,123.37)}, rotate = 21.41] [fill={rgb, 255:red, 0; green, 0; blue, 0 }  ][line width=0.08]  [draw opacity=0] (10.72,-5.15) -- (0,0) -- (10.72,5.15) -- (7.12,0) -- cycle    ;
\draw    (459.84,52.26) -- (575.52,99.44) ;
\draw [shift={(578.3,100.57)}, rotate = 202.19] [fill={rgb, 255:red, 0; green, 0; blue, 0 }  ][line width=0.08]  [draw opacity=0] (10.72,-5.15) -- (0,0) -- (10.72,5.15) -- (7.12,0) -- cycle    ;
\draw    (459.84,170.33) -- (575.51,124.57) ;
\draw [shift={(578.3,123.47)}, rotate = 158.42] [fill={rgb, 255:red, 0; green, 0; blue, 0 }  ][line width=0.08]  [draw opacity=0] (10.72,-5.15) -- (0,0) -- (10.72,5.15) -- (7.12,0) -- cycle    ;

\end{tikzpicture}

	\caption{Diagram showing the relation between states of Alice's and Bob's perspectives for a reduced setup consistently embedded in an extended setup. 
    The diagram highlights two properties. (1) The commutativity of the central square shows that transforming from one QRF perspective to another and then removing a subsystem is equivalent to first removing that subsystem and then transforming in the reduced setup. (2) The triangles on the right- and left-hand sides of the diagram show that the reduced state of a relative subsystem is independent of whether it is described as part of an extended setup and can be unambiguously obtained by tracing out the other systems in a given perspective. } \label{FIG:EmbeddingDiagramStates}
\end{figure}
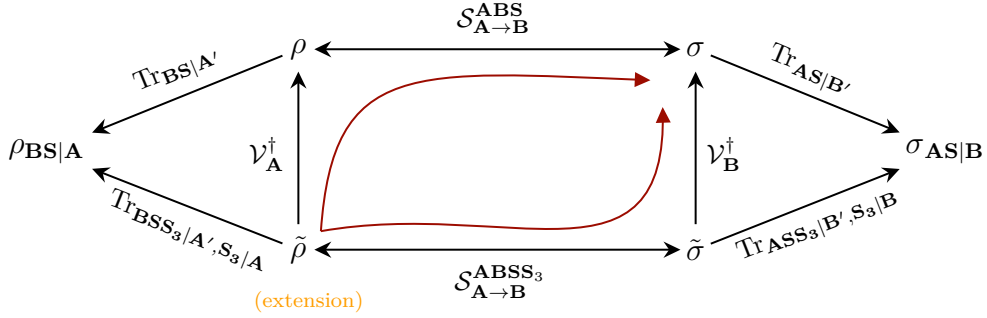

The diagram in Fig.~\ref{FIG:EmbeddingDiagramStates} illustrates 
the general situation. We have a reduced setup, with systems $\mathbf{ABS}$, which can be embedded in an extended setup with an additional system $\mathbf{S_3}$. Suppose that, from Alice's perspective, the state in the reduced setup is given by $\rho$ and in the extended setup is given by $\tilde{\rho}$. Because the addition of $\mathbf{S_3}$ entails no physical change and is merely an update of Alice's description, $\tilde{\rho}$ should be a valid extension of $\rho$ in the sense of~\eqref{ExtensionCondition}. In particular, $\tilde{\rho}$ can be an extension formed by adding a state of $\mathbf{S_3|A}$ in a tensorised form, as in Eq.~\eqref{eq:extendedstate}. 

We can jump from $\tilde{\rho}$ (bottom left of Fig.~\ref{FIG:EmbeddingDiagramStates}) to the state $\tilde{\sigma}$  relative to Bob in the extended setup (bottom right of Fig.~\ref{FIG:EmbeddingDiagramStates}) and apply $\mathcal{V}^\dagger_\mathbf{B}$ to get the reduced description for Bob (top right of Fig.~\ref{FIG:EmbeddingDiagramStates}).
Conversely, we can first  get the reduced description $\rho$ for Alice through $\mathcal{V}^\dagger_\mathbf{A}$ (top left of Fig.~\ref{FIG:EmbeddingDiagramStates}) and then jump from $\mathbf A$ to $\mathbf{B}$ in the reduced setup (top right of Fig.~\ref{FIG:EmbeddingDiagramStates}).
Both paths lead to the same final state $\sigma$, meaning that jumping to $\mathbf{B}$ and then tracing out a subsystem gives the same result as tracing it out first
and then jumping, provided the subsystem is correctly identified. If we look at the scenario from the perspective of the external observer ($\mathbf E$), this subsystem corresponds to $\mathbf{S_3}^{(\mathbf E)}$. By the results of Section~\ref{SEC:InternalisingExternalFrames}, the same conclusions hold for any other external observer. Importantly, the correct partial trace, which removes subsystem $\mathbf{S_3}^\mathbf{(E)}$, is implemented by the maps $\mathcal{V}^\dagger_\mathbf{A}$ and
$\mathcal{V}^\dagger_\mathbf{B}$, not by the partial trace of a relative subsystem with respect to $\mathbf A$ or $\mathbf B$. The same structure was studied in~\cite{Palumbo2026}, where the term `perspective
relational trace' was coined.

The map $\mathcal{V}^\dagger_\mathbf{B}$ acts only on the extra particle of $\mathbf B$ and on $\mathbf{S_3|\textbf{B}}$, leaving other relative subsystems untouched. Consequently, the state of any relative subsystem (e.g. $\mathbf{AS|B}$) of Bob's \emph{reduced} perspective equals the standard partial trace, in Bob's \emph{extended} perspective, over all the subsystems except the subsystem of interest. This is represented by the triangle on the right hand side of Fig.~\ref{FIG:EmbeddingDiagramStates}. A similar statement holds for Alice, as shown by the triangle on the left hand side. Therefore, our transformations satisfy the compositionality requirement introduced in Subsection~\ref{SEC:TheProblem}. More generally, the framework is compositional in the sense that it obeys the structure of the diagram in Fig.~\ref{FIG:EmbeddingDiagramStates}.

In the specific scenario of Subsection~\ref{SEC:TheProblem}, $\mathbf{A}$ is a reference frame in a definite position according to Eve. Therefore, the extra particle of the state from Alice's perspective factorises in the maximally mixed state. 
We compare the situation in which the third particle is not included, represented in the perspective of Alice by the state
\begin{equation} \label{paradox1}
\rho=\left(\frac{\mathds{1}}{N} \right)_\mathbf{\overline{BS|A}}\otimes\left[\left[ \frac{1}{\sqrt 2} (\ket{b_1}+e^{\iu \theta}\ket{b_2})\right]\right]_\mathbf{B|A}\otimes\ket{s}\bra{s}_\mathbf{S|A},
\end{equation} 
with the situation in which the third particle is present:
\begin{equation}\label{paradox2}
\tilde{\rho}=\left(\frac{\mathds{1}}{N} \right)_\mathbf{\overline{BSS_3|A}} \otimes \left[\left[ \frac{1}{\sqrt 2}(\ket{b_1}+e^{\iu \theta}\ket{b_2})\right]\right]_\mathbf{B|A}\otimes\ket{s}\bra{s}_\mathbf{S|A}\otimes \ket{s_3}\bra{s_3}_\mathbf{S_3|A},
\end{equation} where the double brackets on both expressions indicate the projector on the vector that lies inside. The update from state~(\ref{paradox1}) to state (\ref{paradox2}) is indeed a valid extension and the corresponding states in Bob's perspective are $\sigma$ and its valid extension $\tilde{\sigma}$, as represented in the diagram of Fig.~\ref{FIG:EmbeddingDiagramStates}. The reduced state of $\mathbf{AS|B}$ obtained after jumping to $\mathbf{B}$ in the extended setup is thus guaranteed to be the same as the one obtained if one jumps to $\mathbf{B}$'s perspective in the reduced setup\footnote{Note that, according to this framework, the information about the phase $\theta$ in both setups is not in the reduced state of $\mathbf{AS|B}$, but in the correlations between $\mathbf{AS|B}$ and the extra particle $\mathbf{\overline{AS|B}}$.}. Moreover, Bob can use the standard partial trace of quantum mechanics to obtain the state of $\mathbf{AS|B}$ both in the reduced and in the extended setup.
Therefore, the paradox of the third particle does not emerge by construction. 

It is also useful to examine the structure of perspectives given in Subsection~\ref{SEC:TheProblem} at the level of operators, since it provides the underlying mechanism for how the paradox arises.
Let us consider perspectives composed solely of relative subsystems. The critical subsystem is $\mathbf{A|B}$. From Bob's perspective, this subsystem  is given by operators of the form
\begin{equation}\label{eq:operatorformereduced}
M_{\mathbf{A|B}} \otimes \mathds{1}_{\mathbf{S|B}}.
\end{equation}
If we embed the system $\mathbf{A|B}$ canonically in the extended setup, the same operators after the extension are represented by
\begin{equation}\label{eq:operatorformextended}
M_{\mathbf{A|B}} \otimes \mathds{1}_{\mathbf{S|B}} \otimes \mathds{1}_{\mathbf{S_3|B}}.
\end{equation}
There is, however, another representation of these operators in the extended setup that should be equally valid. Namely, if we transform~\eqref{eq:operatorformereduced} to Alice's perspective according to the QRF transformation of Refs.~\cite{Giacomini2019, Hamette2020} in the reduced setup, then canonically embed the resulting operator, and finally apply the (extended) QRF transformation to go back to Bob's perspective, we obtain
\begin{equation}\label{eq:ambiguouslabels}
\sum_{x, x'} \bra{x}M\ket{x'}\ketbra{x}{x'} _{\mathbf{A|B}} \otimes \mathds{1}_{\mathbf{S|B}} \otimes (U^\dagger_{x}U_{x'})_{\mathbf{S_3|B}}.
\end{equation}  
 Eqs.~(\ref{eq:operatorformextended}, \ref{eq:ambiguouslabels}) therefore point to an ambiguity in the identification of operators that characterise a subsystem: if we believe that Alice and Bob are on the same footing regarding how they describe the extension of a setup, then~\eqref{eq:operatorformextended} must be correct; if we believe in the adequacy of the QRF transformation, then~\eqref{eq:ambiguouslabels} should correspond to the same operators. Moreover, Eq.~\eqref{eq:ambiguouslabels} presents a mismatch between the label `$\mathbf{A|B}$' and operators on Alice's frame from Bob's perspective, whose representation depends on any potential new subsystem one may choose to include. The same analysis from Alice's perspective leads to an ambiguity for operators and the label of system `$\mathbf{B|A}$'. It is important to note that this analysis does not rely on specific states of the system or the frames.

By contrast, the QRF transformation used in this work, first introduced in Ref.~\cite{CastroRuiz2025a}, induces consistent embeddings for all perspectives, leading to subsystem labels in the extended setup that can always be traced back to the corresponding subsystems in the reduced setup. That is, applying (reduced) QRF transformation from Bob to Alice to 
\begin{equation}\label{eq:operatorformextraparticlered}
\mathds 1_{\mathbf{\overline{AS|B}}} \otimes M_{\mathbf{A|B}} \otimes \mathds{1}_{\mathbf{S|B}} \otimes \mathds{1}_{\mathbf{S_3|B}},
\end{equation}
embedding the resulting operator from Alice's perspective via $\mathcal V_{\mathbf A}$, and transforming back to Bob via the (extended) QRF transformation yields
\begin{equation}\label{eq:operatorformextraparticleext}
\mathds 1_{\mathbf{\overline{ASS_3|B}}} \otimes M_{\mathbf{A|B}} \otimes \mathds{1}_{\mathbf{S|B}} \otimes \mathds{1}_{\mathbf{S_3|B}},
\end{equation}
which coincides with applying $\mathcal V_{\mathbf B}$ to~\eqref{eq:operatorformextraparticlered}.The resulting operator still acts on one particular subsystem identified by the corresponding label. The label therefore has an operational meaning in terms of the operator algebras. Importantly, the system labelled by $\mathbf{\overline{AS|B}}$ is no longer a part of the perspective, consistently with~\eqref{EQnoncommutativity}, and gives way to the new extra particle $\mathbf{\overline{ASS_3|B}}$. This result is the `dual' of the central square in Fig.~\ref{FIG:EmbeddingDiagramStates}.

\section{Adding and removing relative subsystems}\label{SEC:Adding and removing relative subsystems}

\noindent 
In ordinary quantum theory, where the frames are treated implicitly, the tensor product postulate specifies how a system can be extended by additional degrees of freedom without changing its physical state. In particular, if the quantum state of a given system is $\rho$ and $\gamma$ is an arbitrary state of a new system, the product state $\rho\otimes\gamma$ is always admissible. Extensions of this form are considered in Subsection~\ref{SEC:TheProblem} and used routinely in quantum theory to model the measurement process.

When quantum systems are used as reference frames, the situation is more subtle. In our framework, when a system $\mathbf{S_3}$ is added to a setup $\mathbf{AS}$, what determines the compatibility between the extended and the reduced setup is Eq.~\eqref{ExtensionCondition}. The extra-particle degrees of freedom become fundamental to understand how composition extends to QRF perspectives. 
Indeed, as noted in Section~\ref{SEC:BipToTripEmbedding}, Eq.~\eqref{EQnoncommutativity} implies that the subsystems $\mathbf{\overline{S|A}}$ and $\mathbf{S_3|A}$ cannot coexist within the same tensor product decomposition. Thus, the appearance of a new \emph{relative} subsystem in Alice's description is structurally tied to her extra particle. 

As a consequence, Alice's perspective is not in general compatible with the existence of a relative subsystem in an arbitrary tensorised state $\gamma_\mathbf{S_3|A}$. 
Determining when this is the case allows us to understand more concretely how the composition of relative subsystems in a perspective compares with that of a standard quantum description.

\subsection{Which states can be added from a QRF?}
\noindent
Here we ask which states admit a compatible extension by a new system  in a given tensorised relative state $\gamma$. We characterise them and find that compatibility depends only on the diagonal coefficients of 
 $\gamma$ in the momentum basis.

Since the general form of a translation invariant operator is~\eqref{MomentumSuperselectedOp}, let us write the initial and updated states as
\begin{equation}\label{PreAddConstraintState}
    \rho=\sum_p \mu_p \ket{p}\bra{p}_\mathbf{\overline{S|A}}\otimes \left(\phi_p\right)_\mathbf{S|A}\,
\end{equation}
and
\begin{equation}\label{PostAddConstraintState}
   \hspace{1.4cm} \tilde{\rho}=\sum_p \lambda_p \ket{p}\bra{p}_\mathbf{\overline{SS_{3}|A}}\otimes \left(\varphi_p\right)_\mathbf{S|A} \otimes \gamma_\mathbf{S_{3}|A},
\end{equation}
where the kets $\ket{p}$ are eigenstates of the operators $\hat{p}_\mathbf{\overline{S|A}}$ and $\hat{p}_\mathbf{\overline{SS_3|A}}$, respectively, $\phi_p$, $\varphi_p$ are normalised states of $\mathbf{S|A}$, $\mu_p, \ \lambda_p$ are probability distributions and the sum on $p$ goes from $0$ to $N-1.$ Our goal now is to check when  $\rho\mapsto\tilde{\rho}$ is a valid extension in the sense of~\eqref{ExtensionCondition}. Setting $\gamma_p := \bra{p}\gamma \ket{p}$ and plugging~(\ref{PreAddConstraintState}, \ref{PostAddConstraintState}) into~(\ref{ExtensionCondition}), we find
\begin{align}\label{GeneralAdmissibilityCondition}
\sum_p\ket{p}\bra{p}\otimes \left(\mu_p\phi_p -\sum_q \gamma_{(-q+p)} \lambda_{q}\varphi_q\right)\overset{!}{=}0 \quad \implies \quad \mu_p\phi_p \,\overset{!}{=} \, \sum_q \gamma_{(-q+p)} \lambda_{q}\varphi_q
\end{align} 
for every momentum $p$. The above breaks down further into two conditions:
\begin{itemize}
    \item[1.] Taking the trace on both sides of~(\ref{GeneralAdmissibilityCondition}), we have that the coefficients must in particular obey the relation:
\begin{equation} \label{ConvolutionCondition}
\mu_p = \sum_q \gamma_{(-q+p)}\lambda_{q}.
\end{equation}
\item[2.] For $\mu_p \neq 0$, Eq.~\eqref{GeneralAdmissibilityCondition} yields
 \begin{equation}\label{RelativeStatesCondition}
     \phi_p =\sum_q P^{p}_q\varphi_q,
 \end{equation}
for $P^{p}_q:=\gamma_{(-q+p)} \lambda_{q}/\mu_p$, which defines a probability distribution whenever~\eqref{GeneralAdmissibilityCondition} holds. If $\mu_p =0$ for some $p$ we must have $\gamma_{(-q+p)}\lambda_{q} = 0$ as all terms in~\eqref{ConvolutionCondition} are positive. In this case~\eqref{ConvolutionCondition} is trivially satisfied. 
\end{itemize}

Conditions~(\ref{ConvolutionCondition}, \ref{RelativeStatesCondition}) characterise state extensions. Whenever a valid extension $\tilde{\rho}$ exists for a fixed $\rho$, we say that $\gamma$ is \emph{admissible} by the initial state $\rho$. As anticipated, only the momentum distribution $\gamma_p$ is relevant for the admissibility of $\gamma$. We observe that, if the initial state from Alice's perspective factorises, $\rho = \chi_{\mathbf{\overline{S|A}}}\otimes \phi_\mathbf{\mathbf{S|A}},$~\eqref{GeneralAdmissibilityCondition} reduces to~\eqref{ConvolutionCondition}, which then becomes a necessary and sufficient condition. 

Let us analyse two extreme cases. Consider first the case where $\gamma$ is a momentum eigenstate, which we take to be $\ketbra{p=0}{p=0}_{\mathbf{S_3|A}}$ without loss of generality. Then, we have $\gamma_p  = \delta_{0,p}$, and conditions (\ref{ConvolutionCondition}, \ref{RelativeStatesCondition}) can always be satisfied by setting $\lambda_{p}=\mu_{p}$ and $\varphi_{p}=\phi_{p}$ $\forall p$. It is therefore always possible to add states of definite momentum. The opposite extreme occurs when $\gamma_p = 1/N$ $ \forall p$, which is consistent with adding a position eigenstate $\ketbra{x}{x}_{\mathbf{S_3|A}}$ or convex combinations thereof (including the maximally mixed state). In this case, \eqref{ConvolutionCondition} leads to $\mu_p =1/N$ $\forall p$ and~(\ref{RelativeStatesCondition}) implies that $\phi_p =: \phi$ is the same for all $p$. We thus reach the important conclusion that, for Alice’s description to be compatible with a third system in a convex combination of relative classical states, $\mathbf{\overline{S|A}}$ must factor out in the maximally mixed state,
\begin{equation} \label{AddPositionReqsIdentityEP}
    \rho = \left(\frac{\mathds{1}}{N}\right)_\mathbf{\overline{S|A}}\otimes \phi_\mathbf{S|A}.
\end{equation} 
Therefore, a relative system can only be added in \emph{any} tensorised state if the perspective is in a classical-like state  (see~\eqref{classical-quantum-statewrtalice}). 

The distribution $\lambda_p$ and the states $\varphi_q$ are given in implicit form in Eqs.~(\ref{ConvolutionCondition}, \ref{RelativeStatesCondition}). We now discuss how to obtain $\lambda_p$ in explicit form, which is sufficient to solve the case where $\rho$ factorises. The explicit form of $\varphi_q$ is left for further work. Given the distributions $\mu_p$ and $\gamma_p$, condition~\eqref{ConvolutionCondition} is a finite discrete convolution, so we can solve for $\lambda_p$
explicitly by passing to the Fourier domain, where convolution becomes multiplication. Define the discrete Fourier transform (DFT) of a distribution
$f$ as
\begin{equation}
\widehat{f}_j = \sum_{p=0}^{N-1} f_p \, e^{-2\pi \iu  \frac{jp}{N}},
\qquad \text{with inverse} \qquad 
f_p = \frac{1}{N} \sum_{j=0}^{N-1} \widehat{f}_j \, e^{2\pi \iu  \frac{pj}{N}}.
\end{equation}
Transforming~\eqref{ConvolutionCondition} then gives
$
\widehat{\mu}_j = \widehat{\gamma}_j\, \widehat{\lambda}_j.
$
If $\widehat{\gamma}_j \neq 0$ for all $j$, we can write
\begin{equation}\label{eq:fourierinversion}
\widehat{\lambda}_j = \frac{\widehat{\mu}_j}{\widehat{\gamma}_j},
\end{equation}
and $\lambda_p$ can be obtained by inverting the DFT. Eq.~\eqref{eq:fourierinversion} determines a unique candidate $\lambda_p$, which is automatically real and normalised. However, the candidate may not be a valid probability distribution as its entries can be negative. We must therefore impose $\lambda_p \geq 0$ for all $p$ as an additional requirement, which makes $\gamma$ inadmissible if it fails. If $\widehat\gamma_j =0$ for some $j$, then~\eqref{ConvolutionCondition} requires $\widehat\mu_j =0$ and $\widehat{\lambda}_j$ is not unique.

We now give condition~\eqref{ConvolutionCondition} a geometric interpretation. Writing $\gamma_p$ as a vector $\vec \gamma = (\gamma_0, \gamma_1, \dots, \gamma_{N-1})$, we define its cyclic shift $\operatorname S$ by
\begin{equation}
\operatorname{S}\vec \gamma =  \left(\gamma_{N-1},\gamma_0, \gamma_{1},...,\gamma_{N-2}\right),
\end{equation}
so that~(\ref{ConvolutionCondition}) can be written using powers of $\operatorname{S}$,
\begin{equation}\label{eq:cyclicshiftcondition}
\vec \mu= \sum_q \lambda_q \operatorname{S}^q\vec \gamma,
\end{equation} where $\vec{\mu}$ is the vector corresponding to distribution $\mu_p$. In this form, Eq.~\eqref{eq:cyclicshiftcondition} says that $\mu_k$ (initial state information) must lie in the convex hull of the cyclic shifts of $\gamma_k$ (information on the state of the additional system). To build intuition, let us study a concrete example.

\subsubsection*{Simple example (\(N=3\))}

 \begin{figure}
\centering
    \hspace{-1.5cm}\includegraphics[scale=0.4]{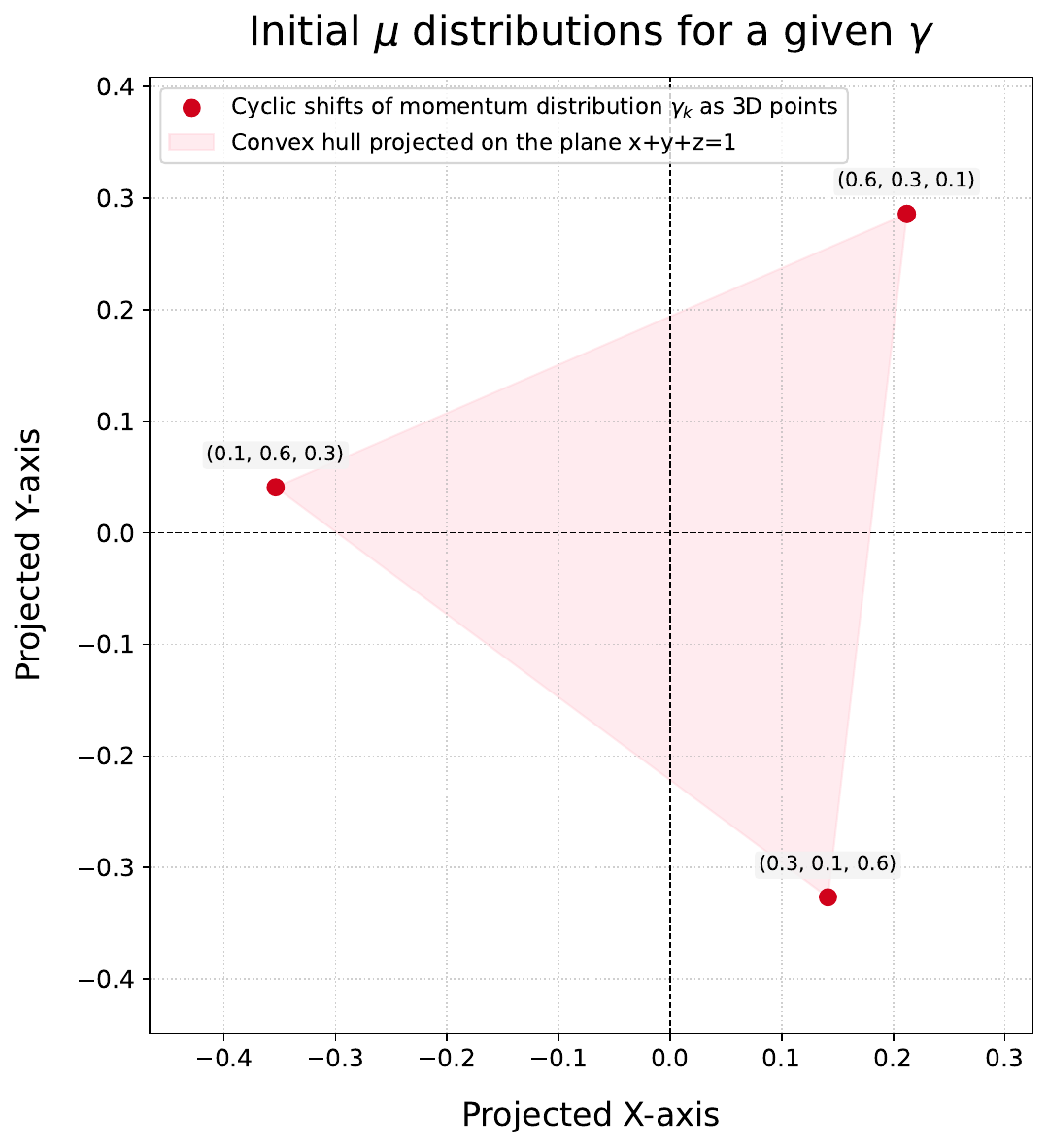}
    \caption{Convex hull defining which initial state distributions $\vec \mu$ obey the condition~\eqref{ConvolutionCondition} for admissibility of a state $\gamma$ with momentum diagonal $(0.6,0.3,0.1)$. If $\vec \mu$ lies outside the triangle, an initial state $\rho=\sum_k\mu_k\ket{k}\bra{k}\otimes \phi_k$ does not admit $\gamma$ as a valid extension.}
    \label{TriangleConvexHull}
\end{figure}
Let $N=3$. Given $\vec \gamma=(\gamma_0,\gamma_1,\gamma_2)$, its three cyclic shifts are three points in $\mathbb{R}^3$ lying in the probability simplex, the positive part of the plane $x+y+z=1$. Condition~\eqref{eq:cyclicshiftcondition} reads
\begin{equation}
\vec \mu = \lambda_0 \vec\gamma + \lambda_1 S \vec \gamma + \lambda_2 S^2 \vec\gamma.
\end{equation}
As $\vec\lambda=(\lambda_0,\lambda_1,\lambda_2)$
ranges over the simplex, the reachable $\vec\mu$ fill in the convex hull of these three points. For a fixed initial distribution $\vec\mu$ the state $\gamma$ is admissible precisely when $\vec \mu$ lies in the triangle spanned by $\vec \gamma$, $\operatorname{S}\vec\gamma$, and $\operatorname{S}^2\vec \gamma$. Take, for instance, a state $\gamma$ whose momentum diagonal is
\begin{equation}
\vec \gamma = (0.6,\, 0.3,\, 0.1).
\end{equation}
Its cyclic shifts are
\begin{equation}
S^0\gamma = (0.6,\, 0.3,\, 0.1), \qquad
S^1\gamma = (0.1,\, 0.6,\, 0.3), \qquad
S^2\gamma = (0.3,\, 0.1,\, 0.6),
\end{equation}
and~\eqref{eq:cyclicshiftcondition} becomes
\begin{equation}
\begin{pmatrix}
\mu_0 \\[4pt] \mu_1 \\[4pt] \mu_2
\end{pmatrix}
=
\begin{pmatrix}
0.6 & 0.1 & 0.3 \\[2pt]
0.3 & 0.6 & 0.1 \\[2pt]
0.1 & 0.3 & 0.6
\end{pmatrix}
\begin{pmatrix}
\lambda_0 \\[2pt] \lambda_1 \\[2pt] \lambda_2
\end{pmatrix},
\end{equation}
for a probability vector \(\vec \lambda = (\lambda_0, \lambda_1, \lambda_2)\). As \(\vec \lambda\) ranges over the probability simplex,
the possible corresponding $\vec \mu$ fill the triangle shown in Fig.~\ref{TriangleConvexHull}.

\textbf{Extreme cases.}
Consider the two extreme cases studied above for this example. The momentum eigenstate $\ketbra{p=0}{p=0}_{\mathbf{S_3|A}}$ gives $\vec \gamma=(1,0,0)$, and the three cyclic shifts are the canonical basis vectors
$(1,0,0)$, $(0,1,0)$, $(0,0,1)$. Their convex hull is the entire simplex, so~(\ref{eq:cyclicshiftcondition}) can hold for any $\vec \mu$. This is consistent with the fact that states of definite momentum can always be added. Convex combinations of position eigenstates $\ketbra{x}{x}_{\mathbf{S_3|A}}$ give $\vec\gamma = (1/3, 1/3, 1/3)$, which is invariant under cyclic shifts. The convex hull collapses to the barycentre of the simplex, leaving $\vec \mu = (1/3,1/3,1/3)$ as the only possibility, in agreement with the general statement that only a frame compatible with an external `classical' state can add a relative subsystem in a convex combination of position eigenstates. 

We can thus summarise the picture that has emerged. In standard quantum theory, one may use a fixed tensor product structure to capture the notion of subsystem independence, and the local description is always compatible with another subsystem existing in an arbitrary tensorised state. From a QRF perspective this no longer holds: not all local descriptions are generally compatible with the existence of a new \emph{relative} subsystem in a tensorised state. Relative subsystem independence is instead captured by Eq.~\eqref{ExtensionCondition}, which leads to conditions~\eqref{ConvolutionCondition} and~\eqref{RelativeStatesCondition} for states. In the special case where the state of the internal perspective is classical-like, we recover the description of standard quantum theory.

\subsection{Removing relative subsystems}

\noindent
 In the previous subsection we studied when adding the tensorised state of a relative subsystem
 $\mathbf{S_3|A}$ defines a valid extension, i.e., preserves the original state of the reduced setup. This naturally suggests asking the reverse: when does removing the relative subsystem $\mathbf{S_3|A}$ from  $\tilde{\rho} \in L(\mathcal H_{\mathbf{SS_3|A'}} \otimes \mathcal H_{\mathbf{S|A}} \otimes  \mathcal H_{\mathbf{S_3|A}})$ preserve the original reduced state 
\(
\rho=\mathcal{V}^\dagger_\mathbf{A} \left(\tilde{\rho}\right)
\)?  
However, we cannot directly make sense of this question, as $\operatorname{Tr}_\mathbf{S_3|A}(\tilde{\rho})$ and $\mathcal{V}^\dagger_\mathbf{A}(\tilde{\rho})$ live in different spaces. 

Nevertheless, operationally, Alice may ask when losing all information about $\mathbf{S_3|A}$ and then removing $\mathbf{S_3}$ is equivalent to just removing $\mathbf{S_3}$. More precisely, she may ask  when the following condition holds:
\begin{equation}\label{ReductionCondition}
     \mathcal{V}^\dagger_\mathbf{A}(\tilde{\rho})=  \mathcal{V}^\dagger_\mathbf{A}\left(\operatorname{Tr}_\mathbf{S_3|A}(\tilde{\rho})\otimes ( \mathds{1}/N)_\mathbf{S_3|A}\right),
\end{equation}
or equivalently, when $\operatorname{Tr}_\mathbf{S_3|A}(\tilde{\rho})\otimes ( \mathds{1}/N)_\mathbf{S_3|A}$ is a valid extension of $\rho$. Since $(\mathds{1}/N)_\mathbf{S_3|A}$ is a convex combination of position eigenstates,  Eq.~\eqref{AddPositionReqsIdentityEP} implies that~\eqref{ReductionCondition} can only be satisfied if 
the extra particle $\mathbf{\overline{S|A}}$ in state $\rho$ factorises out in the  maximally mixed state. Therefore, that the system $\mathbf{S|A}$ is left in the same reduced state whether one removes $\mathbf{S_3}$ or its relative counterpart $\mathbf{S_3|A}$ is a non-trivial requirement which only holds in the perspective of a frame in a classical-like state.

Now imagine a setup including another QRF $\mathbf B$ so that the states $\rho$ and $\tilde \rho$ now include this QRF. In Section~\ref{SEC:BipToTripEmbedding}, we studied how one can jump between QRF perspectives if a system $\mathbf{S_3}$ is removed or not from the description. We may now ask whether Alice can `jump' to Bob's perspective after losing access to the \emph{relative} subsystem $\mathbf{S_3|A}$. Strictly speaking, the state $\operatorname{Tr}_\mathbf{S_3|A}(\tilde{\rho})$ would not belong to any perspective, as it is identified with an element of $\mathbf{\overline{BSS_3|A} \vee \mathbf{BS|A}}$ instead of $\mathbf{\overline{BS|A} \vee \mathbf{BS|A}}$. However, we can still make operational sense of Alice losing access to $\mathbf{S_3|A}$ and then jumping to Bob's perspective. She can do this by applying the QRF transformation to the state $ \mathcal{V}^\dagger_\mathbf{A} \left(\operatorname{Tr}_\mathbf{S_3|A}(\tilde{\rho})\otimes ( \mathds{1}/N)_\mathbf{S_3|A}\right)$. The state thus obtained will in general differ from $\mathcal{S}_{\mathbf{A} \to \mathbf B}(\rho)$. This is not surprising, as $\mathbf{A|B}$ is not independent of $\mathbf{S_3|A}$.

\section{Classicalised frames: an unrestricted adding rule for QRFs}\label{SEC:AddingWClassicalisation}
\noindent
The preceding sections show that the extra particle determines how closely the description of an internal observer, Alice, resembles that of standard quantum theory with an implicit frame. When her extra particle factorises out in the maximally mixed state, her frame $\mathbf{A}$ admits a classical-quantum external description as in~\eqref{classical-quantum-statewrteve}. In this case, the state of the relative subsystems reproduces the external description up to a mixture of translations as in~\eqref{classical-quantum-statewrtalice}. Crucially, the standard composition rules of quantum theory hold: we can always find valid extensions in which a relative subsystem is added in an arbitrary tensorised state. Conversely, tracing out the relative subsystem $\mathbf{S_3|A}$ does not affect the reduced setup where $\mathbf{S_3}$ is removed. When the extra particle departs from the maximally mixed state, however, Alice's description differs from the standard one.

Given the special status of the tensorised maximally mixed state, a natural question arises: is there a transformation that maps a general state in a QRF perspective to a state in which the extra particle factorises in the maximally mixed state? In this section we show that such a transformation exists, provided the external frame is first internalised. We call this procedure the \emph{classicalisation} of a QRF perspective. Classicalisation yields an update rule for adding subsystems in arbitrary tensorised states. Although the rule violates the extension condition~(\ref{ExtensionCondition}), it admits a physically meaningful interpretation.

\subsection{Classicalisation}

\noindent

Consider a general initial state for the bipartite setup $\mathbf{AS}$, as described by Eve: 
\begin{equation}\label{preAddingInitialStateForE}
\psi= \frac{1}{N}\sum_{aa'ss'}\psi_{asa's'}\ket{a}\bra{a'}_\mathbf{A}\otimes \ket{s}\bra{s'}_\mathbf{S},
\end{equation} 
where $a,a',s,s'\in \mathbb{Z}_N$. Jumping to QRF $\mathbf{A}$, we obtain
\begin{equation}\label{preAddingInitialStateForA}
\rho=\frac1N\left[\sum_{aa'ss'}\psi_{asa's'}R^{\dagger}_{a}R_{a'}\otimes U^{\dagger}_{a}\ket{s}\bra{s'}U_{a'}\right]_\mathbf{\overline{S|A}, S|A}
\end{equation}
where $R_x$ denotes an element of the right-regular representation, acting as $R_x\ket{y}=\ket{yx^{-1}}= \ket{y-x}$. 

The state~(\ref{preAddingInitialStateForA}) is subject to the restrictions on adding derived in Section~\ref{SEC:Adding and removing relative subsystems}. Consider, however, the extended setup in which Eve's frame is internalised as a subsystem $\mathbf{E}$ in a sharply localised state. The state in the perspective of QRF $\mathbf{E}$ is then given by {\begin{equation}\label{EveIncluded}
\left(\frac{\mathds{1}}{N}\right)_\mathbf{\overline{AS|E}}\otimes \psi_\mathbf{AS|E}, \quad  \text{where} \quad \psi_\mathbf{AS|E}\cong \psi.
\end{equation} Applying the jumping map $\mathcal{S}^\mathbf{EAS}_\mathbf{E\to A}$ to~\eqref{EveIncluded} yields
\begin{equation}
\frac1N    \Bigl[\sum_{aa'ss'}\psi_{asa's'} \ket{-a}\bra{-a'}\otimes R^{\dagger}_{a}R_{a'}\otimes \ket{-a+s}\bra{-a'+s'}\Bigr]_\mathbf{E|A, \overline{ES|A}, S|A}. \label{InternalisedEStatefromA}
\end{equation}
We now show that the subsystem $\mathbf{E|A}$ enters~(\ref{InternalisedEStatefromA}) in such a way that a global unitary can bring the extra particle to the maximally mixed state for any initial state~\eqref{preAddingInitialStateForE}. Define $\mathcal{C}_\mathbf{A}$ as
\begin{equation}\label{ClassicalisationMap}
    \mathcal{C}_\mathbf{A}= C_\mathbf{A}(\cdot)C_\mathbf{A}^{\dagger} \qquad C_\mathbf{A}= \sum_x \ket{x}\bra{x}_\mathbf{E|A} \otimes \left(R_x^{\dagger}\right)_\mathbf{\overline{ES|A}}\otimes \mathds{1}_\mathbf{S|A}.
\end{equation}
The map $\mathcal C_{\mathbf A}$ commutes with the action of the group, since $[R_x, L_{x'}] = 0$ for all $x$ and $x'$ (even in the non-Abelian case). We thus do not interpret it as a refactorisation in the sense of~(\ref{jumpingmap}, \ref{VUnitary}) but rather as a change of coordinates within Alice's factorisation. Applying $\mathcal{C}_\mathbf{A}$ to~(\ref{InternalisedEStatefromA}) yields
\begin{equation}
    \left(\frac{\mathds{1}}{N}\right)_\mathbf{\overline{ES|A}}  \otimes \Bigl[\sum_{aa'ss'}\psi_{asa's'} \ket{-a}\bra{-a'}\otimes \ket{-a+s}\bra{-a'+s'}\Bigr]_\mathbf{ES|A}. \label{classtatePreAdd}
\end{equation}
Therefore, for a general state~\eqref{preAddingInitialStateForE}, internalising $\mathbf{E}$ as above and applying $\mathcal{C}_\mathbf{A}$ always returns a tensorised maximally mixed extra particle for $\mathbf{A}$. More generally, $\mathcal{C}_\mathbf{A}$ copies the state of $\mathbf{E}$'s extra particle onto $\mathbf{A}$'s extra particle. We refer to the procedure above as the \emph{classicalisation} of QRF $\mathbf{A}$, since it brings its perspective to a classical-like state. The procedure can be reversed: jumping to QRF $\mathbf{E}$ from~(\ref{classtatePreAdd}) and applying the map analogous to~\eqref{ClassicalisationMap} for Eve returns the state~\eqref{EveIncluded}.

Classicalisation defines a map from Eve's perspective to Alice's, taking~\eqref{EveIncluded} to~\eqref{classtatePreAdd}. Namely
\begin{equation}
\mathcal{S}^{\text{class.}}_{\mathbf{E\to A}} = C_\mathbf{A}\circ \mathcal{S}_\mathbf{E\to A}.
\end{equation}
In Appendix~\ref{appendixB}, we show that $\mathcal{S}^{\text{class.}}_{\mathbf{E\to A}}$ acts on the relative subsystems exactly as the transformations of Refs.~\cite{Giacomini2019, Hamette2020}, providing a link between the present formalism and `purely internal' approaches.  

In the classicalised frame, the restrictions on adding disappear. It is therefore natural to add an arbitrary tensorised state $\gamma_\mathbf{S_{3}|A}$ and then reverse the classicalisation procedure in the enlarged setup. As shown in Section~\ref{SEC:Adding and removing relative subsystems}, the state~(\ref{classtatePreAdd}) admits a valid extension of the form 
\begin{equation}\label{ClassicalisedExtendedStateGamma}
\left(\frac{\mathds{1}}{N}\right) _\mathbf{\overline{ESS_{3}|A}} \otimes  \Bigl[\sum_{aa'ss'}\psi_{asa's'} \ket{-a}\bra{-a'}\otimes \ket{-a+s}\bra{-a'+s'}\Bigr]_\mathbf{ES|A}\otimes \gamma_\mathbf{S_{3}|A}
\end{equation} for any $\gamma_\mathbf{S_{3}|A}$. We may now jump to QRF $\mathbf{E}$ and classicalise it through the map
\begin{equation}
     \mathcal{C}_\mathbf{E} = C_\mathbf{E}(\cdot)C_\mathbf{E}^{\dagger} \qquad C_\mathbf{E}= \sum_x \ket{x}\bra{x}_\mathbf{A|E} \otimes \left(R_x^{\dagger}\right)_\mathbf{\overline{ASS_3|E}} \otimes \mathds{1}_{\mathbf{S|E}}\otimes\mathds{1}_\mathbf{S_3|E}.  
\end{equation}
The resulting state is
\begin{equation}
\left(\frac{\mathds{1}}{N}\right)_\mathbf{\overline{ASS_{3}|E}}\otimes \Bigl[ \sum_{aa'ss'} \psi_{asa's'} \ket{a}\bra{a'} \otimes \ket{s}\bra{s'}\otimes U_{a}\gamma U^{\dagger}_{a'}\Bigr]_\mathbf{A|E, S|E, S_{3}|E}. \label{ClassicalisationOfEve} 
\end{equation}

Finally, we trace out the trivial extra particle, promote Eve to an external observer, and transform to Alice's perspective. The initial state~\eqref{preAddingInitialStateForA} of Alice's perspective is thus transformed as:
\begin{equation}\label{ModifiedUpdateRuleforA}
\rho \longmapsto \tilde{\rho}=\rho_{\mathbf{\overline{SS_3|A}, \mathbf{S|A}}} \otimes \gamma_{\mathbf{S_3|A}}.
\end{equation}
That is, $\rho$ retains the same form but the extra particle is reassigned as $\mathbf{\overline{S|A}} \to \mathbf{\overline{SS_3|A}}$. Eq.~\eqref{ModifiedUpdateRuleforA} defines a new \emph{update rule} relating the states in $\mathbf{A}$'s perspective before and after a system is added in a tensorised relative state. The update rule~\eqref{ModifiedUpdateRuleforA} does not in general define a valid extension in the sense of~\eqref{ExtensionCondition}.
Nevertheless, it has a meaningful physical interpretation, as we explain next.

\subsection{Interpretation of composing in classicalised frames}
\noindent 
The rule~(\ref{ModifiedUpdateRuleforA}) may be interpreted as the `internal' preparation of a new state $\gamma_\mathbf{S_3|A}$. In agreement with momentum conservation, the extra particle $\mathbf{\overline{S|A}}$ is transferred to the new extra particle $\mathbf{\overline{SS_3|A}}$.
To illustrate this interpretation, consider a classical scenario in which a cannon fires cannonballs, as shown in Fig.~(\ref{FIG:ClassicalCannon}). Suppose a cannonball is described as a separate subsystem only after it is fired; from that moment onward, the momentum of the new cannonball is singled out as one of the contributions to the total momentum. From the external point of view, when the first cannonball $\mathbf{S}$ is shot, the cannon $\mathbf{A}$ recoils in such a way that the total momentum $p_{\text{tot}}$
is conserved. When an additional cannonball $\mathbf{S_3}$ is fired, the cannon recoils once more. The total momentum is again conserved, but the momentum of the subsystem composed of the cannon $\mathbf{A}$ and the first cannonball $\mathbf{S}$ is not: once $\mathbf{S_3}$ is singled out as a new subsystem, the subsystem $\mathbf{AS}$ receives a recoil, thereby changing its momentum.

\begin{figure}[h]
    \centering
\includegraphics[width=0.9\linewidth]{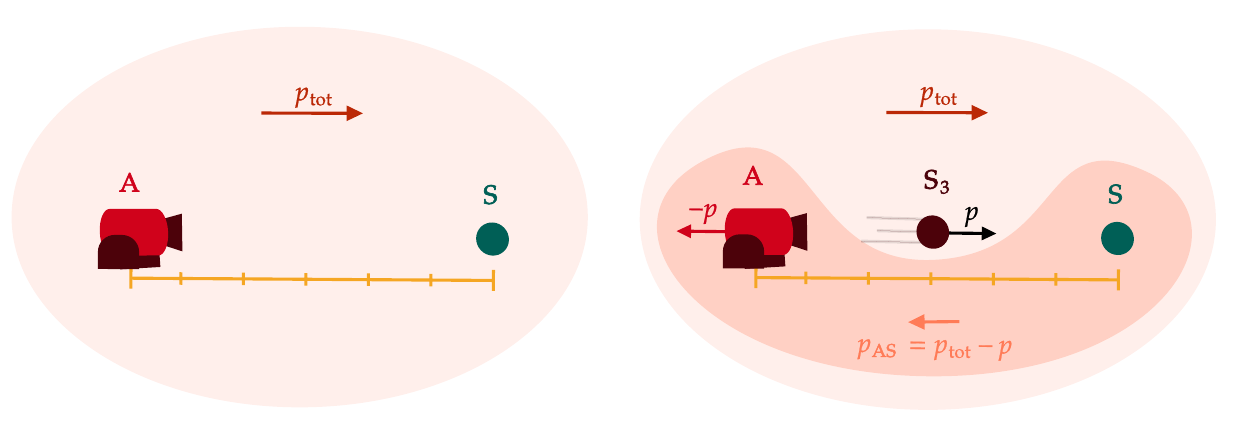}
    \caption{Cannon analogy for the classicalised adding rule~\eqref{ModifiedUpdateRuleforA}. Left: the bipartite setup $\mathbf{AS}$ with total momentum $p_\text{tot}$. Right: firing the cannonball $\mathbf{S_3}$. By momentum conservation, the firing of $\mathbf{S_3}$ causes a recoil on $\mathbf{A}$, effectively changing the momentum of the old subsystem $\mathbf{AS}$ from $p_\text{tot}$ to $p_\text{tot}−p$.  The updated state of the subsystem in~\eqref{MomentumRecoilExample} shows a generalisation of this classical behaviour, where the `recoil' appears term by term in the momentum-superselected state.
}
    \label{FIG:ClassicalCannon}
\end{figure}

Let us now describe an analogous situation using our formalism. Suppose that, as seen by Eve, the bipartite setup $\mathbf{AS}$
is in an eigenstate of total momentum, with $\mathbf{S}$ is in a definite position relative to $\mathbf{A}$: 
\begin{equation} \label{DefTotalMomentumPreaddingStateForEve}
    \ket{\psi} = \frac{1}{\sqrt{N}}\sum_{a}\left(e^{\frac{2\pi \iu}{N} a p_\text{tot}}\right)\ket{a}_\mathbf{A}\ket{a+d}_\mathbf{S},
\end{equation} where $p_\text{tot}$ is the value of total momentum of $\mathbf A$ and $\mathbf S$.
The state above is already invariant under discrete translations and can be written in Alice's perspective as
\begin{equation}\label{eq:originalstate} 
    \ket{\rho}=\mathcal{U}^\mathbf{AS}_\mathbf{\to A}\ket{\psi}=\ket{p_\text{tot}}_\mathbf{S|A'}\otimes \ket{d}_{\mathbf{S|A}}=\ket{-p_\text{tot}}_\mathbf{\overline{S|A}} \otimes \ket{d}_{\mathbf{S|A}},
\end{equation} where the sign appears due to the relation $\ket{p}_\mathbf{\overline{S|A}} = \ket{-p}_\mathbf{S|A'}$ $\forall$ $p.$
 Suppose the relative subsystem $\mathbf{S_3}$ is now added from $\mathbf{A}$'s perspective in a pure state corresponding to $\ket{\gamma}_\mathbf{S_3|A}$, according to update rule~(\ref{ModifiedUpdateRuleforA}). The new state from Alice's perspective reads
\begin{equation} \label{PostAddingStateExample}
\ket{\tilde{\rho}}= \ket{-p_\text{tot}}_\mathbf{\overline{SS_3|A}} \otimes \ket{d}_{\mathbf{S|A}}\otimes \ket{\gamma}_\mathbf{S_3|A}.
\end{equation}
What does the state of the original systems look like after this update? We can obtain it by rewriting~(\ref{PostAddingStateExample}) in the factorisation $\mathcal{H}_\mathbf{S|A'}\otimes \mathcal{H}_{\mathbf{S|A}}\otimes\mathcal{H}_\mathbf{S_3}$ by means of~\eqref{VUnitary} and tracing out the factor $\mathbf{S_3}$. This amounts to applying the operation $\mathcal{V}^\dagger_\mathbf{A}$, resulting in
\begin{equation}\label{MomentumRecoilExample}
    \mathcal{V}^\dagger_\mathbf{A}\left(\ket{\tilde{\rho}}\bra{\tilde{\rho}}\right)=\sum_p \left|\bra{p}\gamma \rangle \right|^2 \ket{-(p_\text{tot}-p)}\bra{-(p_\text{tot}-p)}_\mathbf{\overline{S|A}} \otimes \ket{d}\bra{d}_{\mathbf{S|A}}.
\end{equation} The state~\eqref{MomentumRecoilExample} is still invariant, but it is different from the original state~\eqref{eq:originalstate}. Now $\mathbf{AS}$ does not have momentum $p_\text{tot}$, but is rather in a mixture of momenta weighted by the distribution $\left|\bra{p}\gamma \rangle \right|^2$. The change of momentum can be attributed to a `recoil' received by $\mathbf A$ depending on the diagonal momentum coefficients of $\gamma$. This is analogous to the situation depicted in Fig.~\ref{FIG:ClassicalCannon}.

The link between the update and the cannon's momentum shift is most natural for eigenstates of total momentum. However, because quantum systems cannot simultaneously have sharply defined position and momentum, `recoil' is not the most intuitive way of understanding what happens in position basis. Let us then study one more example. Consider the state 
\begin{equation} \label{InvariantPartpreaddinglocalised}
    \ket{\psi} = \frac{1}{\sqrt{2}}\left( \ket{0} \ket{d}+e^{\iu\theta}\ket{1}\ket{1+d}\right)_\mathbf{AS}.
\end{equation} As in~(\ref{DefTotalMomentumPreaddingStateForEve}), the relative position of $\mathbf S$ with respect to $\mathbf A$ is definite and given by $d$. However, the total momentum is not definite anymore. Jumping to the perspective of $\mathbf A$ gives
\begin{equation} \label{LocalisedExampleClassicalisedAdding}
   \rho=\mathcal{U}^\mathbf{AS}_\mathbf{\to A}\circ \mathcal{G}_\mathbf{A,S} \left(\ket{\psi}\bra{\psi}\right)= \frac{1}{N}\left(\mathds{1}+ \cos \hat{\Theta}\right)_\mathbf{\overline{S|A}} \otimes \ket{d}\bra{d}_{\mathbf{S|A}}, \qquad \hat{\Theta}:= \theta \mathds{1} + \frac{2\pi}{N} \hat{p}_\mathbf{\overline{S|A}}.
\end{equation}
\begin{figure}[h]
    \centering
    \includegraphics[width=0.5\linewidth]{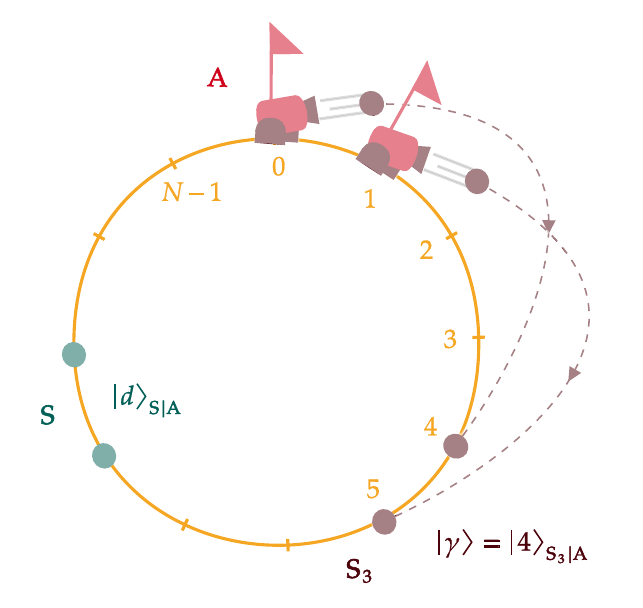}
    \caption{The effect of adding in a classicalised frame. When the system $\mathbf{AS}$ is in a superposition of sharply localised states according to an external observer, the addition of a localised state $\gamma_\mathbf{S_3|A}$ as per~\eqref{ModifiedUpdateRuleforA} 
leads to a tripartite entangled state and to the decoherence of the original system in position basis.
  } 
    \label{FIG:AddingEntangled}
\end{figure}
Adding a pure state $\gamma_\mathbf{S_3|A} = \ketbra{\gamma}{\gamma}_{\mathbf{S_3|A}}$ through the rule~(\ref{ModifiedUpdateRuleforA}) updates~\eqref{LocalisedExampleClassicalisedAdding} to
\begin{equation}  \label{LocalisedExamplePostAdding}
  \tilde{\rho}=\frac{1}{N}\left(\mathds{1}+ \cos \hat{\Theta} \right)_\mathbf{\overline{SS_3|A}} \otimes \ket{d}\bra{d}_{\mathbf{S|A}}\otimes \ket{\gamma}\bra{\gamma}_\mathbf{S_3|A}.
\end{equation} 
From~(\ref{LocalisedExamplePostAdding}) we can calculate the reduced state of the original setup, obtaining
\begin{equation} \label{DecoheredStateClassicalisedAdding}
\mathcal{V}^\dagger_\mathbf{A}\left(\ket{\tilde{\rho}}\bra{\tilde{\rho}}\right)=\frac{1}{N}\left(\mathds{1}+  \sum_{xx'}\operatorname{Tr}\left(L_x \gamma L^{\dagger}_{x'}\right)\bra{x} \cos\hat{\Theta} \ket{x'}\ket{x}\bra{x'}\right)_\mathbf{\overline{S|A}} \otimes \ket{d}\bra{d}_{\mathbf{S|A}}.
\end{equation}
Comparing~(\ref{LocalisedExampleClassicalisedAdding}) with~(\ref{DecoheredStateClassicalisedAdding}) reveals how the initial state is affected by the addition of $\mathbf{S_3}$. Specifically,  we see that each position off-diagonal element is dampened by a factor $\operatorname{Tr}\left(L_x \gamma L^{\dagger}_{x'}\right)$.

It is instructive to consider this process from Eve's external perspective. According to her, a possible state giving rise to~\eqref{LocalisedExamplePostAdding} is given by
\begin{equation}\label{LocalisedExamplePostAddingExternal}
    \ket{\tilde{\psi}}= \frac{1}{\sqrt{2}}\left( \ket{0}_\mathbf{A} \ket{d}_\mathbf{S}\ket{\gamma}_\mathbf{S_3}+e^{\iu\theta}\ket{1}_\mathbf{A}\ket{1+d}_\mathbf{S}L_1\ket{\gamma}_\mathbf{S_3}\right).
\end{equation} The update from~\eqref{InvariantPartpreaddinglocalised} to~\eqref{LocalisedExamplePostAddingExternal} is intuitive: from Eve's perspective, the new state $\ket \gamma$ is added `branch by branch' to the entangled state $(\ket{0,d} + \mathrm{exp}(\iu \theta) \ket{1,1+d})/\sqrt 2$ with the corresponding translation dictated by the state of QRF $\mathbf A$. Therefore, the suppression of coherences in~\eqref{DecoheredStateClassicalisedAdding} is compatible with the decoherence of~\eqref{LocalisedExamplePostAddingExternal} after tracing out $\mathbf{S_3}$, as depicted in Fig.~\ref{FIG:AddingEntangled}. Both~\eqref{LocalisedExamplePostAddingExternal} and its decohered reduction are, however, specific choices among the external states compatible with the relevant invariant states; the update itself is fully captured by the invariant~(\ref{LocalisedExamplePostAdding}).

\section{Conclusions}
\noindent
The search for a quantum generalisation of reference frame transformations faces several challenges. In particular, consistently combining quantum reference frame transformations with the composition of subsystems is of crucial importance. In this work, we have argued that the formalism of Ref.~\cite{CastroRuiz2025a} provides a satisfactory solution to this problem, and our analysis allows us to explore a number of interesting scenarios.

By construction, the framework is compatible with the potential existence of external reference frames, with respect to which subsystems are composed as in standard quantum theory. We show that external frames can be treated on the same footing as internal ones, leading to a consistent hierarchy of quantum reference frame perspectives. The description of a quantum reference frame perspective is independent of the potential external reference frame. It is also independent of the layer in the hierarchy one decides to describe it and of whether the subsystem might be embedded into larger portions of the universe. Our analysis generalises the way subsystems are composed in ordinary quantum theory. In particular, there are restrictions to adding relative subsystems in an arbitrary factorised state, and we have characterised the cases where these extensions are possible. The case of standard quantum theory is recovered for a specific type of perspectives, which we call `classical-like'. There is a sense, however, in which the restrictions for adding can be circumvented. This is done through  a procedure we call `classicalisation’, which leads to a rule for adding relative subsystems that has a meaningful physical interpretation.  

Several open questions remain. First, our construction relies on the extra-particle degrees of freedom, whose operational interpretation is not fully clear, although some recent works have made progress in this direction~\cite{delaHamette2026, Doat2026}. Relatedly, the external-frame independence of the extra particle suggests the puzzling idea that there may be an `absolute' notion of a reference frame being in a quantum state. Second, we have focused only on the finite Abelian case. While the main structural aspects of our work seem to be directly generalisable, this is not so clear when it comes to our more specific results. It would be specially interesting to characterise the restrictions to adding relative tensorised states for continuous and non-Abelian groups, as well as for non-ideal quantum reference frames~\cite{Garmier2025}. 

Finally, it would be very interesting to understand in depth how the compositional structure of the present formalism relates to broader research directions and to other approaches to quantum reference frames.  In particular, it might be worth investigating whether the embedding maps developed here are related to recent efforts in understanding potential generalisations of the compositional structure of quantum theory~\cite{Erba2026}. Moreover, it would be important to investigate the precise connection between the extra particle and the `edge-mode’ degrees of freedom, which play a key role in the compositional structure of gauge theories and gravity (see, e.g.~\cite{donnelly2016local, carrozza2022edge}). On a similar vein, it would be interesting to explore further connections between the present  approach and the perspective-neutral framework~\cite{Vanrietvelde2020, Hamette2021}, which is closely related to Dirac quantisation and has recently been employed to study subsystems in gravitational scenarios~\cite{DeVuyst2025, DeVuyst2025a}. The `classicalisation' procedure introduced here may be an important point of contact. Moreover, it would be insightful to understand how this procedure relates to perspectival approaches~\cite{Giacomini2019, Hamette2020}, and might be connected to the concept of alignable states~\cite{Krumm2021}. 

More broadly, we hope that the analysis presented here contributes to overcoming the conceptual challenges required to an understanding of physics `as seen' by a quantum reference frame.

\vspace{1cm}
\noindent \textbf{Note added.} Before the completion of this work we became aware of the recent work by A. Palumbo and L. Apadula~\cite{Palumbo2026}, which analyses the paradox of the third particle and the notion of frame-dependent traces in different approaches to QRFs, including the formalism used in this work. 
\section*{Acknowledgements}
\noindent We thank Ognyan Oreshkov for insightful discussions on several topics of this work. B.S. thanks S.C. Garmier, N. S. Móller, A. Spalvieri,  M. Mekonnen, A. Giovanakis, D. Mattei, G. Doat for useful discussions and T. Ehrensberger for feedback on the draft.

\printbibliography

@Article{Bartlett2007,
  author    = {Bartlett, Stephen D. and Rudolph, Terry and Spekkens, Robert W.},
  journal   = {Rev. Mod. Phys.},
  title     = {Reference frames, superselection rules, and quantum information},
  year      = {2007},
  pages     = {555--609},
  volume    = {79},
  doi       = {10.1103/RevModPhys.79.555},
  issue     = {2},
  publisher = {American Physical Society},
}

@Misc{Hardy2001,
  title={Quantum Theory From Five Reasonable Axioms}, 
      author={Lucien Hardy},
      year={2001},
      eprint={quant-ph/0101012},
      archivePrefix={arXiv},
      primaryClass={quant-ph},
      url={https://arxiv.org/abs/quant-ph/0101012}, 
      %doi = {10.48550/arXiv.quant-ph/0101012}
}

@Article{Aharonov1967,
  author  = {Aharonov, Yakir and Susskind, Leonard},
  journal = {Phys. Rev.},
  title   = {Charge Superselection Rule},
  year    = {1967},
  pages   = {1428--1431},
  volume  = {155},
  doi     = {10.1103/PhysRev.155.1428},
  issue   = {5},
}

@Article{Giacomini2019,
  author  = {Giacomini, Flaminia and Castro-Ruiz, Esteban and Brukner, {\v{C}}aslav},
  journal = {Nat. Commun.},
  title   = {Quantum mechanics and the covariance of physical laws in quantum reference frames},
  year    = {2019},
  issn    = {2041-1723},
  number  = {1},
  pages   = {494},
  volume  = {10},
  doi     = {10.1038/s41467-018-08155-0},
}

@Article{Aharonov1984,
  author  = {Aharonov, Yakir and Kaufherr, Tirzah},
  journal = {Phys. Rev. D},
  title   = {Quantum frames of reference},
  year    = {1984},
  pages   = {368--385},
  volume  = {30},
  doi     = {10.1103/PhysRevD.30.368},
  issue   = {2},
}

@Article{Kitaev2004,
  author  = {Kitaev, Alexei and Mayers, Dominic and Preskill, John},
  journal = {Phys. Rev. A},
  title   = {Superselection rules and quantum protocols},
  year    = {2004},
  number  = {5},
  pages   = {052326},
  volume  = {69},
  doi     = {10.1103/PhysRevA.69.052326},
}

@Article{Loveridge2017,
  author  = {Loveridge, Leon and Busch, Paul and Miyadera, Takayuki},
  journal = {Europhysics Letters},
  title   = {Relativity of quantum states and observables},
  year    = {2017},
  number  = {4},
  pages   = {40004},
  volume  = {117},
  doi     = {10.1209/0295-5075/117/40004},
}

@Misc{Glowacki2023,
  author        = {Jan G{\l}owacki},
  title         = {Operational Quantum Frames: An operational approach to quantum reference frames},
  year          = {2023},
  archiveprefix = {arXiv},
  doi           = {10.48550/arXiv.2304.07021},
  eprint        = {2304.07021},
  primaryclass  = {quant-ph},
}

@Article{Carette2025,
  author    = {Carette, Titouan and Glowacki, Jan and Loveridge, Leon},
  journal   = {{Quantum}},
  title     = {Operational {Q}uantum {R}eference {F}rame {T}ransformations},
  year      = {2025},
  pages     = {1680},
  volume    = {9},
  doi       = {10.22331/q-2025-03-27-1680},
  publisher = {{Verein zur F{\"{o}}rderung des Open Access Publizierens in den Quantenwissenschaften}},
}

@Article{Vanrietvelde2020,
  author  = {Vanrietvelde, Augustin and H{\"o}hn, Philipp A. and Giacomini, Flaminia and Castro-Ruiz, Esteban},
  journal = {Quantum},
  title   = {A change of perspective: switching quantum reference frames via a perspective-neutral framework},
  year    = {2020},
  pages   = {225},
  volume  = {4},
  doi     = {10.22331/q-2020-01-27-225},
}

@Article{CastroRuiz2020,
  author  = {Castro-Ruiz, Esteban and Giacomini, Flaminia and Belenchia, Alessio and Brukner, {\v{C}}aslav},
  journal = {Nat. Commun.},
  title   = {Quantum clocks and the temporal localisability of events in the presence of gravitating quantum systems},
  year    = {2020},
  number  = {1},
  pages   = {2672},
  volume  = {11},
  doi     = {10.1038/s41467-020-16013-1},
}

@Article{Hoehn2021,
  author  = {H{\"o}hn, Philipp A. and Smith, Alexander R. H. and Lock, Maximilian P. E.},
  journal = {Phys. Rev. D},
  title   = {Trinity of relational quantum dynamics},
  year    = {2021},
  number  = {6},
  pages   = {066001},
  volume  = {104},
  doi     = {10.1103/PhysRevD.104.066001},
}

@Article{Hamette2020,
  author  = {de la Hamette, Anne-Catherine and Galley, Thomas D.},
  journal = {Quantum},
  title   = {Quantum reference frames for general symmetry groups},
  year    = {2020},
  issn    = {2521-327X},
  pages   = {367},
  volume  = {4},
  doi     = {10.22331/q-2020-11-30-367},
}

@Article{Krumm2021,
  author  = {Krumm, Marius and H{\"o}hn, Philipp A. and M{\"u}ller, Markus P.},
  journal = {Quantum},
  title   = {Quantum reference frame transformations as symmetries and the paradox of the third particle},
  year    = {2021},
  issn    = {2521-327X},
  pages   = {530},
  volume  = {5},
  doi     = {10.22331/q-2021-08-27-530},
}

@Misc{Hamette2021,
  author        = {de la Hamette, Anne-Catherine and Galley, Thomas D. and H{\"o}hn, Philipp A. and Loveridge, Leon and M{\"u}ller, Markus P.},
  title         = {Perspective-neutral approach to quantum frame covariance for general symmetry groups},
  year          = {2021},
  archiveprefix = {arXiv},
  doi           = {10.48550/arXiv.2110.13824},
  eprint        = {2110.13824},
  primaryclass  = {quant-ph},
}

@Article{AhmadAli2022,
  author  = {Ahmad Ali, Shadi and Galley, Thomas D. and H{\"o}hn, Philipp A. and Lock, Maximilian P. E. and Smith, Alexander R. H.},
  journal = {Phys. Rev. Lett.},
  title   = {Quantum Relativity of Subsystems},
  year    = {2022},
  number  = {17},
  pages   = {170401},
  volume  = {128},
  doi     = {10.1103/PhysRevLett.128.170401},
}

@Article{Hoehn2022,
  author  = {H{\"o}hn, Philipp A. and Krumm, Marius and M{\"u}ller, Markus P.},
  journal = {J. Math. Phys.},
  title   = {Internal quantum reference frames for finite Abelian groups},
  year    = {2022},
  number  = {11},
  pages   = {112207},
  volume  = {63},
  doi     = {10.1063/5.0088485},
}

@Article{Ballesteros2021,
  author  = {Ballesteros, Angel and Giacomini, Flaminia and Gubitosi, Giulia},
  journal = {Quantum},
  title   = {The group structure of dynamical transformations between quantum reference frames},
  year    = {2021},
  pages   = {470},
  volume  = {5},
  doi     = {10.22331/q-2021-06-08-470},
}

@Article{DeVuyst2025,
  author  = {De Vuyst, Julian and Eccles, Stefan and H{\"o}hn, Philipp A. and Kirklin, Josh},
  journal = {Journal of High Energy Physics},
  title   = {Gravitational entropy is observer-dependent},
  year    = {2025},
  number  = {7},
  pages   = {146},
  volume  = {2025},
  doi     = {10.1007/JHEP07(2025)146},
}

@Article{Zanardi2004,
  author    = {Zanardi, Paolo and Lidar, Daniel A. and Lloyd, Seth},
  journal   = {Phys. Rev. Lett.},
  title     = {Quantum Tensor Product Structures are Observable Induced},
  year      = {2004},
  pages     = {060402},
  volume    = {92},
  doi       = {10.1103/PhysRevLett.92.060402},
  issue     = {6},
  publisher = {American Physical Society},
}

@Article{Knill2000,
  author    = {Knill, Emanuel and Laflamme, Raymond and Viola, Lorenza},
  journal   = {Phys. Rev. Lett.},
  title     = {Theory of Quantum Error Correction for General Noise},
  year      = {2000},
  pages     = {2525--2528},
  volume    = {84},
  doi       = {10.1103/PhysRevLett.84.2525},
  issue     = {11},
  publisher = {American Physical Society},
}

@Article{Zanardi2001,
  author    = {Zanardi, Paolo},
  journal   = {Phys. Rev. Lett.},
  title     = {Virtual Quantum Subsystems},
  year      = {2001},
  pages     = {077901},
  volume    = {87},
  doi       = {10.1103/PhysRevLett.87.077901},
  issue     = {7},
  publisher = {American Physical Society},
}

@Misc{Rio2015,
  author        = {del Rio, Lidia and Kraemer, Lea and Renner, Renato},
  title         = {Resource theories of knowledge},
  year          = {2015},
  archiveprefix = {arXiv},
  doi           = {10.48550/arxiv.1511.08818},
  eprint        = {1511.08818},
  eprinttype    = {arxiv},
  primaryclass  = {quant-ph},
}

@Article{Chiribella2018,
  author         = {Chiribella, Giulio},
  journal        = {Entropy},
  title          = {Agents, Subsystems, and the Conservation of Information},
  year           = {2018},
  number         = {5},
  volume         = {20},
  article-number = {358},
  %doi            = {10.3390/e20050358},
  url   = {https://doi.org/10.3390/e20050358}
}

@Misc{Vanrietvelde2025,
   title={Partitions in quantum theory}, 
      author={Augustin Vanrietvelde and Octave Mestoudjian and Pablo Arrighi},
      year={2025},
      eprint={2506.22218},
      archivePrefix={arXiv},
      primaryClass={quant-ph},
      url={https://arxiv.org/abs/2506.22218}, 
}

@Misc{CastroRuiz2025,
  author        = {Castro-Ruiz, Esteban and Galley, Thomas D. and Loveridge, Leon},
  title         = {Interpreting quantum reference frame transformations through a simple example},
  year          = {2025},
  archiveprefix = {arXiv},
  doi           = {10.48550/arxiv.2508.09540},
  eprint        = {2508.09540},
  eprinttype    = {arxiv},
  primaryclass  = {quant-ph},
}

@Article{CastroRuiz2025a,
  author  = {Castro-Ruiz, Esteban and Oreshkov, Ognyan},
  journal = {Communications Physics},
  title   = {Relative subsystems and quantum reference frame transformations},
  year    = {2025},
  number  = {1},
  pages   = {187},
  volume  = {8},
  doi     = {10.1038/s42005-025-02036-x},
}

@Article{Hausmann2025,
  author    = {Hausmann, Ladina and Schmidhuber, Alexander and Castro-Ruiz, Esteban},
  journal   = {{Quantum}},
  title     = {Measurement events relative to temporal quantum reference frames},
  year      = {2025},
  pages     = {1616},
  volume    = {9},
  doi       = {10.22331/q-2025-01-30-1616},
  publisher = {{Verein zur F{\"{o}}rderung des Open Access Publizierens in den Quantenwissenschaften}},
}

@Article{DeVuyst2025a,
  author  = {De Vuyst, Julian and Eccles, Stefan and H{\"o}hn, Philipp A. and Kirklin, Josh},
  journal = {Journal of High Energy Physics},
  title   = {Crossed products and quantum reference frames: on the observer-dependence of gravitational entropy},
  year    = {2025},
  number  = {7},
  pages   = {63},
  volume  = {2025},
  doi     = {10.1007/JHEP07(2025)063},
}

@Article{Barnum2004,
  author    = {Barnum, Howard and Knill, Emanuel and Ortiz, Gerardo and Somma, Rolando and Viola, Lorenza},
  journal   = {Phys. Rev. Lett.},
  title     = {A Subsystem-Independent Generalization of Entanglement},
  year      = {2004},
  pages     = {107902},
  volume    = {92},
  doi       = {10.1103/PhysRevLett.92.107902},
  issue     = {10},
  publisher = {American Physical Society},
}

@Article{Kabel2025,
  author  = {Kabel, Viktoria and de la Hamette, Anne-Catherine and Apadula, Luca and Cepollaro, Carlo and Gomes, Henrique and Butterfield, Jeremy and Brukner, {\v{C}}aslav},
  journal = {Communications Physics},
  title   = {Quantum coordinates, localisation of events, and the quantum hole argument},
  year    = {2025},
  number  = {1},
  pages   = {185},
  volume  = {8},
  doi     = {10.1038/s42005-025-02084-3},
}

@Book{Eddington1949,
  author    = {Eddington, Arthur S.},
  publisher = {Cambridge University Press},
  title     = {Fundamental Theory},
  year      = {1949},
}

@Misc{Merriam2005,
  author        = {Merriam, Paul},
  title         = {Quantum Relativity: Physical Laws Must be Invariant Over Quantum Systems},
  year          = {2005},
  archiveprefix = {arXiv},
  doi           = {10.48550/arxiv.quant-ph/0506228},
  eprint        = {quant-ph/0506228},
  eprintclass   = {quant-ph},
  eprinttype    = {arxiv},
  primaryclass  = {quant-ph},
}

@Misc{Adlam2025,
    title={Wigner's Frame}, 
      author={Emily Adlam},
      year={2025},
      eprint={2512.07101},
      archivePrefix={arXiv},
      primaryClass={quant-ph},
      url={https://arxiv.org/abs/2512.07101}, 
}

@Article{Chiribella2011,
   title={Informational derivation of quantum theory},
   volume={84},
   ISSN={1094-1622},
   url={http://dx.doi.org/10.1103/PhysRevA.84.012311},
   %DOI={10.1103/physreva.84.012311},
   number={1},
   journal={Physical Review A},
   publisher={American Physical Society (APS)},
   author={Chiribella, Giulio and D’Ariano, Giacomo Mauro and Perinotti, Paolo},
    month=Jul,
   year={2011}
   }

@Article{Masanes2011,
   title={A derivation of quantum theory from physical requirements},
   volume={13},
   ISSN={1367-2630},
   url={http://dx.doi.org/10.1088/1367-2630/13/6/063001},
   DOI={10.1088/1367-2630/13/6/063001},
   number={6},
   journal={New Journal of Physics},
   publisher={IOP Publishing},
   author={Masanes, Lluís and Müller, Markus P},
   year={2011},
   month=Jun, pages={063001} }

@article{Angelo2011,
   title={Physics within a quantum reference frame},
   volume={44},
   ISSN={1751-8121},
   url={http://dx.doi.org/10.1088/1751-8113/44/14/145304},
   DOI={10.1088/1751-8113/44/14/145304},
   number={14},
   journal={Journal of Physics A: Mathematical and Theoretical},
   publisher={IOP Publishing},
   author={Angelo, Renato M and Brunner, Nicolas and Popescu, Sandu and Short, Anthony J and Skrzypczyk, Paul},
   year={2011},
   month=mar, pages={145304} }

@book{NielsenChuang,
  author = {Nielsen, Michael A. and Chuang, Isaac L.},
  publisher = {Cambridge University Press},
  title = {Quantum Computation and Quantum Information},
  year = 2000
}

@misc{Garmier2025,
         title={The Perspectives of Non-Ideal Quantum Reference Frames}, 
      author={Sébastien Christophe Garmier and Ladina Hausmann and Esteban Castro-Ruiz},
      year={2026},
      eprint={2512.19343},
      archivePrefix={arXiv},
      primaryClass={quant-ph},
      url={https://arxiv.org/abs/2512.19343},  
}

@article{Doat2026,
   title={What can we do in a symmetry-constrained perspective? The importance of the total charge's status in quantum reference frame frameworks},
   volume={10},
   ISSN={2521-327X},
   url={http://dx.doi.org/10.22331/q-2026-06-08-2126},
   DOI={10.22331/q-2026-06-08-2126},
   journal={Quantum},
   publisher={Verein zur Forderung des Open Access Publizierens in den Quantenwissenschaften},
   author={Doat, Guilhem and Vanrietvelde, Augustin},
   year={2026},
   month=jun, pages={2126} }

@misc{delaHamette2026,
      title={Accessibility of Global Properties from Internal Quantum Reference Frame Perspectives}, 
      author={Anne-Catherine de la Hamette and Viktoria Kabel and Časlav Brukner},
      year={2026},
      eprint={2510.09100},
      archivePrefix={arXiv},
      primaryClass={quant-ph},
      url={https://arxiv.org/abs/2510.09100}, 
}

@article{Poulin2006,
   title={Toy Model for a Relational Formulation of Quantum Theory},
   volume={45},
   ISSN={1572-9575},
   url={http://dx.doi.org/10.1007/s10773-006-9052-0},
   DOI={10.1007/s10773-006-9052-0},
   number={7},
   journal={International Journal of Theoretical Physics},
   publisher={Springer Science and Business Media LLC},
   author={Poulin, David},
   year={2006},
   month=jun, pages={1189–1215} 
   }

@Article{Carcassi2021,
   title={Four Postulates of Quantum Mechanics Are Three},
   volume={126},
   ISSN={1079-7114},
   url={http://dx.doi.org/10.1103/PhysRevLett.126.110402},
   %DOI={10.1103/physrevlett.126.110402},
   number={11},
   journal={Physical Review Letters},
   publisher={American Physical Society (APS)},
   author={Carcassi, Gabriele and Maccone, Lorenzo and Aidala, Christine A.},
   year={2021},
   month=Mar }

@misc{Erba2026,
      title={The composition rule for quantum systems is not the only possible one}, 
      author={Marco Erba and Paolo Perinotti},
      year={2024},
      eprint={2411.15964},
      archivePrefix={arXiv},
      primaryClass={quant-ph},
      url={https://arxiv.org/abs/2411.15964}, 
}

@misc{Renato2026,
      title={The Paradox of the Third Particle is classical}, 
      author={Ladina Hausmann and Renato Renner},
      year={2026},
      eprint={2607.15351},
      archivePrefix={arXiv},
      primaryClass={quant-ph},
      url={https://arxiv.org/abs/2607.15351}, 
}

@misc{Brukner2026,
      title={An Operational Resolution of the Third-Particle Paradox}, 
      author={Časlav Brukner and Esteban Castro-Ruiz and Marius Krumm},
      year={2026},
      eprint={2607.20815},
      archivePrefix={arXiv},
      primaryClass={quant-ph},
      url={https://arxiv.org/abs/2607.20815}, 
}

@misc{Palumbo2026,
      title={Frame-Dependent Traces and the Third-Particle Paradox}, 
      author={Alessandro Palumbo and Luca Apadula},
      year={2026},
      eprint={2607.21703},
      archivePrefix={arXiv},
      primaryClass={quant-ph},
      url={https://arxiv.org/abs/2607.21703}, 
}

@article{Mekonnen2026,
	title={Invariance under quantum permutations rules out parastatistics},
	volume={17},
	ISSN={2041-1723},
	url={http://dx.doi.org/10.1038/s41467-026-73064-6},
	%DOI={10.1038/s41467-026-73064-6},
	number={1},
	journal={Nature Communications},
	publisher={Springer Science and Business Media LLC},
	author={Mekonnen, Manuel and Galley, Thomas D. and Müller, Markus P.},
	year={2026},
	month=May }

@article{LoveridgeThesis2018,
	title={Symmetry, Reference Frames, and Relational Quantities in Quantum Mechanics},
	volume={48},
	ISSN={1572-9516},
	url={http://dx.doi.org/10.1007/s10701-018-0138-3},
	DOI={10.1007/s10701-018-0138-3},
	number={2},
	journal={Foundations of Physics},
	publisher={Springer Science and Business Media LLC},
	author={Loveridge, Leon and Miyadera, Takayuki and Busch, Paul},
	year={2018},
	month=Feb, pages={135–198} }

@article{brukner2018no,
   title={A No-Go Theorem for Observer-Independent Facts},
   volume={20},
   ISSN={1099-4300},
   url={http://dx.doi.org/10.3390/e20050350},
   DOI={10.3390/e20050350},
   number={5},
   journal={Entropy},
   publisher={MDPI AG},
   author={Brukner, Časlav},
   year={2018},
   month=May, pages={350} }

@article{frauchiger2018quantum,
   title={Quantum theory cannot consistently describe the use of itself},
   volume={9},
   ISSN={2041-1723},
   url={http://dx.doi.org/10.1038/s41467-018-05739-8},
   %DOI={10.1038/s41467-018-05739-8},
   number={1},
   journal={Nature Communications},
   publisher={Springer Science and Business Media LLC},
   author={Frauchiger, Daniela and Renner, Renato},
   year={2018},
   month=Sep }

@article{bong2020strong,
   title={A strong no-go theorem on the Wigner’s friend paradox},
   volume={16},
   ISSN={1745-2481},
   url={http://dx.doi.org/10.1038/s41567-020-0990-x},
   DOI={10.1038/s41567-020-0990-x},
   number={12},
   journal={Nature Physics},
   publisher={Springer Science and Business Media LLC},
   author={Bong, Kok-Wei and Utreras-Alarcón, Aníbal and Ghafari, Farzad and Liang, Yeong-Cherng and Tischler, Nora and Cavalcanti, Eric G. and Pryde, Geoff J. and Wiseman, Howard M.},
   year={2020},
   month=Aug, pages={1199–1205} }

@misc{vanrietvelde2026specifying,
      title={Specifying the operational meaning of quantum reference frames}, 
      author={Augustin Vanrietvelde},
      year={2026},
      eprint={2607.03417},
      archivePrefix={arXiv},
      primaryClass={quant-ph},
      url={https://arxiv.org/abs/2607.03417}, 
}

@misc{di2026adlam,
      title={Adlam's Frame: comment on "Wigner's Frame"}, 
      author={Andrea Di Biagio and Anne-Catherine de la Hamette},
      year={2026},
      eprint={2607.18383},
      archivePrefix={arXiv},
      primaryClass={quant-ph},
      url={https://arxiv.org/abs/2607.18383}
}

@article{donnelly2016local,
   title={Local subsystems in gauge theory and gravity},
   volume={2016},
   ISSN={1029-8479},
   url={http://dx.doi.org/10.1007/JHEP09(2016)102},
   DOI={10.1007/jhep09(2016)102},
   number={9},
   journal={Journal of High Energy Physics},
   publisher={Springer Science and Business Media LLC},
   author={Donnelly, William and Freidel, Laurent},
   year={2016},
   month=Sep }

@article{carrozza2022edge,
   title={Edge modes as reference frames and boundary actions from post-selection},
   volume={2022},
   ISSN={1029-8479},
   url={http://dx.doi.org/10.1007/JHEP02(2022)172},
   %DOI={10.1007/jhep02(2022)172},
   number={2},
   journal={Journal of High Energy Physics},
   publisher={Springer Science and Business Media LLC},
   author={Carrozza, Sylvain and Höhn, Philipp A.},
   year={2022},
   month=Feb }

\newpage

\appendix

\section{Using the QRF formalism in basic examples} \label{appendixA}
\noindent Let us consider some basic examples within the setups introduced in Sections~\ref{SEC:TheProblem} and~\ref{SEC:IntroToFormalism}. 
\subsection{Superpositions of two sharp position configurations}
Suppose Eve assigns the following pure external state for the setup $\mathbf{AS}$:
\begin{equation} \label{thetaphasestate} 
    \ket{\psi}^\mathbf{(E)}_\mathbf{AS}=\frac{1}{\sqrt{2}}\left(\ket{a_1}\ket{s_1}+e^{\iu \theta}\ket{a_2}\ket{s_2}\right)_\mathbf{A,S},
\end{equation}  
where $a_1,a_2,s \in \mathbb{Z}_N$. Using the formalism developed in Section~\ref{SEC:IntroToFormalism}, the state in the perspective of QRF $\mathbf{A}$ is 
\begin{align}\label{PerspectiveStateEx1}
&\rho= \mathcal{U}^\mathbf{AS}_{\to \mathbf{A}} \circ\mathcal{G}_\mathbf{A,S}\Big(\ket{\psi}\bra{\psi}^\mathbf{(E)}_\mathbf{AS}\Big)= \frac{1}{2N}\Bigg[\mathds{1}\otimes \left(\ket{-a_1+s_1}\bra{-a_1+s_1}+ \ket{-a_2+s_2}\bra{-a_2+s_2}\right)+ \notag
\\&+e^{\iu \hat{\Theta}(-a_1+a_2,\theta)} \otimes\ket{-a_2+s_2}\bra{-a_1+s_1} +e^{-\iu \hat{\Theta}(-a_1+a_2,\theta)}\otimes\ket{-a_1+s_1}\bra{-a_2+s_2} \Bigg]_\mathbf{\overline{S|A}, S|A},
\end{align} where 
\begin{equation} \label{ThetaDefinition}
    \hat{\Theta}(r,\theta):=  \left(\theta \mathds{1}+\frac{2\pi r}{N}\hat{p}_\mathbf{\overline{S|A}}\right)
\end{equation} and the convention $\hat{p}_\mathbf{\overline{S|A}}:= -\hat{p}_\mathbf{S|A'}$ is being used for the momentum operator of the extra particle. As expected of any invariant operator written in the factorisation preferred to $\mathbf{A}$ (see Section~\ref{SEC:IntroToFormalism}), the state~\eqref{PerspectiveStateEx1} only has elements that are diagonal in the momentum basis of the space $\mathbf{S|A'}$, which actually corresponds to the total momentum basis of the full setup $\mathbf{AS}$. 

A special case of the state above is the scenario in which $\mathbf{A}$ is in a quantum superposition of two positions and $\mathbf{S}$ is in a sharp position state, depicted in Fig.~\ref{FIG:CircleDiscreteTranslations} with $a_1=1, a_2=2$ and $s_1=s_2=s=5$. In that case, the state in Alice's perspective becomes
\begin{equation}\label{PerspectiveStateEx1withnumbers}
    \rho = \frac{1}{2N}\left[\mathds{1}\otimes \left(\ket{4}\bra{4}+\ket{3}\bra{3}\right)+ e^{\iu \hat{\Theta}(1,\theta)}\otimes\ket{3}\bra{4}+ e^{-\iu \hat{\Theta}(1,\theta)}\otimes\ket{4}\bra{3}\right]_\mathbf{\overline{S|A},S|A}.
\end{equation} From the state~\eqref{PerspectiveStateEx1withnumbers}, it is clear that part of the external information about the frame $\mathbf{A}$ is available in the perspective of Alice. Namely, if the frame $\mathbf{A}$ is in a pure quantum superposition with the phase $\theta$, this phase is accessible. Interestingly, however, the phase appears in the correlations between the systems $\mathbf{\overline{S|A}}$ and $\mathbf{S|A}$, and is therefore available only through joint measurements. If one exclusively measures the relative degrees of freedom, the statistics is described by the reduced state
\begin{equation}
    \operatorname{Tr}_\mathbf{S|A'}\left(\rho\right) =\frac{1}{2}\left(\ket{4}\bra{4}+ \ket{3}\bra{
3}\right)_\mathbf{S|A}.
\end{equation}

Another special case of~\eqref{thetaphasestate} is when $\mathbf{A}$ and $\mathbf{S}$ are perfectly entangled. That is, when $s_1-a_1=s_2-a_2=:d$. Then, the state~\eqref{PerspectiveStateEx1} in Alice's perspective becomes
\begin{equation}
    \rho=\frac{1}{2N} \left(\mathds{1}+ e^{\iu \hat{\Theta}(-a_1+a_2,\theta)}+ e^{-\iu \hat{\Theta}(-a_1+a_2,\theta)}\right)_\mathbf{\overline{S|A}}\otimes \ket{d}\bra{d}_\mathbf{S|A}.
\end{equation}
Therefore, if Eve describes a global state in which $\mathbf{AS}$ are entangled with phase $\theta$, while their separation is the same in both branches, this phase is also internally accessible. For this case, such information is completely contained in the extra particle subsystem. Indeed, the phases in both examples are invariant information and should not be discarded as they are internally available physical quantities in this translation-invariant setting (see \cite{delaHamette2026, Doat2026} for an operational example in which this information can be obtained internally through a collaborative strategy between observers).

\subsection{Superposition of two sharp total momentum configurations}
Let us now consider an example of external state different from~\eqref{thetaphasestate}, where the setup is very delocalised:
\begin{equation}\label{ExternalStateEx3}
    \ket{\psi}^\mathbf{(E)}_\mathbf{AS} = \frac{1}{\sqrt{2N}}\sum_{a=0}^{N-1}\left(e^{\frac{2\pi \iu}{N} a P_1}+e^{\frac{2\pi \iu}{N} a P_2}\right)\ket{a}_\mathbf{A}\ket{a+d}_\mathbf{S}, 
\end{equation} where $a,d,P_1,P_2 \in \mathbb{Z}_N$ and $P_1 \neq P_2$. This is a superposition state of two values of total momentum for $\mathbf{A}$ and $\mathbf{S}$. Indeed, if we apply the map $\mathcal{U}^\mathbf{AS}_\mathbf{\to A}$ and express the first factor in the momentum basis, we get:
\begin{equation}
\frac{1}{\sqrt{2}}\big(\ket{P_1}+\ket{P_2}\big)_\mathbf{S|A'}\otimes \ket{d}_{\mathbf{S|A}}.
\end{equation} When we proceed to apply the G-twirl, the resulting state is 
\begin{equation}
    \rho = \frac{1}{2}\big(\ket{P_1}\bra{P_1}+\ket{P_2}\bra{P_2}\big)_\mathbf{\overline{S|A}} \otimes \ket{d}\bra{d}_{\mathbf{S|A}}.
\end{equation} 
Here it becomes clear that the refactorisation realised by $\mathcal{U}^\mathbf{AS}_\mathbf{\to A}$ has the property of `isolating' the momentum superselection rule for translation invariant operators by making it act on the first factor only. Namely, the momentum of the first factor coincides with total momentum of $\mathbf{AS}$ and the information lost after the G-twirl is exactly the coherences of the state on the first factor in the momentum basis. Examples 1 and 2 are useful for highlighting what is the accessible information and how it appears in a perspective in the present formalism (see, for instance, ~\cite{Doat2026,delaHamette2026,Palumbo2026,CastroRuiz2025} for comparisons of different QRF approaches regarding internally accessible information).

\subsection{Jumping between a sharply localised and a superposed frame}
 To conclude, let us work an example involving two QRFs $\mathbf{A}$ and $\mathbf{B}$ used by observers Alice and Bob, as well as a system $\mathbf{S}.$ If $\mathbf{A}$ and $\mathbf{S}$ are in the fixed positions 1 and 5, respectively, while $\mathbf{B}$ is in a superposition of being at 3 and 4, the states in the perspectives of Alice and Bob and the transformation $\mathcal{S}_\mathbf{A\to B}$ between them are given below, where $\hat{\Theta}$ is defined in~\eqref{ThetaDefinition}.

\begin{center}

\tikzset{every picture/.style={line width=0.75pt}} 

\begin{tikzpicture}[x=0.5pt,y=0.5pt,yscale=-1,xscale=1]

\draw [line width=1.5]    (188,278.44) -- (448,278.44) -- (449,278.44) ;
\draw [shift={(453,278.44)}, rotate = 180] [fill={rgb, 255:red, 0; green, 0; blue, 0 }  ][line width=0.08]  [draw opacity=0] (11.61,-5.58) -- (0,0) -- (11.61,5.58) -- cycle    ;
\draw [shift={(184,278.44)}, rotate = 0] [fill={rgb, 255:red, 0; green, 0; blue, 0 }  ][line width=0.08]  [draw opacity=0] (11.61,-5.58) -- (0,0) -- (11.61,5.58) -- cycle    ;
\draw [color={rgb, 255:red, 208; green, 2; blue, 27 }  ,draw opacity=1 ][line width=1.5]    (318,90.25) -- (176.6,255.83) ;
\draw [shift={(174,258.88)}, rotate = 310.5] [fill={rgb, 255:red, 208; green, 2; blue, 27 }  ,fill opacity=1 ][line width=0.08]  [draw opacity=0] (11.61,-5.58) -- (0,0) -- (11.61,5.58) -- cycle    ;
\draw [color={rgb, 255:red, 166; green, 77; blue, 234 }  ,draw opacity=1 ][line width=1.5]    (318,90.25) -- (464.34,254.89) ;
\draw [shift={(467,257.88)}, rotate = 228.37] [fill={rgb, 255:red, 166; green, 77; blue, 234 }  ,fill opacity=1 ][line width=0.08]  [draw opacity=0] (11.61,-5.58) -- (0,0) -- (11.61,5.58) -- cycle    ;
\draw [color={rgb, 255:red, 245; green, 166; blue, 35 }  ,draw opacity=1 ][line width=1.5]    (677.25,219.47) .. controls (565.62,217.43) and (568.37,54.57) .. (679.08,56.15) .. controls (789.79,57.72) and (786.99,219.08) .. (677.25,219.47)(625.76,207.91) -- (630.34,201.35)(594.26,162.92) -- (601.96,160.73)(595,110) -- (602.63,112.4)(626.51,66.79) -- (631.01,73.4)(679.13,52.15) -- (679.02,60.15)(730.15,67.38) -- (725.58,73.94)(761.03,110.73) -- (753.37,113.02)(760.65,164.63) -- (753.01,162.24)(728.31,208.8) -- (723.83,202.17)(677.26,223.47) -- (677.23,215.47) ;
\draw  [draw opacity=0][fill={rgb, 255:red, 0; green, 95; blue, 86 }  ,fill opacity=1 ] (674.55,213.73) .. controls (677.55,212.26) and (681.19,213.64) .. (682.68,216.81) .. controls (684.16,219.98) and (682.94,223.74) .. (679.94,225.21) .. controls (676.94,226.68) and (673.3,225.3) .. (671.81,222.13) .. controls (670.33,218.96) and (671.55,215.2) .. (674.55,213.73) -- cycle ;
\draw  [color={rgb, 255:red, 166; green, 77; blue, 234 }  ,draw opacity=1 ][fill={rgb, 255:red, 166; green, 77; blue, 234 }  ,fill opacity=1 ] (756.84,157.9) .. controls (760.03,157.3) and (763.13,159.56) .. (763.75,162.95) .. controls (764.37,166.34) and (762.28,169.58) .. (759.08,170.18) .. controls (755.88,170.79) and (752.78,168.52) .. (752.16,165.13) .. controls (751.55,161.74) and (753.64,158.5) .. (756.84,157.9) -- cycle ;
\draw [color={rgb, 255:red, 166; green, 77; blue, 234 }  ,draw opacity=1 ][fill={rgb, 255:red, 166; green, 77; blue, 234 }  ,fill opacity=1 ][line width=1.5]    (762.67,165.81) -- (790.35,176.16) ;
\draw  [color={rgb, 255:red, 166; green, 77; blue, 234 }  ,draw opacity=1 ][fill={rgb, 255:red, 166; green, 77; blue, 234 }  ,fill opacity=1 ][line width=1.5]  (800.4,179.71) -- (786.66,186.59) -- (790.35,176.16) -- cycle ;

\draw  [draw opacity=0][fill={rgb, 255:red, 255; green, 255; blue, 255 }  ,fill opacity=0.3 ] (756.59,156.55) .. controls (760.4,155.83) and (764.1,158.6) .. (764.86,162.74) .. controls (765.61,166.88) and (763.14,170.81) .. (759.33,171.53) .. controls (755.52,172.25) and (751.81,169.48) .. (751.06,165.34) .. controls (750.3,161.2) and (752.78,157.27) .. (756.59,156.55) -- cycle ;
\draw  [draw opacity=0][fill={rgb, 255:red, 255; green, 255; blue, 255 }  ,fill opacity=0.3 ] (764.86,164.74) -- (803.46,178.05) -- (798.04,193.75) -- (759.44,180.44) -- cycle ;

\draw  [draw opacity=0][fill={rgb, 255:red, 208; green, 2; blue, 27 }  ,fill opacity=1 ] (720.29,70.92) .. controls (720.38,67.65) and (723.23,65.08) .. (726.66,65.17) .. controls (730.09,65.27) and (732.8,68) .. (732.71,71.27) .. controls (732.61,74.54) and (729.76,77.11) .. (726.33,77.01) .. controls (722.9,76.92) and (720.19,74.19) .. (720.29,70.92) -- cycle ;
\draw [color={rgb, 255:red, 208; green, 2; blue, 27 }  ,draw opacity=1 ][fill={rgb, 255:red, 208; green, 2; blue, 27 }  ,fill opacity=1 ][line width=1.5]    (729.22,66.83) -- (745.14,41.77) ;
\draw  [color={rgb, 255:red, 208; green, 2; blue, 27 }  ,draw opacity=1 ][fill={rgb, 255:red, 208; green, 2; blue, 27 }  ,fill opacity=1 ][line width=1.5]  (750.73,32.62) -- (754.5,47.56) -- (745.14,41.77) -- cycle ;

\draw  [color={rgb, 255:red, 166; green, 77; blue, 234 }  ,draw opacity=1 ][fill={rgb, 255:red, 166; green, 77; blue, 234 }  ,fill opacity=1 ] (730.04,200.56) .. controls (733.02,201.85) and (734.33,205.46) .. (732.96,208.63) .. controls (731.59,211.8) and (728.06,213.33) .. (725.08,212.04) .. controls (722.09,210.75) and (720.78,207.14) .. (722.15,203.97) .. controls (723.53,200.8) and (727.06,199.27) .. (730.04,200.56) -- cycle ;
\draw [color={rgb, 255:red, 166; green, 77; blue, 234 }  ,draw opacity=1 ][fill={rgb, 255:red, 166; green, 77; blue, 234 }  ,fill opacity=1 ][line width=1.5]    (730.48,210.41) -- (747.66,234.5) ;
\draw  [color={rgb, 255:red, 166; green, 77; blue, 234 }  ,draw opacity=1 ][fill={rgb, 255:red, 166; green, 77; blue, 234 }  ,fill opacity=1 ][line width=1.5]  (754.01,243.08) -- (738.8,241.12) -- (747.66,234.5) -- cycle ;

\draw  [draw opacity=0][fill={rgb, 255:red, 255; green, 255; blue, 255 }  ,fill opacity=0.3 ] (730.59,199.3) .. controls (734.14,200.83) and (735.67,205.21) .. (733.99,209.08) .. controls (732.32,212.94) and (728.08,214.84) .. (724.53,213.3) .. controls (720.98,211.77) and (719.45,207.39) .. (721.12,203.53) .. controls (722.79,199.66) and (727.03,197.77) .. (730.59,199.3) -- cycle ;
\draw  [draw opacity=0][fill={rgb, 255:red, 255; green, 255; blue, 255 }  ,fill opacity=0.3 ] (733.99,209.08) -- (757.92,240.87) -- (744.71,250.9) -- (720.78,219.11) -- cycle ;

\draw (16,7.2) node [anchor=north west][inner sep=0.75pt]  [font=\normalsize]  {$\ \left[\left[\ket{1} \otimes \frac{\ket{3} +e^{\mathrm{i} \theta }\ket{4}}{\sqrt{2}} \otimes \ket{5}\right]\right]_{\mathbf{ABS}}^{\mathbf{( E)}}$};
\draw (93.67,158.07) node [anchor=north west][inner sep=0.75pt]  [font=\normalsize]  {${\displaystyle \mathcal{\textcolor[rgb]{0.82,0.01,0.11}{U}}\textcolor[rgb]{0.82,0.01,0.11}{_{\mathbf{\rightarrow A}}^{\mathbf{ABS}}} \circ \mathcal{G}_{\mathbf{A,B,S}}}$};
\draw (423.67,158.07) node [anchor=north west][inner sep=0.75pt]  [font=\normalsize]  {$\mathcal{\textcolor[rgb]{0.65,0.3,0.92}{U}}\mathrm{\textcolor[rgb]{0.65,0.3,0.92}{_{\mathbf{\rightarrow B}}^{\mathbf{ABS}}}} \circ \mathcal{G}_{\mathbf{A,B,S}}$};
\draw (294,250.79) node [anchor=north west][inner sep=0.75pt]  [font=\normalsize]  {$\mathcal{S}\mathrm{_{\mathbf{A\rightarrow B}}}$};
\draw (3,329.68) node [anchor=north west][inner sep=0.75pt]  [font=\normalsize]  {$\left(\frac{\mathds{1}}{\mathnormal{N}}\right)_{\overline{\mathbf{BS|A}}} \otimes \left[\left[\frac{\ket{2} +e^{\mathrm{i} \theta }\ket{3}}{\sqrt{2}} \otimes \ket{4}\right]\right]_{\ \mathbf{BS|A}}$};
\draw (310.67,59.86) node [anchor=north west][inner sep=0.75pt]  [font=\large]  {$\psi $};
\draw (151,267.68) node [anchor=north west][inner sep=0.75pt]  [font=\large]  {$\ \rho $};
\draw (656.83,225.81) node [anchor=north west][inner sep=0.75pt]  [font=\normalsize,color={rgb, 255:red, 65; green, 117; blue, 5 }  ,opacity=1 ,rotate=-359.99] [align=left] {$\displaystyle \mathbf{\textcolor[rgb]{0,0.37,0.34}{S}}$};
\draw (714.72,80.35) node [anchor=north west][inner sep=0.75pt]  [color={rgb, 255:red, 245; green, 166; blue, 35 }  ,opacity=1 ,rotate=-359.99]  {$1$};
\draw (734.2,149.86) node [anchor=north west][inner sep=0.75pt]  [color={rgb, 255:red, 245; green, 166; blue, 35 }  ,opacity=1 ,rotate=-359.99]  {$3$};
\draw (671.91,192.55) node [anchor=north west][inner sep=0.75pt]  [color={rgb, 255:red, 245; green, 166; blue, 35 }  ,opacity=1 ,rotate=-359.99]  {$5$};
\draw (712.75,179.04) node [anchor=north west][inner sep=0.75pt]  [color={rgb, 255:red, 245; green, 166; blue, 35 }  ,opacity=1 ,rotate=-359.99]  {$4$};
\draw (672.61,63.02) node [anchor=north west][inner sep=0.75pt]  [color={rgb, 255:red, 245; green, 166; blue, 35 }  ,opacity=1 ,rotate=-359.99]  {$0$};
\draw (743.88,60.01) node [anchor=north west][inner sep=0.75pt]  [font=\normalsize,color={rgb, 255:red, 208; green, 2; blue, 27 }  ,opacity=1 ,rotate=-359.99] [align=left] {$\displaystyle \mathbf{A}$};
\draw (767.8,202.44) node [anchor=north west][inner sep=0.75pt]  [font=\normalsize,color={rgb, 255:red, 166; green, 77; blue, 234 }  ,opacity=1 ,rotate=-1.88] [align=left] {$\displaystyle \mathbf{B}$};
\draw (734.2,109.86) node [anchor=north west][inner sep=0.75pt]  [color={rgb, 255:red, 245; green, 166; blue, 35 }  ,opacity=1 ,rotate=-359.99]  {$2$};
\draw (379,313.84) node [anchor=north west][inner sep=0.75pt]  [font=\normalsize]  {$ \begin{array}{l}
\Biggl[\frac{\mathds{1}}{N} \otimes \left(\frac{\ket{-2}\bra{-2} \otimes \ket{2}\bra{2} +\ket{-3}\bra{-3} \otimes \ket{1}\bra{1}}{2}\right) +\\
\ \ \ \ \ \ \ \ \ \ \ \ \ \ +\frac{\mathrm{e^{\mathrm{i}\hat{\Theta }( 1,\theta )}}}{2N} \otimes \left(\ket{-3}\bra{-2} \otimes \ket{1}\bra{2}\right) \ \ +c.c.\Biggr]_{\mathbf{\overline{AS|B} ,AS|B}} \ 
\end{array}$};
\draw (463,263.84) node [anchor=north west][inner sep=0.75pt]  [font=\large]  {$\sigma \ $};
\draw    (148,280.77) .. controls (121.26,284.35) and (124.3,311.42) .. (130.21,322.83) ;
\draw [shift={(131.69,325.28)}, rotate = 234.28] [fill={rgb, 255:red, 0; green, 0; blue, 0 }  ][line width=0.08]  [draw opacity=0] (7.14,-3.43) -- (0,0) -- (7.14,3.43) -- cycle    ;
\draw    (484,272.97) .. controls (505.35,270.01) and (531.88,281.48) .. (543.02,307.03) ;
\draw [shift={(544,309.44)}, rotate = 248.97] [fill={rgb, 255:red, 0; green, 0; blue, 0 }  ][line width=0.08]  [draw opacity=0] (7.14,-3.43) -- (0,0) -- (7.14,3.43) -- cycle    ;
\draw    (314.18,55.46) .. controls (314.49,40.35) and (299.45,37.06) .. (267.97,40.28) ;
\draw [shift={(265,40.59)}, rotate = 353.66] [fill={rgb, 255:red, 0; green, 0; blue, 0 }  ][line width=0.08]  [draw opacity=0] (7.14,-3.43) -- (0,0) -- (7.14,3.43) -- cycle    ;

\end{tikzpicture}

\end{center}

\section{Connection to `internal  approaches' to QRFs} \label{appendixB}

Here we show that the classicalisation map presented in Section~\ref{SEC:AddingWClassicalisation} acts like the QRF transformation of the approach~\cite{Giacomini2019} when restricted to the relative subsystems, while acknowledging and preserving the extra particle subsystem. Classicalisation begins with the internalisation of the description given by the external observer, Eve. Her implicit frame becomes an explicit system $\mathbf{E}$ according to the new external observer, Francis. This procedure induces a reinterpretation of the subsystem $\mathbf{AS}^\mathbf{(E)}$ as the relative subsystem $\mathbf{AS|E}^\mathbf{(F)}$, as described in Section~\ref{SEC:InternalisingExternalFrames}. We then apply $C_\mathbf{A}\circ \mathcal{S}_\mathbf{E\to A}$, where $\mathcal{S}_\mathbf{E\to A}$ is the jumping map from $\mathbf{E}$ to $\mathbf{A}$ in the setup $\mathbf{EAS}$      and $\mathcal{C}_\mathbf{A}$ is defined as
\begin{equation}\label{Ca}
    \mathcal{C}_\mathbf{A} = C_\mathbf{A}(\cdot)C_\mathbf{A}^\dagger, \qquad C_\mathbf{A}= \sum_g \ket{g}\bra{g}_\mathbf{E|A}  \otimes \mathds{1}_\mathbf{S|A}\otimes R_g^{\dagger}.
\end{equation}
In exponential form, we know that
\begin{equation}
    \mathcal{S}_\mathbf{E\to A} = S_\mathbf{E\to A} (\cdot) S^{\dagger}_\mathbf{E\to A}, \qquad S_\mathbf{E\to A}=P_\mathbf{E\to A} e^{\frac{2\pi i}{N}\hat{x}_\mathbf{A|E}\otimes\left(\hat{p}_\mathbf{S|E}+\hat{p}_\mathbf{\overline{AS|E}}\right)},
\end{equation}
where $e^{-\frac{2\pi \iu}{N}x\hat{p}_\mathbf{\overline{AS|E}}}= R_x=L^\dagger_x$. By definition, we have
\begin{equation}e^{+\frac{2\pi \iu}{N} \hat{x}_\mathbf{A|E}\otimes\left(\hat{p}_\mathbf{S|E}+\hat{p}_\mathbf{\overline{AS|E}}\right)}= \sum_g \left[\ket{g}\bra{g}\otimes U^{\dagger}_g\otimes R^{\dagger}_g\right]_\mathbf{A|E, S|E, \overline{AS|E}}\end{equation}
Therefore, the unitary in~\eqref{Ca} can be written as
\begin{equation}
    C_\mathbf{A} = e^{\frac{2\pi \iu}{N}\hat{x}_\mathbf{E|A}\otimes \hat{p}_\mathbf{\overline{ES|A}}}\otimes\mathds{1}_\mathbf{\mathbf{S|A}},
\end{equation}
and the resulting total map can be written as
\begin{align}
    C_\mathbf{A}\circ \mathcal{S}_\mathbf{E\to A}&= e ^{\frac{2\pi \iu}{N}\hat{x}_\mathbf{E|A}\otimes \hat{p}_\mathbf{\overline{ES|A}}} \circ P_\mathbf{E\to A} \circ e^{\frac{2\pi \iu}{N}\hat{x}_\mathbf{A|E}\otimes\left(\hat{p}_\mathbf{S|E}+\hat{p}_\mathbf{\overline{AS|E}}\right)}\nonumber
    \\ 
    &= P_\mathbf{E\to A} e^{\frac{2\pi \iu}{N}\hat{x}_\mathbf{A|E}\otimes\hat{p}_\mathbf{S|E}} \otimes \mathds{1}_\mathbf{\overline{AS|E}},
\end{align}
where $P_\mathbf{E\to A}$ is the parity-swap operator defined in~\cite{Giacomini2019}. The transformation on the spaces labelled $\mathbf{A|E, S|E}$ therefore corresponds to the internal transformation between perspectives of Ref.~\cite{Giacomini2019} between 2 agents $\mathbf{E}$ and $\mathbf{A}$ in a setup with 3 systems, $\mathbf{E},$ $\mathbf A$ and $\mathbf S$. 
For example, if we have an initial pure state
\[\left(\frac{\mathds{1}}{N}\right)_\mathbf{\overline{AS|E}}\otimes \left[\!\left[\frac{\ket{a_1}\ket{s_1}+ e^{\iu \theta}\ket{a_2}\ket{s_2}}{\sqrt{2}}\right]\!\right]_\mathbf{AS|E},\] the procedure maps it to the following state:
\[\left(\frac{\mathds{1}}{N}\right)_\mathbf{\overline{ES|A}} \otimes \left[\!\left[\frac{\ket{-a_1}\ket{-a_1+s_1}+ e^{i\theta}\ket{-a_2}\ket{-a_2+s_2}}{\sqrt{2}}\right]\!\right]_\mathbf{E|A, S|A}.\] For the special case in which $s_1=s_2=s$, $\mathbf A$ is in a separable superposition state for Eve, while Alice assigns an entangled state to $\mathbf E$ and $\mathbf S$, showcasing the predicted frame dependence of superpositions and entanglement of the approach~\cite{Giacomini2019}. 

In the present work, we are interested in including $\mathbf{E}$ in a way that the state of its extra particle is  $\left(\mathds{1}/N\right)_\mathbf{\overline{AS|E}}$ because this state has a special status (cf. Sections~\ref{SEC:Adding and removing relative subsystems} and~\ref{SEC:AddingWClassicalisation}), which is copied into $\mathbf{A}$'s extra particle after $\mathcal{C}_\mathbf{A}$ is applied. This gives a new unambiguous state-independent notion of adding and removing relative subsystems from $\mathbf A$'s perspective. However, we note that even if Eve is internalised in other states, the reduced state in $\mathbf{A|E},\mathbf{S|E}$ still transforms as in~\cite{Giacomini2019}.

\section{Observers' agreement on perspective states and operators}\label{appendixInternalisation}

This appendix proves the statements of Section~\ref{SEC:InternalisingExternalFrames}. Given any internalisation $\tilde{\phi}^\mathbf{(F)}_\mathbf{E_1E_2AS}$ consistent with the state Eve 1 assigns to Alice's perspective,  $\rho^\mathbf{(E_1)}_\mathbf{\overline{S|A},S|A}$, it is true that the states
\begin{equation}
   \rho^\mathbf{(F)}_\mathbf{\overline{S|A},S|A},  \quad \rho^\mathbf{(E_2)}_\mathbf{\overline{S|A},S|A}
\end{equation}
assigned to Alice's perspective according to the internal observer Eve 2 and to the external observer Francis, respectively, have identical matrix representations which coincide with those assigned by Eve 1. 

An internalisation $\tilde{\phi}^\mathbf{(F)}_\mathbf{E_1E_2AS}$ is consistent with the state assignment $\rho^\mathbf{(E_1)}_\mathbf{\overline{S|A},S|A}$ by Eve 1 if the state
\begin{equation}\label{DefPsiTilde}
\tilde{\psi}_\mathbf{\overline{E_2AS|E_1},E_2AS|E_1}:= \mathcal{U}^\mathbf{E_1E_2AS}_\mathbf{\to E_1}\circ \mathcal{G}_\mathbf{E_1,E_2,A,S} \left(\tilde{\phi}^\mathbf{(F)}_\mathbf{E_1E_2AS}\right)
\end{equation} of $\mathbf{E_1}$'s perspective leads to $\rho^\mathbf{(E_1)}_\mathbf{\overline{S|A},S|A}$ when the formalism is applied. That is, when
\begin{equation} \label{internalisationConsistency}
\mathcal{U}^\mathbf{AS|E_1}_\mathbf{\to A|E_1}\circ \mathcal{G}_\mathbf{A|E_1,S|E_1} \circ \operatorname{Tr}_\mathbf{E_2AS|E_1', E_2|E_1}\left(\tilde{\psi}_\mathbf{\overline{E_2AS|E_1},E_2AS|E_1}\right) \cong \rho^\mathbf{(E_1)}_\mathbf{\overline{S|A},S|A}
\end{equation} holds given the identification~\eqref{eq:identification}.

\subsection{Matrix elements of states}
Let us begin by computing the matrix elements of the state assigned by Francis, which is obtained by his use of the formalism in the subsystem $\mathbf{AS}^\mathbf{(F)}$. We can write
\begin{align}
&\bra{x',y'}\ \rho^\mathbf{(F)}_\mathbf{\overline{S|A},S|A}\ket{x, y}= \nonumber
\\ \nonumber
&=\operatorname{Tr}\left[\mathcal{U}^\mathbf{AS}_\mathbf{\to A}\circ\mathcal{G}_\mathbf{A,S}\circ\operatorname{Tr}_\mathbf{E_1E_2}\left(\tilde{\phi}^\mathbf{(F)}_\mathbf{E_1E_2AS}\right) \cdot\left(\ket{x}\bra{x'}\otimes\ket{y}\bra{y'}\right)\right]
\\ \nonumber
&=\operatorname{Tr}\left[\left[\mathcal{I}_\mathbf{E_1,E_2} \otimes \left(\mathcal{U}^\mathbf{AS}_\mathbf{\to A}\circ\mathcal{G}_\mathbf{A,S}\right)\right]\left(\tilde{\phi}^\mathbf{(F)}_\mathbf{E_1E_2AS}\right)\cdot\left(\mathds{1}_\mathbf{E_1,E_2}\otimes\ket{x}\bra{x'}\otimes\ket{y}\bra{y'}\right)\right] \\ \nonumber
&=\operatorname{Tr}\left[\tilde{\phi}^\mathbf{(F)}_\mathbf{E_1E_2AS}\cdot \left[\mathcal{I}_\mathbf{E_1,E_2} \otimes \left(\mathcal{G}_\mathbf{A,S} \circ \left(\mathcal{U}^\mathbf{AS}_\mathbf{\to A}\right)^{\dagger}\right)\right]\left(\mathds{1}_\mathbf{E_1,E_2}\otimes\ket{x}\bra{x'}\otimes\ket{y}\bra{y'}\right)\right]\\ \nonumber
&=\operatorname{Tr}\left[\tilde{\phi}^\mathbf{(F)}_\mathbf{E_1E_2AS}\cdot \left(\mathds{1}_\mathbf{E_1,E_2}\otimes \mathcal{G}_\mathbf{A,S}\left(\ket{x}\bra{x'}\otimes\ket{x+y}\bra{x'+y'}\right)\right)\right]\\ 
&=\operatorname{Tr}\left[\mathcal{G}_\mathbf{E_1,E_2,A,S}\left(\tilde{\phi}^\mathbf{(F)}_\mathbf{E_1E_2AS}\right)\cdot \left(\mathds{1}_\mathbf{E_1,E_2}\otimes \mathcal{G}_\mathbf{A,S}\left(\ket{x}\bra{x'}\otimes\ket{x+y}\bra{x'+y'}\right)\right)\right]
\end{align}
where we are allowed to apply $\mathcal{G}_\mathbf{E_1,E_2,A,S}$ to the state of Francis without changing the result because the matrix on the right-hand side is globally invariant. We can also use the cyclicity of the trace to apply the unitary operation $\mathcal{U}^\mathbf{E_1E_2AS}_\mathbf{\to \mathbf{E_1}}$ on both terms and obtain an expression with the state~\eqref{DefPsiTilde}:
\begin{align}
&\bra{x',y'}\ \rho^\mathbf{(F)}_\mathbf{\overline{S|A},S|A}\ket{x, y}= \nonumber
\\ \nonumber
&=\operatorname{Tr}\left[\tilde{\psi}_\mathbf{\overline{E_2AS|E_1},E_2AS|E_1}\cdot \mathcal{U}^\mathbf{E_1E_2AS}_\mathbf{\to \mathbf{E_1}}\left(\mathds{1}_\mathbf{E_1,E_2}\otimes \mathcal{G}_\mathbf{A,S}\left(\ket{x}\bra{x'}\otimes\ket{x+y}\bra{x'+y'}\right)\right)\right]
\\ 
&=\operatorname{Tr}\left[\tilde{\psi}_\mathbf{\overline{E_2AS|E_1},E_2AS|E_1}\cdot \left(\mathds{1}_\mathbf{E_2AS|E_1',E_2|E_1}\otimes \mathcal{G}_\mathbf{A|E_1,S|E_1}\left(\ket{x}\bra{x'}\otimes\ket{x+y}\bra{x'+y'}\right)\right)\right] \label{E1directToE2}
\end{align}
\begin{align} \nonumber
&=\operatorname{Tr}\Bigg[\tilde{\psi}_\mathbf{\overline{E_2AS|E_1},E_2AS|E_1} \\ 
&\hspace{1.5cm}\cdot \left[\mathcal{I}_\mathbf{E_2AS|E_1',E_2|E_1}\otimes \mathcal{G}_\mathbf{A|E_1,S|E_1}\circ\left(\mathcal{U}^\mathbf{AS|E_1}_\mathbf{\to A|E_1}\right)^\dagger\right]
\left(\mathds{1}_\mathbf{E_2AS|E_1',E_2|E_1}\otimes\ket{x}\bra{x'}\otimes\ket{y}\bra{y'}\right)\Bigg]  \nonumber
\\
&=\operatorname{Tr}\left[\mathcal{U}^\mathbf{AS|E_1}_\mathbf{\to A|E_1}\circ \mathcal{G}_\mathbf{A|E_1,S|E_1} \circ \operatorname{Tr}_\mathbf{E_2AS|E_1', E_2|E_1}\left(\tilde{\psi}_\mathbf{\overline{E_2AS|E_1},E_2AS|E_1}\right)\cdot 
\left(\ket{x}\bra{x'}\otimes\ket{y}\bra{y'}\right)\right] \nonumber
\\
&=\operatorname{Tr}\left[\rho^\mathbf{(E_1)}_\mathbf{\overline{S|A},S|A}\cdot 
\left(\ket{x}\bra{x'}\otimes\ket{y}\bra{y'}\right)\right] =\bra{x',y'}\rho^\mathbf{(E_1)}_\mathbf{\overline{S|A},S|A}\ket{x,y}, \label{EquivalenceObserversPerspA}
\end{align}
where the last step follows directly from~\eqref{internalisationConsistency}. Because the above holds for Francis and the analogous derivation can be made by exchanging the indices $\mathbf{E_1}$ and $\mathbf{E_2}$, we have the desired result,~\eqref{AgreementStatement}. Note that the result for the state according to $(\mathbf{E_2})$ can also be obtained directly from $\mathbf{E_1}$'s perspective. It suffices to use the cyclicity of the trace to repeatedly apply the unitary QRF transformation $\mathcal{S}_\mathbf{E_1\to E_2}$, for instance at the step~\eqref{E1directToE2}:
\begin{align}
&\operatorname{Tr}\left[\mathcal{S}_\mathbf{E_1 \to E_2} \left(\tilde{\psi}_\mathbf{\overline{E_2AS|E_1},E_2AS|E_1}\right)\cdot \mathcal{S}_\mathbf{E_1 \to E_2} \left(\mathds{1}_\mathbf{E_2AS|E_1',E_2|E_1}\otimes \mathcal{G}_\mathbf{A|E_1,S|E_1}\left(\ket{x}\bra{x'}\otimes\ket{x+y}\bra{x'+y'}\right)\right)\right] \nonumber\\ \nonumber
&= \operatorname{Tr}\left[\tilde{\eta}_\mathbf{\overline{E_1AS|E_2},E_1AS|E_2}\cdot \left(\mathds{1}_\mathbf{E_1AS|E_2',E_1|E_2}\otimes \mathcal{G}_\mathbf{A|E_2,S|E_2}\left(\ket{x}\bra{x'}\otimes\ket{x+y}\bra{x'+y'}\right)\right)\right]\\
&=\operatorname{Tr}\left[\mathcal{U}^\mathbf{AS|E_2}_\mathbf{\to A|E_2}\circ \mathcal{G}_\mathbf{A|E_2,S|E_2} \circ \operatorname{Tr}_\mathbf{E_1AS|E_2', E_1|E_2}\left(\tilde{\eta}_\mathbf{\overline{E_1AS|E_2},E_1AS|E_2}\right)\cdot 
\left(\ket{x}\bra{x'}\otimes\ket{y}\bra{y'}\right)\right] \nonumber
\\
&=:\operatorname{Tr}\left[\rho^\mathbf{(E_2)}_\mathbf{\overline{S|A},S|A}\cdot 
\left(\ket{x}\bra{x'}\otimes\ket{y}\bra{y'}\right)\right] =\bra{x',y'}\rho^\mathbf{(E_2)}_\mathbf{\overline{S|A},S|A}\ket{x,y}. \label{EquivalenceObserversPerspAAlt}
\end{align}

\subsection{Matrix elements of operators}

The same agreement is true for the representation of operators. When defined by Eve 1, an operator acting on $\mathbf{A}$'s perspective can always be written as
\begin{equation}
O^\mathbf{(E_1)}_\mathbf{\overline{S|A}, S|A} = \mathcal{U}^\mathbf{AS}_\mathbf{\to A}\circ\mathcal{G}_\mathbf{A,S}(Q^\mathbf{(E_1)}_\mathbf{AS})
\end{equation}
for some general operator $Q^\mathbf{(E_1)}_\mathbf{AS}$ in the description of Eve 1.
We can then write this operator as acting on a subsystem relative to the internalised frame $\mathbf{E_1}$ in Francis' description:
\begin{equation}
O^\mathbf{(E_1)}_\mathbf{\overline{S|A}, S|A} \cong 
\left(\mathcal{U}^\mathbf{AS|E_1}_\mathbf{\to A|E_1}\circ \mathcal{G}_\mathbf{A|E_1,S|E_1}\right) (Q_\mathbf{AS|E_1}),
\end{equation} where $Q_\mathbf{AS|E_1}\cong Q^\mathbf{(E_1)}_\mathbf{AS}$ through an identification analogous to~\eqref{eq:identification}. Therefore, the total operator for Francis before Eve 1 transforms to the perspective of $\mathbf{A}$ is $\mathds{1}_\mathbf{E_2AS|E_1',E_2|E_1}\otimes Q_\mathbf{AS|E_1}$. We can write that global operator in Francis' factorisation and only then transform to Alice's perspective in $\mathbf{AS}^\mathbf{(F)}$, obtaining
\begin{equation}
O^\mathbf{(F)}_\mathbf{\overline{S|A},S|A} := \mathcal{U}^\mathbf{AS}_\mathbf{\to A}\circ\mathcal{G}_\mathbf{A,S}\circ\operatorname{Tr}_\mathbf{E_1E_2}\circ\left(\mathcal{U}^\mathbf{E_1E_2AS}_\mathbf{\to E_1}\right)^\dagger\left(\mathds{1}_\mathbf{E_2AS|E_1',E_2|E_1}\otimes Q_\mathbf{AS|E_1}\right).
\end{equation}

Similarly, we can instead jump directly to the internal perspective of $\mathbf{E}_2$
and apply the formalism from there to find the form of the same operator in Alice's perspective as assigned by Eve 2. The operator thus obtained is
\begin{equation}
O^\mathbf{(E_2)}_\mathbf{\overline{S|A},S|A} \cong \mathcal{U}^\mathbf{AS|E_2}_\mathbf{\to A|E_2}\circ\mathcal{G}_\mathbf{A|E_2,S|E_2}\circ\operatorname{Tr}_\mathbf{E_1AS|E_2', E_1|E_2}\circ \mathcal{S}_\mathbf{E_1 \to E_2}\left(\mathds{1}_\mathbf{E_2AS|E_1',E_2|E_1}\otimes Q_\mathbf{AS|E_1}\right)
\end{equation}

Then, by the same sequence of manipulations as in~\eqref{EquivalenceObserversPerspA}, one shows that
\begin{equation}
\bra{x',y'}O^\mathbf{(E_1)}_\mathbf{\overline{S|A}, S|A}\ket{x,y}= \bra{x',y'}O^\mathbf{(F)}_\mathbf{\overline{S|A}, S|A}\ket{x,y}= \bra{x',y'}O^\mathbf{(E_2)}_\mathbf{\overline{S|A}, S|A}\ket{x,y}.
\end{equation}

\end{document}